\documentclass[12pt]{article}

\usepackage{scicite}

\usepackage{lscape}

\usepackage{times}
\usepackage{amsmath}
\usepackage{morefloats}
\usepackage{ctable}
\usepackage{rotating}
\usepackage{changepage}
\usepackage{graphicx}
\usepackage{setspace} 
\usepackage{lscape} 
\newenvironment{sciabstract}{%
\begin{quote} \bf}
{\end{quote}}

\renewcommand\refname{References and Notes}

\newcounter{lastnote}

\title{Using digital traces to build prospective and real-time county-level early warning systems to anticipate COVID-19 outbreaks in the United States}

\author
{Lucas M. Stolerman,$^{\dagger 1,2,3}$ Leonardo Clemente,$^{\dagger 1}$ Canelle Poirier,$^{1,2}$ Kris V. Parag,$^{4}$ \\ Atreyee Majumder,$^{5}$ Serge Masyn,$^{5}$ Bernd Resch,$^{7,8}$ Mauricio Santillana,$^{1,2,6^\ast}$ \\
\\
\normalsize{$^{1}$Computational Health Informatics Program, Boston Children's Hospital,Boston, MA, USA}\\

\normalsize{$^{2}$Department of Pediatrics, Harvard Medical School, Boston, MA, USA}\\
\normalsize{$^{3}$Department of Mathematics, Oklahoma State University, Stillwater, OK, USA}\\

\normalsize{$^{4}$NIHR Health Protection Research Unit,} 
\\
\normalsize{Behavioural Science and Evaluation, University of Bristol, Bristol, UK.}\\

\normalsize{$^{5}$Global Public Health, Janssen R\&D}\\

\normalsize{$^{6}$Harvard University, T.H. Chan School of Public Health, Boston, MA, USA}\\

\normalsize{$^{7}$Department of Geoinformatics - Z-GIS, University of Salzburg, Salzburg, Austria.}\\

\normalsize{$^{8}$ Center for Geographic Analysis, Harvard University, Cambridge, MA, USA.}\\

\\
\normalsize{$^\dagger$ These authors contributed equally to this study.}
\\
\normalsize{$^\ast$To whom correspondence should be addressed; E-mail:  msantill@g.harvard.edu.}
}

\date{}

\begin{document} 
\baselineskip24pt
\maketitle 

\begin{sciabstract}
The ongoing COVID-19 pandemic continues to affect communities around the world. To date, almost 6 million people have died as a consequence of COVID-19, and more than one-quarter of a billion people are estimated to have been infected worldwide. The design of appropriate and timely mitigation strategies to curb the effects of this and future disease outbreaks requires close monitoring of their spatio-temporal trajectories. We present machine learning methods to anticipate sharp increases in COVID-19 activity in US counties in real-time. Our methods leverage Internet-based digital traces -- e.g., disease-related Internet search activity from the general population and clinicians, disease-relevant Twitter micro-blogs, and outbreak trajectories from neighboring locations-- to monitor potential changes in population-level health trends. Motivated by the need for finer spatial-resolution epidemiological insights to improve local decision-making, we build upon previous retrospective research efforts originally conceived at the state level and in the early months of the pandemic. Our methods --tested in real-time and in an out-of-sample manner on a subset of 97 counties distributed across the US-- frequently anticipated sharp increases in COVID-19 activity 1-6 weeks before the onset of local outbreaks (defined as the time when the effective reproduction number $R_t$ becomes larger than $1$ consistently). Given the continued emergence of COVID-19 variants of concern --such as the most recent one, Omicron-- and the fact that multiple countries have not had full access to vaccines, the framework we present, while conceived for the county-level in the US, could be helpful in countries where similar data sources are available.
\end{sciabstract}

\section*{Introduction}

With more than 6 million deaths worldwide as of March 2022, the COVID-19 pandemic has become a global catastrophic event \cite{lipsitch2019enhancing}. The United States (US) alone has reported more than 80 million infections, and nearly 1 million deaths \cite{Worldometers}. While COVID-19 vaccination strategies have been deployed in the US since the early months of 2021, the proportion of fully vaccinated individuals is still low, at around 64\%. With the emergence of new variants of SARS-CoV-2 --the virus responsible for infecting people with COVID-19-- such as Omicron, the observed waning of immunity conferred by vaccines \cite{goldberg2021waning}, and the fact that many non-pharmaceutical interventions (NPIs), such as mask mandates and social distancing have become less frequently practiced, the US is still highly vulnerable to the effects of the COVID-19 outbreaks \cite{kissler2020projecting}. Thus, our best line of defense against uncontrolled outbreaks remains to be vaccinated and to adjust our social behavior when sharp increases of infections are first detected \cite{dashtbali2021compartmental,buonomo2020effects}. In the context of designing timely and appropriate public health responses to slow down infections and eventual deaths, robust real-time indicators of COVID-19 activity are of great importance as they guide authorities in their decision-making processes.

Tracking COVID-19 in real-time with reliable data sources remains a challenge despite many initiatives led by hospitals, local health authorities, and the research community\cite{covidtracking}. For instance, PCR COVID-19 test results are typically delayed by multiple days and reported with days or weeks of delay. Testing availability may significantly impact the recorded number of positive COVID-19 cases, which may suggest that changes in COVID-19 activity reflect testing volumes rather than the underlying proportions of infections in the population\cite{kaashoek2020covid,lipsitch2019enhancing}. Furthermore, the reliability, consistency, and in general the quality of reported COVID-19 data --such as confirmed cases, hospitalization, and deaths-- varies highly from country to country (and within countries) frequently due to disparities in economic resources locally allocated to monitor and respond to the pandemic\cite{covidtracking}.
 
Statistical models have been proposed to address delays in data collection and ascertainment biases retrospectively and in real-time \cite{jombart2021real,gutierrez2020delays, lu2021estimating, de2021near}. Computational mechanistic (Susceptible-Infected-Recovered, SIR) models, on the other hand, have been used to reconstruct the spatio-temporal patterns of the spread of COVID-19 retrospectively and to forecast likely COVID-19 cases and deaths to occur in the near future \cite{ihme2020modeling,monod2021age,della2020network,vespignani2020modelling, lai2020effect, davis2021cryptic, chang2021mobility}. Many studies characterizing the quality and accuracy of forecasts have emerged from COVID-19 initiatives coordinated by the U.S. Centers for Disease Control and Prevention (CDC) \cite{CDCforecast}. Those models are usually based on mechanistic SIR-like systems \cite{baker2021assessing} and/or Bayesian frameworks \cite{flaxman2020estimating}. Despite their ability to explore potential ``what if" scenarios and their accuracy during periods where the epidemic curves have been monotonically increasing or decreasing, most of these forecasting models have not been very consistent or reliable in anticipating sharp changes in disease activity \cite{ray2020ensemble}.

Several studies have also shown the potential utility of  ``digital" (or Internet-based) data sources as a complementary way to track (and/or confirm) changes in disease activity at the population level \cite{yang2015accurate, mcgough2017forecasting,santillana2015combining, dugas2013influenza,lee2017forecasting, aiken2020real, lu2018accurate, lu2019improved}. In the past, many approaches explored valuable information from search engines \cite{yuan2013monitoring,gluskin2014evaluation,althouse2011prediction, aiken2020real, ginsberg2009detecting}, Twitter microblogs \cite{nagar2014case,aramaki2011twitter,paul2014twitter}, and electronic health records \cite{yang2017using,rangarajan2019forecasting, santillana2016cloud} for real-time estimates of disease prevalence, and characterized the limitations of those non-traditional data sources in the context of influenza  \cite{santillana2014can,lazer2014parable}. In the past two years, statistical and machine learning approaches have explored how to incorporate disease-related Internet search data to track and forecast COVID-19 activity\cite{lampos2021tracking, liu2020real,lu2021internet}, and some of their limitations have been documented as well \cite{bento2020evidence}. The logic behind using disease-related ``digital data'' to monitor disease activity is that user-generated \textit{digital traces} may capture changes in human behavior (human mobility, situational awareness, increases in certain clinical treatments, population-level topic interests, social media trending content) that may have an impact on disease transmission and/or may reflect increases in symptomatic infections \cite{salathe2012digital, santillana2016perspectives}.

 Kogan et al. explored the effectiveness of Google Trends, Twitter microblogs, clinician searches, anonymized human mobility from mobile phones records, and smart thermometers to anticipate increases and decreases in COVID-19 activity at the state level, as reported by healthcare systems \cite{kogan2021early}. By combining multiple data streams, they proposed a Bayesian indicator capable of predicting an impending COVID-19 outbreak with several weeks of anticipation in near-real time. Their methods were successful when tested in a retrospective fashion and during the first half of the year 2020. However, at the time of their study, Kogan and colleagues did not have enough data to perform out of sample validation tests, which is now possible given the higher number of COVID-19 outbreaks. Moreover, Kogan et al. did not explore the feasibility of using their approaches at finer spatial resolutions, such as the county level, where the signal to noise ratio in aggregated digital data streams may be compromised and where most outbreak control strategies are implemented in the US.

\textbf{Our contribution:} Here we present a framework to deploy a prospective real-time machine learning-based early warning system to anticipate or confirm COVID-19 outbreaks at the county level in the United States. Our analysis restricts our choice of counties to those with the infrastructure to conduct vaccine clinical trials or those with more than 1 million inhabitants. Our systems leverage the predictive power of both individual Internet-based data sources and their combined consensus. We quantify their predictive performance in a prospective out-of-sample way from January 2020 to January 2022, including the most recent periods when the highly contagious Omicron variant was detected. By implementing event-detection algorithms on each of these Internet-based time series and employing machine-learning strategies to combine this information, we anticipate the onset of local COVID-19 outbreaks -- defined as the time when the local effective reproductive number, $R_t$, becomes larger than $1$ in a given region \cite{parag2020using}. 

\section*{Results}

We analyzed COVID-19 activity in 97 US counties across the US between January 1, 2020, and January 1, 2022. First, we identify weeks when the local reproductive number (commonly denoted by $R_t$) was  higher than $1$ (with 95\% C.I. as described in \cite{parag_improved_2021}), suggesting that the local number of secondary COVID-19 infections was larger than 1 per index case. We labeled each first week when the local $R_t$ transitioned from a value smaller than 1 to one above 1 as \emph{outbreak onset} for each location (see Data and Methods section for details). Interchangeably, we also refer to these outbreak onsets as \emph{events} in this manuscript.  We identified  464 outbreak onsets at the county level. From this total, 367 events were used to test our methods out-of-sample, after using the first outbreak onset of each location as initial training data.  We replicated this analysis for the 50 US states, where we identified 252 outbreaks onsets at the state level (a total of 202 out-of-sample outbreak onsets). 

We obtained COVID-19-related digital streams for the same time period with the goal of identifying, for example, moments in time when (a) COVID-19 related Internet searches, such as fever or anosmia, showed sharp increases -- perhaps signaling a population-wide increase of symptomatic infections --, (b) when clinicians were looking for dosage information for specific medications to control fever or other COVID-19 symptoms, or (c) when Twitter users expressed that they or their family/friends may have caught COVID-19, among other signatures. We then explored the ability of our methods to extract information from these data sources (individually and as a consensus) to anticipate outbreak onsets for each geographical scale. 

Our results are summarized in Figs \ref{fig:results_county} and \ref{fig:results_state} for the county and state levels, respectively. By dynamically training\footnote{By dynamically training, we refer to the use of all available historical information at a given point in time to produce any prospective future-looking prediction.} our machine learning methods to recognize temporal patterns that precede increases in COVID-19 activity, we tested their ability to anticipate outbreak onsets. Specifically, we quantified how early they could anticipate unseen outbreaks (referred to as \emph{earliness}), the number of times they anticipated, synchronously identified, or lately confirmed a subsequently observed outbreak (referred to as an \emph{early}, \emph{synchronous} or \emph{late warning}). We also quantified the number of times our methods triggered an alarm, but no outbreak onset was subsequently observed (referred to as a \emph{false alarm}), and the number of \emph{missed outbreaks} when no alarm preceded an outbreak onset. 

Specifically, we defined an \emph{early warning} whenever an alarm was triggered up to six weeks before the outbreak onsets. The choice of a six-week window was made to plausibly relate a digital trend change with a potential subsequent infection. For example, if a person uses Google to search for COVID-19 related information due to a likely symptomatic infection, that person's COVID-19 infection may be confirmed in the following week or two, if they are admitted to a health care facility or tested by a provider. Alternatively, such person may search for COVID symptoms not in response to their own symptoms but to someone else's within their close contact network (that may eventually infect them). In that case, the lag between that Internet search and an eventual confirmed infection may be longer, perhaps up to 4-6 weeks. We also defined \emph{synchronous warning} and \emph{late warning} whenever an alarm was triggered on the same date or up to two weeks later than the identified outbreak onset, respectively.

In both Figs \ref{fig:results_county} and \ref{fig:results_state}, panel (A) serves as a graphical representation of the different outcomes observed in our early warning systems, namely, when the system leads to a successful early, synchronous or late warning; a false alarm, and a missed outbreak. Two other scenarios were characterized: one that explores when the system may have suggested a full outbreak would occur but only a mild increase in COVID-19 activity was subsequently observed, labeled as \emph{warning, increase is observed}, and another labeled as a \emph{soft warning}, that is observed when the system almost triggered an alarm and an increase of COVID-19 activity was subsequently seen --this could signal an improper model calibration, perhaps a consequence of the small number of events to train the models in a given location. For simplicity, the COVID-19 cases reported by Johns Hopkins University (JHU) are shown in gray, and an early warning indicator is shown in light orange. The horizontal dotted black line represents an early warning system's decision boundary, i.e., a threshold value used to activate an outbreak alarm. Please refer to our Data and Methods Section for a more detailed description of the event definitions, machine learning methods, and our validation criteria.

\subsection*{County-level Performance}  

Fig \ref{fig:results_county} (B) displays a summary count of all the outbreak onsets observed at the county level. Each horizontal bar is colored, from orange to purple, depending on the event class previously described. In this work, we developed two different machine learning methods: (i) the \emph{Single Source} method that explores the predictive power of individual digital sources by detecting increases in the search volume of a given term, and (ii) the \emph{Multiple Source} method that incorporates many different signals, optimizing on their best individual performances, to produce a single output. We refer the reader to our Data and Methods section for more details on the formulation of our methodologies. We compared the predictive performance of our methods against an intuitive baseline, which we refer to as \emph{Naive method}. The Naive method predicts an outbreak will happen whenever there is an increase in the number of confirmed COVID-19 cases, i.e., if the COVID-19 cases increase on week $t$ compared to week $t-1$, the Naive method triggers an alarm at week $t$.

\textbf{Early warnings:} The alarms produced by the Naive method resulted in early warnings for 337 out of the 367 total events (92\%) and displayed 23 synchronous and one late warning in the remaining 19 events (6\%). The best signal in the Single Source method (Google search term ``How long does covid last?") identified 237 early warnings (65\%), 25 synchronous (7\%), and 30 (8\%) late warnings. Here we use the term ``best signal'' to reference those digital traces with a higher number of early warnings across the 97 counties in our dataset. We obtained a comparable performance for the Google search term  ``side effects of vaccine'', with 227 early warnings (62\%),  19 synchronous (5\%), and no late warnings. The Multiple Source method identified 213 early warnings (61\%), 23 synchronous  (6\%), 37 late (10\%) warnings. \textbf{Soft warnings and missed outbreaks:} The Naive method missed six events (2\%) and the Single Source method for the two displayed Google Trends (``How long does covid last?" and ``side effects of vaccine'') missed 75 and 106 events (20\% and 29\%), respectively. For the Multiple Source method, 61 out of the 87 remaining events were soft warnings (17\% of the total events) and 26 were missed outbreaks (7\% of the total events).

\textbf{False alarms:} Fig \ref{fig:results_county} (C) summarizes the false alarms for the different methods in our analysis. The Naive method registered 617 false alarms (about 1.7 times the number of observed outbreaks). In comparison, the Single Source method led to 374 false alarms for the term ``How long does covid last?" and 479 false alarms for the term ``side effects of vaccine" (1 and 1.3 times the number of outbreaks, respectively). The Multiple Source method produced the lowest amount with only 114 false alarms (0.3 times the number of outbreaks). A minimum number of false alarms should be preferred to avoid alarm fatigue. The Multiple Source method performs best in this aspect. The Naive Method also exhibited the highest number of ``warnings with an increase'' observed (252 registered events), followed by the Single Source method (139 events for ``How long does covid last?'' and 171 for ``side effects of vaccine") and the Multiple Source method with 36 events.

Fig \ref{fig:results_county} (D) shows a graphical representation of the probability of resurgence $P(R_t>1)$, i.e, the probability that the effective reproductive number is higher than one given the data\footnote{Technically, we compute the conditional probability $P(R_t > 1 \vert I_1^t)$ where $I_1$ denotes the initially infected population. For notational simplicity, we write $P(R_t > 1)$ throughout the manuscript.}, along with the weekly confirmed COVID-19 cases (gray-filled curve in the top), and three representative signals for the Naive, Single, and Multiple Source methods for the county of Marion (FL). We chose to depict this specific county as an example where our methods performed well (but similar visualization for all counties and states can be found in the SI). The outbreak onsets were defined when $P(R_t>1)>0.95$ and marked with red vertical lines that extend across the five horizontal panels in Fig \ref{fig:results_county} (D)  containing the COVID-19 case counts and the three early warning methods. Triggered alarms are displayed as vertical tick marks for each method (Naive, Single, and Multiple Source methods in gray, blue and, yellow, respectively).

\textbf{Earliness:} Fig \ref{fig:results_county} (E) shows the \textbf{earliness} (in weeks) for the alarms triggered by each method. The bars represent a count of the number of activated alarms within the out-of-sample time window (between six weeks before and two after the outbreak onset). Triggered alarms that did not precede any increment in the activity of confirmed COVID-19 cases within the six-week observational window were considered false alarms (displayed in red). We observed that most early warning activation counts of the Naive and Single Source methods fell within the 4 and 6 weeks early range (68\% Naive, and 61\% and 62\% for the Single Source). The Multiple Source method's highest count (55 alarms) fell within the six-week early mark (26\%). The rest of the activations were spread across the 5 to 1 week-early mark, with a higher number of activations as we reached the sync warning mark.

\textbf{Omicron-attributable outbreaks:} A total of 62 outbreak onsets were observed at the county level after December 01, 2021. The Single Source method correctly identified 35 and 43 early warnings (56\% and 69\%) for the Google Trends terms ``How long does covid last?" and ``side effects of vaccine", respectively. The Multiple Source method could anticipate 42 (68\%) outbreak onsets.

\subsection*{State-level performance}  
 
Fig \ref{fig:results_state} summarizes the state-level results. 
\textbf{Early warnings:} The alarms produced by the Naive method preceded COVID-19 outbreak onsets in 178 out of the 202 total events (88\%), displayed a synchronous warning in 23 (11\%) events and only one late warning. The best signal in the Single Source method (Google Trends "How long does covid last?") identified 128 early warnings (63\%), 13 synchronous (6\%), and 18 late  warnings (9\%). Comparable results were found for the Google Trends term ``Chest Pain'' with 120 early warnings (59\%), 12 synchronous warnings (6\%), and 12 late warnings (6\%). On the other hand, the Multiple Source method produced 99 early warnings (49\%), 22 synchronous warnings (11\%), and 24 late warnings   (12\%). \textbf{Soft warnings and missed outbreaks:} The Naive method had no missed events, and the Single Source method for the two displayed Google Trends ( ``How Long Does Covid Last" and ``Chest pain") missed 43 and 58 events (21\% and 29\%), respectively. For the Multiple Source method, the remaining 57 events were characterized by 41 soft warnings and 16 missed outbreaks, representing 20\% and 8\% of the total events.

\textbf{False alarms:} The Naive method led to 271 false alarms (Fig \ref{fig:results_state} (C)), about 1.3 times the number of observed outbreaks. The Single Source method produced 171 false alarms for the Google Trends term ``How long does covid Last" and 151 false alarms for the term ``chest pain" (about 0.84 and 0.74 times the number of outbreaks, respectively). 
The Multiple Source method led to 34 false alarms (about 0.24 times the number of outbreaks), the lowest of all methods. For ``warnings with an increase observed'', the Naive method exhibited 74 events, followed by the Single Source method 82 and 63 events for the two best signals) and the Multiple Source method with only 20 events.

Fig \ref{fig:results_state} (D) shows a graphical representation of the probability of resurgence $P(R_t>1)$ and weekly confirmed COVID-19 cases in Florida.  The top signals for the Single Source method and the output of the Multiple Source method are also shown, with tick marks representing the triggered alarms for each method. In Fig \ref{fig:results_state} (E), we observe that the Naive method's highest early warning counts fell within the -6, -3, and -1 week mark (approximately 59\% of total early warning rates). The highest counts for the Single Source method fall between the four and six-week early mark (65\% and 69\% for the best signals, respectively). The majority of early warning activations of the Multiple Source methods fall between the one and three-week early range (60\%).
 
\textbf{Omicron-attributable outbreaks:} We observed 19 state-level outbreak onsets occurring after December 1, 2021. The Single Source methodology preceded 13 and 12 (68\% and 63\%, accordingly) outbreak onsets. The Multiple Source method preceded  17 out of the 19 (89\%).
 
 \subsection*{Geographical County-Level Analysis}
 
 Given the lower number of false alarms and the number early warning rates of the Multiple Source method, we investigated this method more extensively. Fig \ref{fig:clustering figure} shows a detailed breakdown of the performance for each county in our dataset. We implemented a k-means clustering approach to group the counties based on their COVID-19 weekly activity. We separated each set of selected locations into three different groups: the set of counties that experienced their first outbreak at the beginning of 2020 (blue), the set of counties that experienced their first outbreak during summer 2020 (yellow), and a set of counties that experienced their first major outbreak after the summer of 2020 (green). Fig \ref{fig:clustering figure} (A) shows the geographical location of the selected counties across the United States map, where the colors represent each cluster. The clustering analysis highlights how the COVID-19 outbreak dynamics seem heavily dependent on the geographical location. Counties in Cluster 1 were mainly located in the northeast part of the country, while Counties in  Clusters 2 and 3 were scattered throughout the south and north/central regions, respectively.

  Fig \ref{fig:clustering figure} (B) shows a list of counties under the colored timeseries representing each cluster center. Along with its name and corresponding number, we display a set of tick marks at the bottom representing the out-of-sample outbreak onset (varying from 2 to 7). For each outbreak onset, we trained our method on the previous onsets. For this reason, onset 1 is used for training only and was not included in the analysis. We then show the performance of the  Multiple Source method predicting the out-of-sample outbreak onset with different colors. In general, counties in Cluster 1 experienced at most four onset events, while counties in Clusters 2 and 3 experienced up to seven outbreak onsets (we refer to our Data and Method sections for a precise definition of outbreak onsets). The performance of the Multiple Source method was consistent throughout the three groups. In cluster 1, we observed mostly early warnings and soft warnings, while mixed results were obtained for clusters 2 and 3. However, it is worth noting that those clusters contain more than twice the number of group members of cluster 1. The multiple source method successfully improved over time for some counties. That was the case for Shelby (TN, 13) and Tarrant (TX, 14) in Cluster 2, Philadelphia (PA, 2), Lake (IN, 4), Laporte (IN, 5), and Hamilton (OH, 9) in Cluster 3. Counties where our Multiple Source method did not provide early warnings included Middlesex (MA, 8) and Queens (NY, 11) in Custer 1, Jackson (MO, 8) and Cook (IL, 17) in Cluster 2, Cumberland (NC, 3), Palm Beach (FL, 20), and Hillsborough (FL, 33) in Cluster 3.
 
  \subsection*{Geographical State-Level Analysis}
 
 By extending our analysis to the state level (Fig  \ref{fig:clustering figure_state}), we observe that the COVID-19 trajectories generated by the clustering analysis remained consistent with their county-level counterpart (early outbreak states in the northeast, summer, and post-summer outbreaks in the south and north, respectively). States where the Multiple Source method successfully predicted most of the outbreak onsets included Ohio (6), Connecticut (12), Indiana (5) in Cluster 1, Nevada (2), Florida (8), Tennessee (16), and North Carolina (19) in Cluster 2. On the other hand, states where our method did not predict most onsets ahead of time included Michigan (8), Maine (9), and Massachusetts (13) in Cluster 1,
 Arizona (3), South Carolina (9), Virginia (17), Texas (20) in Cluster 2, Missouri (1), Montana (12), and Wyoming (17) in Cluster 3.

\subsection*{Feature Importance Analysis} 

Our Multiple Source method dynamically selected the most predictive Internet-based data streams, as well as historical epidemiological information --both at the state and county levels--, for each location to produce future-looking early warnings. We analyzed which data streams were most informative in our early warning systems across time, individually, and across locations (see Data and Methods section for more details). Tables \ref{tab:word_frequency_county} and \ref{tab:word_frequency_state}  show the top 10 most frequently digital and historical proxies selected by the Multiple Source method, across all locations, to predict each out-of-sample outbreak onset at the county and state levels. In both cases, the number of out-of-sample outbreak onsets varied from 2 to 7. Moreover,  many counties experienced at least two onsets, while fewer counties experienced more than six onsets. Hence, we display six columns with the corresponding number ($n$) of locations for the analysis.  \textbf{County-level:} For the outbreak onsets 2 and 3, our Multiple Source method predominantly selected the Google Trends terms "covid," "covid-19," and "covid symptoms," along with local confirmed COVID-19 cases as its primary selected data sources. For later outbreak onsets,  we observed an increase in the importance of Neighbor Activity (weekly epidemiological reports from geographically close counties) as a top selected source after onsets 4, 5, and 6. Some of the remaining terms in the top 10 list represented general symptoms of respiratory diseases such as ``chest pain'', ``fever'', ``nasal congestion''. \textbf{State-level: } We found consistent performances for the Multiple Source Method at the state level, with COVID-19 related Google trends leading the list of most selected streams for onsets 2 and 3. After onset 4, state-level neighbor activity also appeared as a highly selected source. Other relevant terms in the top 10 list included  ``Acute bronchitis'', ``cough'', and ``Asthma''.

\color{black}

\begin{landscape}

\begin{figure}
    \centering
    \includegraphics[width=1.5\textwidth, angle=0,origin=c]{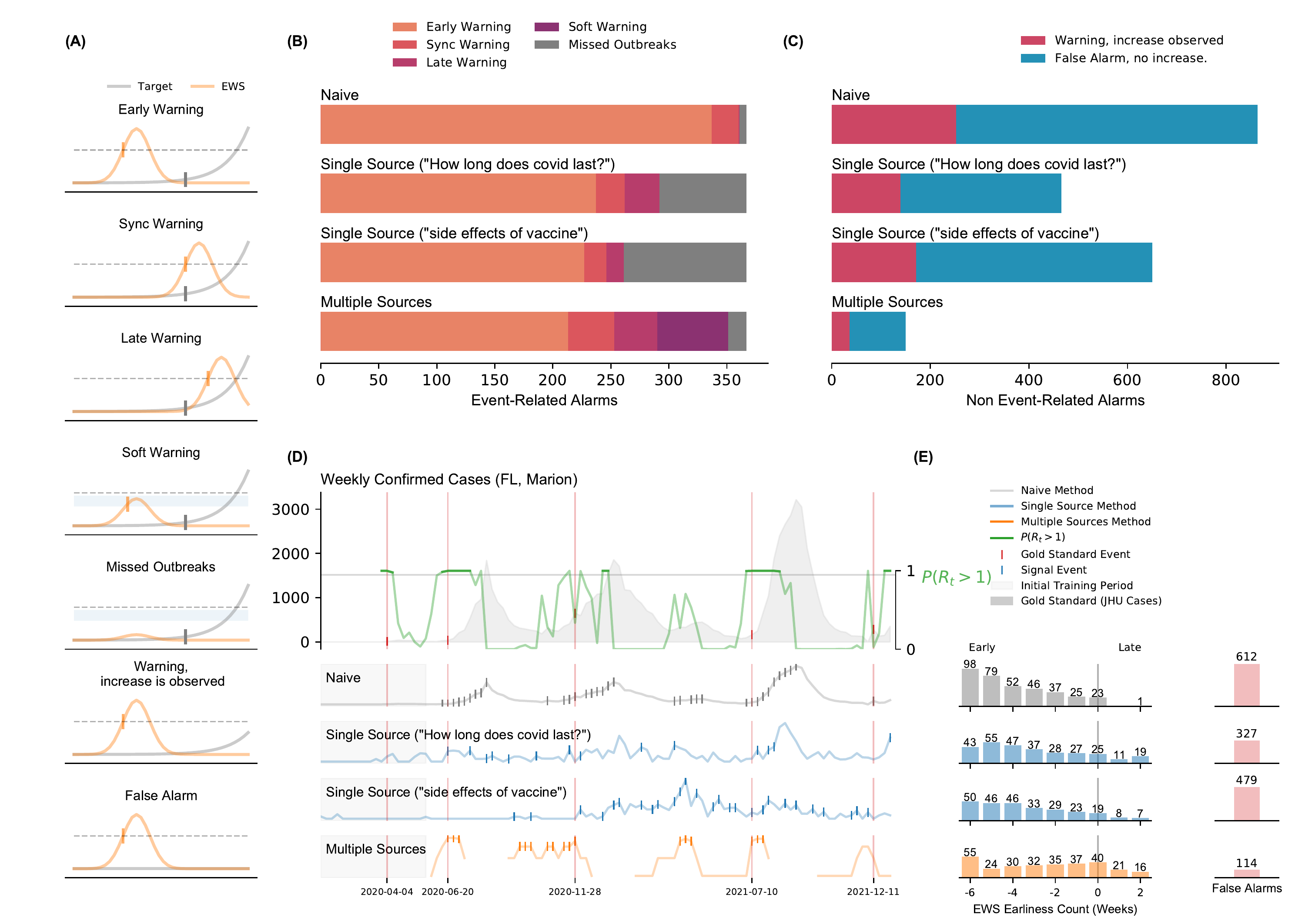}
    \caption{\textbf{A summary of our results at the county level.}}
    \label{fig:results_county}
\end{figure}

\end{landscape}

\begin{landscape}
\begin{figure}
    \centering
    \includegraphics[width=1.5\textwidth, angle=0,origin=c]{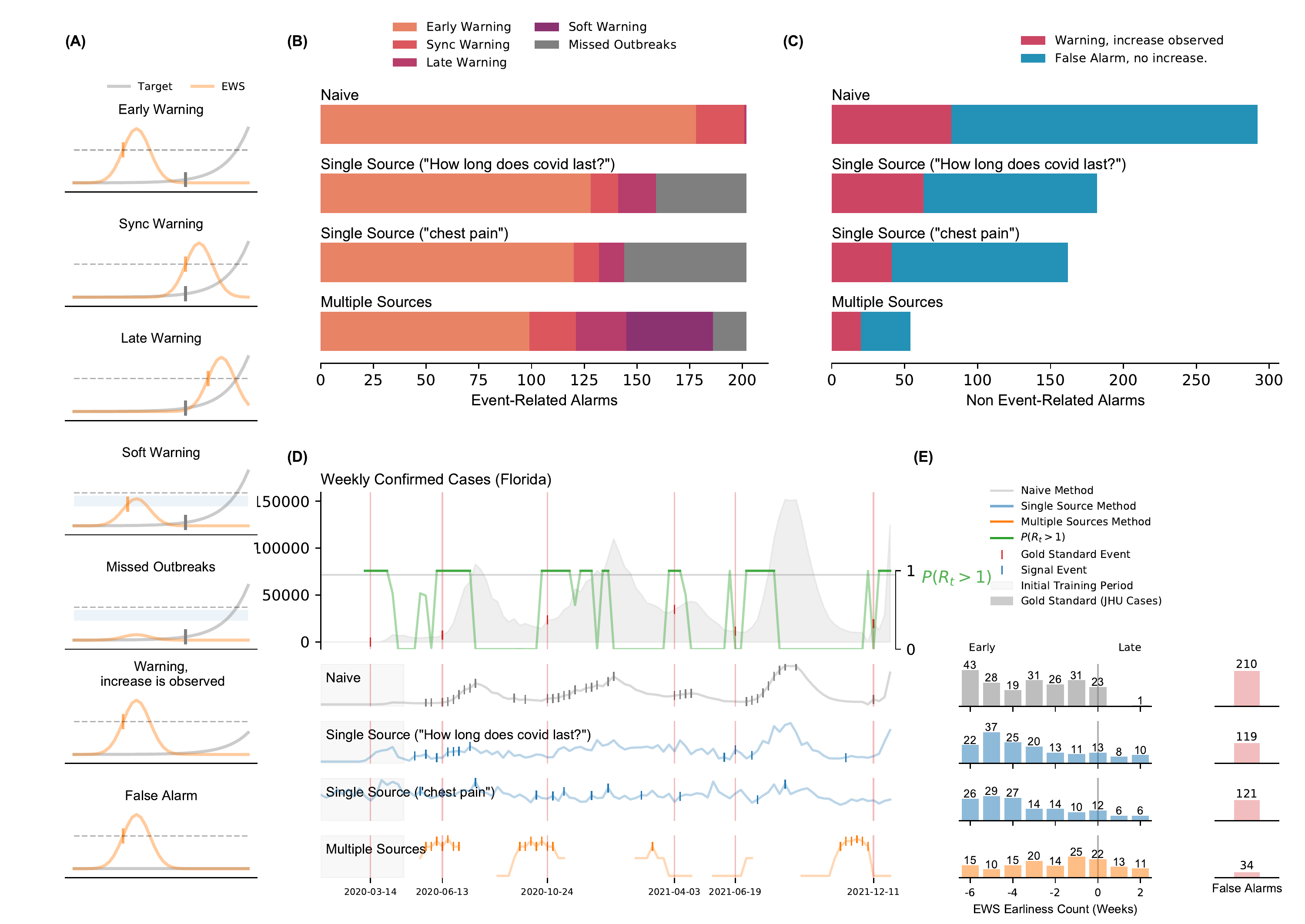}
    \caption{\textbf{A summary of our results at the state level.}}
    \label{fig:results_state}
\end{figure}

\end{landscape}

\begin{figure}
    \centering
    \includegraphics[width=0.83\columnwidth]{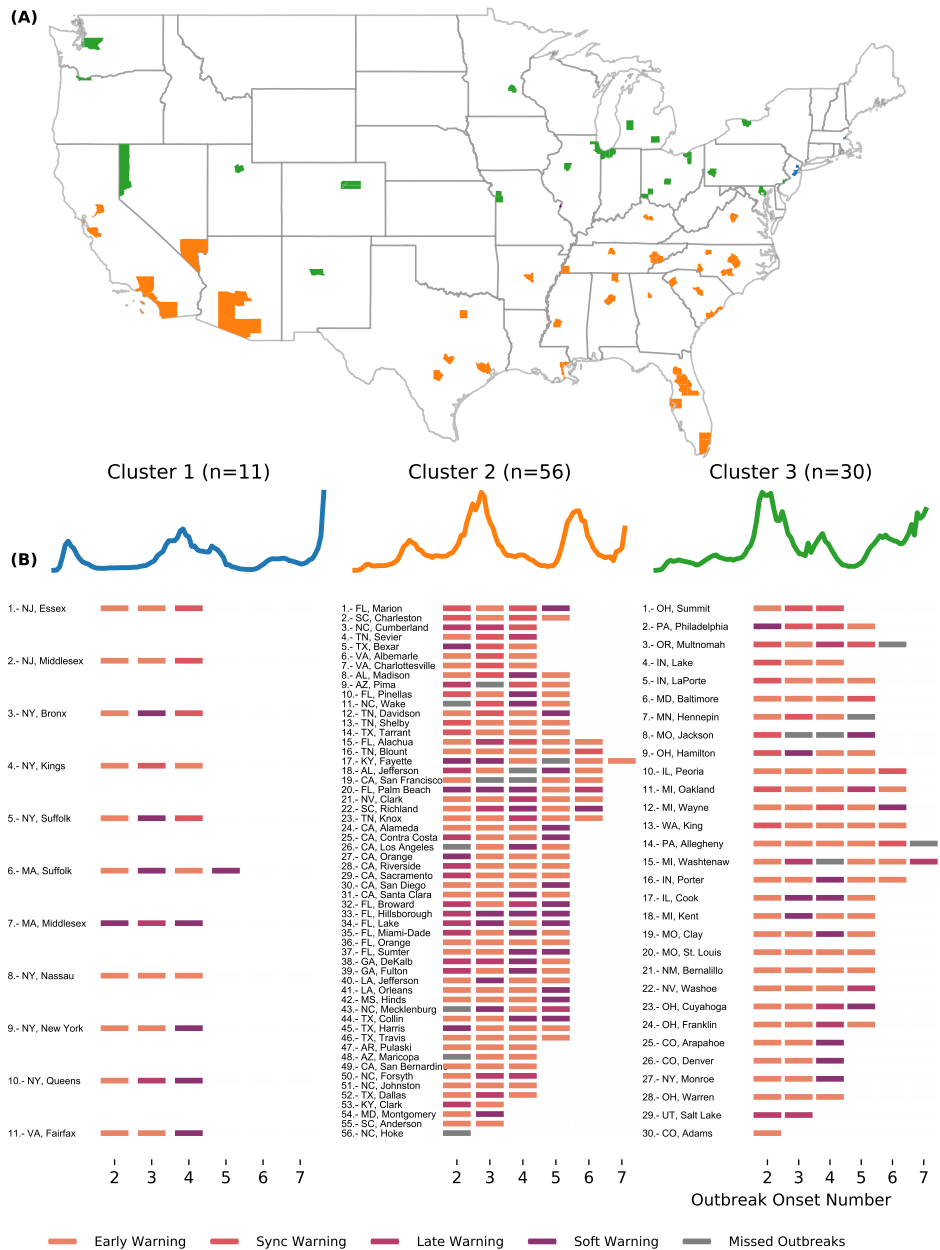}
    \caption{ \small \textbf{Clustering analysis at the county level based on confirmed COVID-19 case trajectories.} (A) The geographical map color-codes each location based on their cluster. A total of three clusters described groups of counties that experienced their first outbreak onset early in 2020 (blue), during summer (yellow) and, late in 2020 (green). (B) Orange-to-purple and gray markers correspond to the performance of the Multiple Source Method for each out-of-sample outbreak onset. For example, the first location in cluster 1, Essex (NJ), experienced three out-of-sample events: the first two  were early warnings, whereas the second event was a synchronous warning.}
    \label{fig:clustering figure}
\end{figure}

\begin{figure}
    \centering
    \includegraphics[width=0.83\columnwidth]{Figures/Figures_leo/cluster_state_2022-02-14.pdf}
    \caption{ \small \textbf{Clustering analysis at the state level based on COVID-19 confirmed cases trajectories.} (A) The geographical map color-codes each location based on their cluster. A total of three clusters describe groups of states that experienced their first outbreak onset early in 2020 (Cluster 1 in blue), during summer (Cluster 2 in yellow) and, late in 2020 (Cluster 3 green). (B) The set of orange-to-purple and gray markers correspond to the performance of the Multiple Source Method for each out-of-sample outbreak onset.}
    \label{fig:clustering figure_state}
\end{figure}

\begin{landscape}

\begin{table}[]
    \centering \scriptsize
        	\begin{adjustwidth}{-0.9in}{0in} 
   \begin{tabular}{lllllll}
\toprule
{} &                      Onset 2 (n=97) &                      Onset 3 (n=95) &               Onset 4 (n=91) &               Onset 5 (n=64) &             Onset 6 (n=17) &               Onset 7 (n=3) \\
\midrule
0 &                     (GT) covid (81) &            (GT) covid symptoms (48) &    (GT) covid symptoms (34) &       Neighbor Activity (28) &      Neighbor Activity (8) &       Neighbor Activity (3) \\
1 &                  (GT) covid-19 (80) &                  (GT) covid-19 (43) &      Neighbor Activity (31) &     (GT) covid symptoms (19) &          (GT) covid-19 (6) &         Confirmed Cases (2) \\
2 &            (GT) covid symptoms (68) &                     (GT) covid (41) &             (GT) covid (26) &         Confirmed Cases (15) &    (GT) covid symptoms (4) &             (GT) Asthma (1) \\
3 &                Confirmed Cases (51) &                Confirmed Cases (31) &        Confirmed Cases (25) &              (GT) covid (14) &           (GT) Anosmia (3) &  (GT) Actinic keratosis (1) \\
4 &                (GT) chest pain (46) &                (GT) chest pain (22) &          (GT) covid-19 (21) &            (GT) Ageusia (14) &             (GT) covid (2) &       (GT) Fibrillation (1) \\
5 &                (GT) quarantine (39) &                      (GT) Acne (20) &           (GT) Anosmia (18) &            (GT) Anosmia (12) &  (GT) Nasal congestion (2) &            (GT) Floater (1) \\
6 &  (GT) how long does covid last (29) &                     (GT) fever (18) &           (GT) Ageusia (18) &           (GT) covid-19 (11) &            (GT) Phlegm (2) &       (GT) Panic attack (1) \\
7 &              (GT) covid-19 who (24) &                   (GT) Ageusia (18) &        (GT) chest pain (13) &        (GT) Night sweats (8) &             (GT) fever (2) &     (GT) covid symptoms (1) \\
8 &           (GT) chest tightness (20) &              Neighbor Activity (14) &              (GT) Acne (10) &          (GT) chest pain (7) &        (GT) Laryngitis (2) &           (GT) covid-19 (1) \\
9 &         (GT) Abdominal obesity (15) &  (GT) how long does covid last (12) &  (GT) Acute bronchitis (10) &  (GT) Autoimmune disease (5) &            (GT) Asthma (2) &            (GT) Anosmia (1) \\
\bottomrule
\end{tabular}
\end{adjustwidth}
    \caption{\textbf{Most selected data streams (ordered by frequency from top to bottom) of the Multiple Source method at the county level.} The number of instances a data stream was selected  (values within parentheses) reduced over time, given some locations experienced fewer outbreak events than others. GT stands for Google Trends.}
    \label{tab:word_frequency_county}
\end{table}

\begin{table}[]
    \centering \scriptsize
        	\begin{adjustwidth}{-0.9in}{0in} 
\begin{tabular}{lllllll}
\toprule
{} &                      Onset 2 (n=50) &              Onset 3 (n=50) &              Onset 4 (n=50) &             Onset 5 (n=36) &            Onset 6 (n=15) &                   Onset 7 (n=1) \\
\midrule
0 &                     (GT) covid (42) &    (GT) covid symptoms (19) &     Neighbor Activity (15) &     Neighbor Activity (16) &     Neighbor Activity (8) &               (GT) Rhinitis (1) \\
1 &                  (GT) covid-19 (41) &          (GT) covid-19 (18) &   (GT) covid symptoms (12) &   (GT) covid symptoms (13) &   (GT) covid symptoms (5) &          (GT) Biliary colic (1) \\
2 &            (GT) covid symptoms (32) &             (GT) covid (16) &  (GT) Acute bronchitis (8) &        Confirmed Cases (8) &            (GT) covid (5) &  (GT) Intracranial pressure (1) \\
3 &                Confirmed Cases (27) &        Confirmed Cases (14) &           (GT) Ageusia (7) &         (GT) Shivering (7) &            (GT) Cough (4) &        (GT) Low-grade fever (1) \\
4 &                (GT) chest pain (25) &      Neighbor Activity (13) &        Confirmed Cases (7) &             (GT) covid (5) &       Confirmed Cases (4) &                  (GT) Rheum (1) \\
5 &                (GT) quarantine (19) &        (GT) chest pain (13) &             (GT) fever (7) &        (GT) Bronchitis (5) &            (GT) Fever (3) &            (GT) Esophagitis (1) \\
6 &  (GT) how long does covid last (16) &             (GT) fever (11) &            (GT) Asthma (7) &  (GT) Nasal congestion (5) &       (GT) Chest pain (3) &                                 \\
7 &         (GT) Abdominal obesity (14) &  (GT) Acute bronchitis (11) &           (GT) Anosmia (7) &  (GT) Hyperventilation (4) &           (GT) Asthma (3) &                                 \\
8 &                     (GT) fever (12) &            (GT) Ageusia (7) &        (GT) Bronchitis (5) &            (GT) Asthma (4) &         (GT) Eye pain (3) &                                 \\
9 &           (GT) chest tightness (10) &         (GT) Chest pain (6) &  (GT) Nasal congestion (5) &         (GT) Toothache (4) &  (GT) Post-nasal drip (2) &                                 \\
\bottomrule
\end{tabular}

\end{adjustwidth}

    \caption{\textbf{Most selected data streams (ordered by frequency from top to bottom) of the Multiple Source method at state level.} The number of instances a data stream was selected (values within parentheses) reduced over time, given some locations experienced less outbreak events than others. GT stands for Google trends.}
    \label{tab:word_frequency_state}
\end{table}

\end{landscape}
\color{black}

\section*{Discussion}

We have presented a set of novel methods that can be deployed in real-time and in prospective mode to anticipate the onset of COVID-19 outbreaks in the United States at the county level. Our proposed methods leverage information from multiple Internet-based data sources, commonly called digital traces, as they are collected when humans navigate the Internet and serve as proxies of human behavior. The early warning system framework presented here extends previous work -- conducted retrospectively and at the state level by Kogan et al.\cite{kogan2021early} -- to the county level, a geopolitical spatial resolution where most outbreak mitigation strategies are designed and deployed in the USA. Specifically, our methods were designed to anticipate sharp increases in COVID-19 transmission, as identified by changes in the effective reproduction number ($R_t$), an outbreak indicator preferred by the community of epidemiologists  \cite{unwin2020state,cowling2010effective, parag_improved_2021, de2021near}. 

We developed two methods that incorporate single or multiple digital signals, namely (a) a Single Source method, which locally identifies the magnitude and the number of uptrends in the digital signals that precede outbreaks,  and (b) a Multiple Source method that dynamically selects a subset of the strongest predictive data streams available at each location historically, and combines them prospectively into a single indicator that quantifies the likelihood of occurrence of an outbreak in the following weeks. Both methods are data-driven techniques that continuously incorporate newly available data as time evolves, making them adaptive and responsive to the frequently changing trends of an emerging disease such as COVID-19. 


 Both Single and Multiple Source methods successfully anticipated most outbreak events between January 2020 and January 2022 for the 97 US counties in our dataset. To compare our methods with a baseline system, we define a Naive method that triggers an alarm whenever there is an increase in COVID-19 cases. As expected, the Naive method had the highest early warning rates, since there was at least one increase in the COVID-19 case counts at least 6-weeks before the outbreak onset events. However, the Naive method typically leads to a significantly high number of undesirable false alarms --it produced 612 alarms when only 367 outbreaks were observed. Our intuitive Single Source method alone dropped the number of false alarms significantly while still producing early or at least synchronous warnings on 237 of the 367 outbreak events (Google Trends term ``How long does covid last?”). From a decision-making perspective, having fewer false alarms is critical to reducing an unnecessary burden of resources --alarm fatigue--, and workforce \cite{lawson2005spatial,surkova2020false}. Moreover, a system with many false alarms may lead to distrust within the end-user community. Remarkably, our Multiple Source method dramatically decreases the number of false alarms compared to our Single Source method while displaying successful early and synchronous warnings in 69\% of observed outbreaks. At the state level, the Naive method also exhibits the highest ability to identify outbreaks at the cost of having more than 200 false alarms. As in the county-level analysis, our methods maintained a high rate of early warnings with a substantial decrease in the number of false alarms. Our Multiple Source method successfully identified  111  (Early/Sync) out of the 202 outbreak events for 50 states, with only 34 false alarms. Tables \ref{tab:table_summary_countylevel}, \ref{tab:table_summary_statelevel}, \ref{tab:table_singlesource_countylevel}, and \ref{tab:table_singlesource_statelevel} in the SI summarize all early percentage rates for the Single and Multiple Source methods at both county and state levels.
 
At the county level, the Single (for the two performing proxies) and Multiple Source methods mainly activated 1-6 weeks before $R_t$. At the state level, most early warnings for the Single and Multiple Source methods preceded the outbreak onset events in 4-6 weeks. This finding can be contrasted with previous work by \cite{kogan2021early} where a 2-3 week anticipation was found. However, it should be noted that Kogan et al. considered a different target quantity to be anticipated by digital traces, namely when  exponential growth in confirmed cases and deaths is observed, not when outbreaks start -- in our case defined as when the reproduction number $R_t$ is  higher than 1. 

\begin{figure}[h!]
    \centering
     \includegraphics[width=0.8\textwidth, angle=0,origin=c]{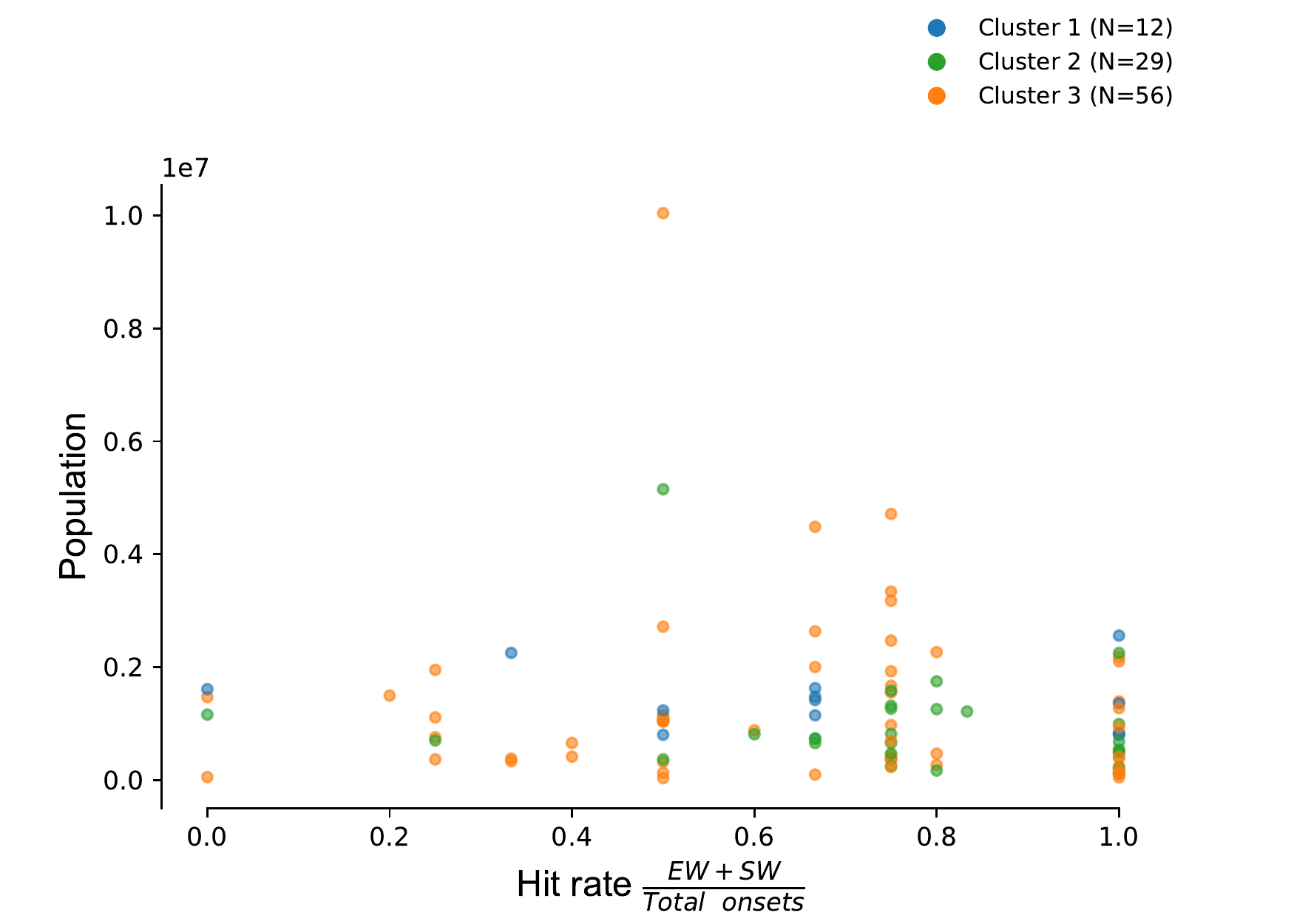}
    \caption{\textbf{Performance of the Multiple Source method vs. population sizes }Scatter plot of the relationship between population size and the out-of-sample Hit rate of the Multiple Source method across the 97 counties (color-coded depending on the assigned cluster group from Fig \ref{fig:clustering figure})}
    \label{fig:my_label_pop_vs_hit}
\end{figure}

Notably, the performance of our early warning systems at the state level was comparable to the county level, as shown in Figs \ref{fig:results_county} and \ref{fig:results_state}. This result is important, given that the signal-to-noise ratio in digital data sources tends to decrease as we \textit{zoom in} to finer spatial resolutions, and thus extracting meaningful signals tends to be more challenging. The spatial-resolution dependency of the signal to noise ration has been documented in multiple studies that have attempted to extend the use of digital streams to monitor disease activity at finer spatial resolutions, and lower population densities \cite{lu2018accurate, baltrusaitis2018comparison, lu2019improved, aiken2021toward}. Our methods seem to overcome this challenge by showing comparable ability and earliness to identify the onset of outbreaks for county and state levels. Moreover, our methods' predictive performance did not show any dependency on total population or population density across counties, as shown in Fig \ref{fig:my_label_pop_vs_hit} in the SI.

The predictive performance of our methods varied across counties, indicating the challenge of accurately detecting COVID-19 outbreaks ahead of time on such a fine spatial resolution. Using a k-means algorithm on the normalized disease activity curves over time, we obtained three different COVID-19 activity clusters for the 97 counties analyzed in our validation experiment. By repeating the analysis at the state level, our results showed similar COVID-19 trajectories and geographic regions (North East for early 2020 outbreaks, south for summer, and north for post-summer outbreaks) generated by the clustering algorithm. As presented in Fig \ref{fig:clustering figure} and Fig \ref{fig:clustering figure_state}, the performance of our methods was consistent throughout each cluster group.

The purely data-driven aspect of our Multiple Source method led to significant differences in the most selected predictive features as time progressed. In the first COVID-19 waves, Google Trends and COVID-19 cases (at both county and state levels) were mainly selected. In later waves, we observed that COVID-19 activity in neighboring counties became a more important predictor. A possible explanation for this alternation of most selected signals might be that Google Searches might have lost their early correlation power as increased awareness of the symptoms COVID-19 was likely in later waves. This preliminary evidence of variability in the chosen signals/features may point to the dynamic nature of how COVID-19 was initially perceived and investigated by the population, as well as the ever-changing trends in COVID-19 outbreaks.

Our Multiple Source method rarely completely missed an outbreak. Instead, we found that the early warning indicator frequently displayed that something was about to happen even when the indicator did not cross the decision threshold. As mentioned before, we refer to this scenario as a ``soft warning" for two reasons. First, a low number of events (for a given location) is not enough to properly calibrate a local predictive system. Second, the changing nature of human behavior and the SARS-CoV-2 virus challenges any prediction system. In these cases, one can always argue that the preferred epidemiological approach (when the effective reproductive number, $R_t$, is larger than 1) would eventually identify these sharp increases in COVID-19 activity. In addition, we find that our systems sometimes suggested that an outbreak would occur, but only a slight increase in cases was subsequently observed. We have referred to these instances as ``Warning, increased observed" on Figs \ref{fig:results_county} and \ref{fig:results_state}. Again, these findings should not be interpreted as a failure but a calibration issue that may be mitigated with more observations in a given location. Alternatively, these results raise the hypothesis that our methods might be more accurate on preceding COVID-19 outbreaks with higher incidence, given that $R_t$ is better inferred at larger case numbers \cite{cori_new_2013}. Future studies could address this question at length.  

Our present study has multiple limitations. First and foremost, our county-level analysis was conducted in a subset of  97 counties, not in all 3,006. We selected a subset of US counties based on the local health care capacity to conduct clinical trials. We also considered populous counties with more than one million inhabitants, totaling 97 counties. Future studies can explore our Single and Multiple Source methods in a larger subset of or all the 3,006 US counties, which would allow us to explore our methods' generalisability across all US geographies. From a methodological viewpoint, our Single Source method has shown good predictive power and earliness in anticipating the outbreak onset events. Future studies could refine our analysis by exploring other nuances of the digital time series, such as uptrend magnitude or downtrends associated with decreased COVID-19 activities. Likewise, the Multiple Source method was designed to identify an outbreak onsets but no other properties, such as magnitude and timing. Further studies could investigate the relationship between those features and the digital signals to build more sophisticated early warning systems. Future efforts could also explore other connections between our results and the probabilistic estimation of $R_t$ adopted in this study, such as cumulative probabilities lower bounds for the false alarm rates, among others. It is also important to acknowledge that digital traces usually exhibit specific statistical properties, such as lack of first and second-order stationarity,  that make results hard to interpret and the application of statistical methods potentially inappropriate \cite{rovetta2021reliability}. From a machine learning perspective, our methods would likely benefit from learning from more outbreaks in a given location  \cite{ardabili2020covid,mohamadou2020review}. These increased datasets will probably improve the robustness and performance of our analysis if the underlying relationship between predictors (Internet-based data streams) and outbreaks maintains temporal coherence.

\section*{Data and Methods}
In the following sections, we present the data sources that have been used for our study, along with a detailed explanation of our methods. In this work, we collected several data sources from January 2020 to January 2022 for 97 counties that are potential locations for vaccination trials or have a population of at least one million inhabitants.

\subsection*{Data sources}
\label{section_data}

In this section, we present and describe the epidemiological COVID-19 reports,  COVID-19 and health-related searches from Google Trends API, UpToDate trends, and Twitter microblogs.

\subsubsection*{Official COVID-19 Reports} 

We collected daily COVID-19 case counts from the Johns Hopkins University database \cite{dong2020interactive}. The serial interval (time from symptomatic primary infection to symptomatic secondary infection) was obtained from \cite{ferguson2020impact}. For each US county, we obtained aggregated weekly time series covering the period between January 1, 2020, and January 1, 2022.


\subsubsection*{UpToDate Trends}

UpTodate is a software developed by UpToDate, Inc, a company in the Wolters Kluwer Health division \cite{marion2015pacing}. UpToDate is utilized by physicians and health centers as a tool to search for medical resources. Unlike Google Trends, all information provided in the UpToDate database is edited by experts using rigorous standards. This study used state-level COVID-19 related UpToDate searches from January 2020 to March 2021. We note that UpToDate stopped providing data after that period, and hence we could not evaluate the signal performance from April to December 2021.


\subsubsection*{Google Trends}

 We utilized the Google Trends Application Programming Interface (API) to obtain daily COVID-19-related search terms. For the Single Source method, we chose the following COVID-19 related terms: ``Covid'', Covid19, How Long Does Covid Last, Covid Symptoms and Covid 19 WHO. To account for common COVID-19 symptoms, we also selected Google Trends ``Fever'' and ``Chest Pain''. To account for searches related to vaccination, we also chose the Google Trends terms  ``After Covid Vaccine'', ``Side Effects Of Covid Vaccine'', and ''Effects Of Covid Vaccine''. For the Multiple Source method, we increased the Google Trends dataset with more health-related terms to allow for a purely data-driven emergence of most relevant terms. A list of all terms can be found at the supplementary materials Table \ref{tab:google_searches_list}.

\subsubsection*{Twitter API:} 
The Twitter data were harvested by an automated crawler connecting to Twitter's APIs in a fully automated fashion. The geo-crawler software collects georeferenced social media posts, i.e., tweets with an explicit geospatial reference. 
The geo-crawler requests data from two Twitter endpoints: the REST and streaming APIs. 
The REST API offers several API functionalities to access tweets, including the “search/tweets” endpoint that enables, the collection of tweets from the last seven days in a moving window. This requires a stringently designed collection procedure in order to harvest all provided tweets within the fast-moving time window of the API with a minimal number of requests. In contrast, the streaming API provides a real-time data stream that can be filtered using multiple parameters, including a post's language, location or user IDs.\vspace{12pt}  

The geo-crawler software requests tweets that include location information either as a point coordinate from the mobile device used for tweeting (e.g., GPS) or a rectangular outline based on a geocoded place name. The combination of the REST and streaming APIs makes crawling robust against interruptions or backend issues that inevitably lead to data holes. For example, if data from the streaming API cannot be stored in time, the missing data can be retrieved via the partly redundant REST API.\vspace{12pt} 

All tweets used for this study are located in the USA. To filter the data for COVID-19-relevant tweets, we used a simple keyword list as shown in Box \ref{tab:twitter_term_list}. We opted for keyword-based filtering because of high performance, filtering in near real time, and its simplicity compared to machine-learning based semantic analysis methods. While a machine learning-based semantic clustering method like Guided Latent Dirichlet Allocation (LDA) may generate more comprehensive results (e.g., through identifying co-occurring and unknown terms), 
controlling the ratio between false positives and false negatives requires extensive experimental work and expert knowledge, which is typically a strong limitation when dealing with large datasets.\vspace{12pt} 

\begin{table}[h!]
\centering
\caption{Search term list for Twitter.}
\label{tab:twitter_term_list}
\begin{tabular}{@{} p{15cm}p{3cm}|p{3cm}|p{3cm} @{}}
\toprule
 covid, corona, epidemic, flu, influenza, face mask, spread, virus, infection, fever, panic buying, state of emergency, masks, quarantine, sars, 2019-ncov \\ \bottomrule
\end{tabular}
\end{table}

\subsubsection*{Apple Mobility} 
Apple mobility data were generated by counting the number of requests made to Apple Maps for directions in selected locations. Data sent from users' devices to Apple Maps service is associated with random, rotating identifiers, so Apple does not profile users' information. Data availability in a particular location is based on several factors, including minimum thresholds for direction requests per day.  More details are available at \\ \texttt{https://www.apple.com/covid19/mobility}.

\subsection*{Addressing delays in data}
Our data streams experience specific availability delays. For example,  the most recent values from Google Searches are available up to 36 hours before the current date. In addition, epidemiological reports suffer from backfilling and reporting delays due to post-processing. Thus for Google searches, data reported at time $t$ was shifted to time $t+2$ to address the 36-hour delay. The following table shows the delay (in the number of days) that we imposed on each data source.

\begin{table}[h!]
    \centering
    \begin{tabular}{c|c}
    \toprule
     Source     &   Delay (days)\\
     \midrule
    Epidemiological Reports     &  7 \\
    Google & 2 \\
    UpToDate & 1 \\
    \bottomrule
    \end{tabular}
    
    \label{tab:my_label}
\end{table}

\subsection*{Estimating the effective reproductive number $R_t$  and defining outbreak onsets in COVID-19 activity}

The effective reproductive number $R_t$ is defined as the expected number of secondary cases generated by a primary case infected at time t. It can be used as a near real-time indicator to track trends and changes during an outbreak or to measure the impacts of public health interventions. When $R_t > 1$, we can expect epidemic growth, whereas when $R_t<1$, the epidemic decreases. However, estimating $R_t$ in near real-time can be challenging due to delays in reporting cases and under-ascertainment. While the latter is difficult to correct for general bias in all Rt estimation methods, we overcome reporting delays by using cases by date of onset.\cite{gostic_practical_2020,parag_improved_2021}.\\
The most popular approach for near real-time estimation of R$_t$ is the EpiEstim method introduced by Cori et al. \cite{cori_new_2013}. This method, while powerful, can suffer from edge effects and unstable inference during periods of low incidence\cite{parag_improved_2021}. A recent approach from Parag et al. (EpiFilter)\cite{parag_improved_2021} circumvents some of these issues by applying Bayesian smoothing theory to improve estimate robustness (or minimize noise), especially in low incidence periods and between outbreaks. As early warning signals are desired in exactly such settings, we use this approach as a ground truth signal for outbreak onset. 

\begin{figure}
    \centering
    \includegraphics[width=\textwidth]{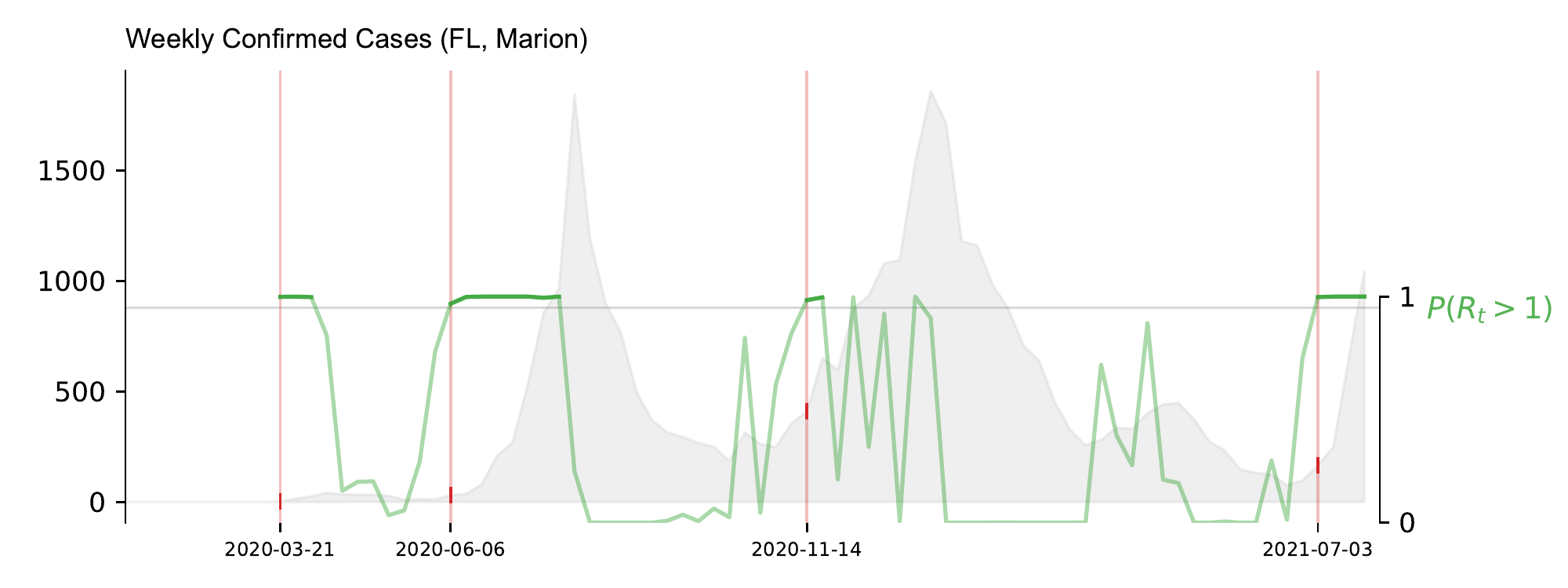}
    \caption{\textbf{Defining Outbreak onsets}. The value of $P(R_t > 1)$ (green line) is calculated for the weekly volume of COVID-19 confirmed cases (gray) in Marion county (FL). A successive increase of COVID-19 confirmed case activity is labeled as an outbreak if $P(R_t > 1) > 0.95$ for 2 weeks or more. Marked events are enclosed within a rectangle.}
    \label{fig:event_definition}
\end{figure}

In this work, we aimed to anticipate sharp increases in COVID-19 activity, using reported by JHU's confirmatory cases to determine outbreak onset events when the effective reproduction number ($R_t$) was probabilistically higher than 1. To this end, we first defined the concept of \textbf{outbreak onset}. In this work, outbreak onsets marked the beginning of an exponential surge for a location's JHU's COVID-19 confirmed cases signal based on the probability $P(R_t>1)$. For details on how this probability is calculated, we refer the reader to the recent work from Parag and Donnelly \cite{parag2021fundamental}. We labeled a given date as the start of an outbreak onset whenever $P(R_t>1)$ crossed the 0.95 threshold for at least two consecutive weeks. An event was considered finished after $P(R_t>1) < 0.05$ (consecutive events that happened within at most 1-month gap were considered a single event).   Figure \ref{fig:event_definition} exhibits a visual explanation of our outbreak onset definition for the COVID-19 confirmed cases activity of Marion county, Florida. We used this definition to identify the onsets that occurred within each the US locations selected for experimentation (97 counties and 50 states), and tested our ability to anticipate such dates utilizing alternative sources of information.
 
\subsection*{Single Source Method}

\begin{figure}
    \centering
    \includegraphics[width=1\textwidth]{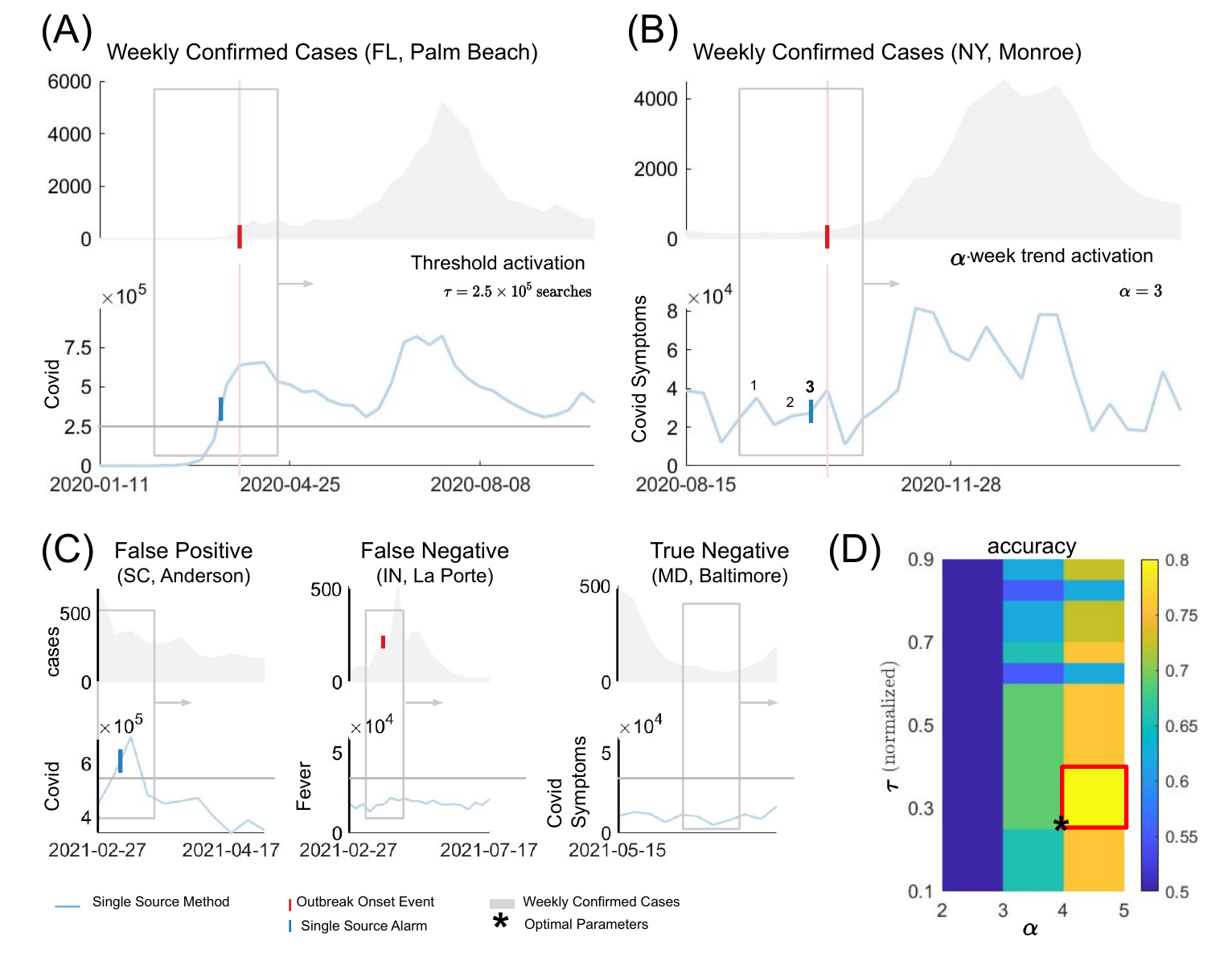}
    \caption{\small{{\bf An Early Warning System for COVID-19 based on single sources (Single Source method).} A moving window spans both predictor (Google trends for the term COVID) and COVID-19 cases. A \emph{Threshold activation} occurs when a predictor signal crosses the value $\tau$. In the example, $\tau = 2.5 \times 10^5$ Google searches for Palm Beach County (FL). The blue tick denotes the week of crossing. The red vertical tick and line denote the week of the outbreak onset event. {\bf(B)} If a number $\alpha$ of increases happen in the predictor signal, an $\alpha$\emph{-week trend activation} takes place. In the example, $\alpha =3$ triggers an alarm for Monroe (NY) county preceding the outbreak onset event. Both {\bf(A)} and {\bf(B)} represent true positive events. {\bf(C)} Definitions of false positive, false negative, and true negative events. {\bf(D)} For the training step, we evaluate the performance of our Single Source method for different thresholds $\tau$ (normalized by the maximum of the signal on the training period) and $\alpha$ values for the $\alpha$-week trend activation.  A colormap with the training accuracy shows the highest rate for $\tau$ and $\alpha$. The example illustrates optimal parameters $\tau$ and $\alpha$ as the lowest values such that accuracy is maximized, indicated by a black star in the lower leftmost corner of the red rectangle.}}
    \label{fig:methods_single_signal}
\end{figure}

As a way to evaluate the predictive power of individual signals, we developed the Single Source method that explores the volume increase of available digital data to generate early warnings of COVID-19 activity. In Fig \ref{fig:methods_single_signal} (A) and (B), we illustrate the two possible alarm events. Given a 6-week time window spanning both digital and COVID-19 cases data, a \emph{Threshold activation} is defined if the digital signal crosses a given threshold $\tau$ (Fig \ref{fig:methods_single_signal} A). Fig \ref{fig:methods_single_signal} (B) shows a different kind of alarm, where a number $\alpha$ of increases happen within the 6-week moving window ($\alpha=3$ in the example). In this case, we define an $\alpha$\emph{-week trend activation}. A true positive occurs when an alarm in the digital signal (either threshold or $\alpha$-week trend) precedes the outbreak onset event  within the six-week moving window. Fig   \ref{fig:methods_single_signal} (C) illustrates other possible outcomes in the Single Source method. We only show threshold activations for simplicity.  A false positive occurs when the digital signal activates (either through threshold or $\alpha$-week trend) but no outbreak onset event occur. Conversely, a false negative may occur when an event occur but no alarm is triggered by the individual signal. Finally, a true negative takes place when no alarms in the individual signal or outbreak onset events occur within the 6-week moving window.

For the training step, we choose multiple threshold values, which are normalized by the maximum of the digital signal in the training period, resulting in a scale from 0.1 to 0.9. We also select possible values for $\alpha$ in the $\alpha-$week trend activation ranging between 2 and 5 signal uptrends within the 6-week moving window. As the window progresses in time, we compute the number of true positives (TP), false positives (FP), false negatives (FN), and true negatives (TN). With that information, we obtain performance metrics such as accuracy as depicted in Fig \ref{fig:methods_single_signal} (D). From the collection of  $\tau$ and $\alpha$ maximizing performance metrics (red rectangle in Fig \ref{fig:methods_single_signal} (D)), we choose those minimum parameter values (black star) to promote the earliness of our prediction method.  In our simulations, we chose to simultaneously optimize accuracy ($\frac{TP+TN}{P+N}$), precision ($\frac{TP}{P}$) and the negative predictive values ($\frac{TN}{N}$) where $P = TP + FP$ and $N = TN + FN$. By doing this, we arrive at the minimum optimal $\tau$ and $\alpha$ maximizing those three quantities at the same time. If there are no such parameter values, we optimize over precision only.      

\subsection*{Multiple Source Method}

Here we describe our Multiple Source method. Given that our earlier work had shown the feasibility of implementing digital streams as alternative proxies to track state-level COVID-19 activity \cite{kogan2021early}, we hypothesized that a county-level early warning system could provide a higher resolution picture of the pandemic at near real-time. In what follows, we define the variables of our early warning system and provide a detailed explanation for each step of our analysis.

\subsubsection*{Definitions}

To describe our Multiple Source method in detail, in this section we start defining the following variables:

\begin{itemize}
    \item $\mathbf{y}$: Target signal to track using our early warning system methodologies. In this work, $\mathbf{y}$ is the given by  the time series $\{P(R_t>1)\}_{t>0}.$  We define $y_t$ as the value of $\mathbf{y}$ at week $t$.
    \item $\mathbf{X} = \left \{ \mathbf{x_i}, i=1,2,...,N \right \}$: The set of all alternative proxies, i.e,  official epidemiological reports, Google Search volumes, social network activity, Twitter microblogs, among others. We refer the reader to our data sources section \ref{section_data} for a detailed description of these datasets.We define $x_{i,t}$ as the value of $\mathbf{x_i}$ at week $t$.
    \item $t_{e}$: Event onset date and can be used to represent outbreak onset events (denoted by $t_{e,\mathbf{y}}$) generated by the time series $\{P(R_t>1)\}_{t>0}$ or proxy events (denoted by $t_{e,\mathbf{X_i}}$) generated by the lambda approach.
    \item $\mathbf{M} = EWS(\mathbf{y}, \mathbf{X})$: An early warning system output as a function of all available data from both target and proxies. We define $M_t$ as the value of $\mathbf{M}$ at week $t$.
\end{itemize}


\subsubsection*{Lambda Approach: capturing increasing activity trends in digital proxies}
To define and label proxy events, we utilized a measure of a proxy trend henceforth called $\lambda$. Here we emphasize that $\lambda$ was calculated for our proxy variables, i.e., the Google Trends Signals, Twitter, and other digital traces. However, we derived this measure from the  classic SIR epidemic model equations

\begin{equation}
\begin{split}
\dot{S} = - \beta SI \\
\dot{I} = \beta SI - \gamma I 
\end{split}
\end{equation}{}

where $S$ and $I$ represent the populations of susceptible and infected individuals and $r = N-S-I$ represents the recovered class in a constant population of size $N$. For an inter-outbreak period, where the number of infected individuals $I$ is very low, the susceptible pool of individuals S can be assumed as a constant parameter $S = S^*$. In this case,  the SIR system of equations reduces to

\begin{equation}
\begin{split}
    \dot{S} = 0 \\
    \dot{I} = (\beta S^{*} - \gamma)I 
\end{split}    
\end{equation}

and the solution to the equation for the number of infected individuals $I$ is then an exponential function with the intrinsic growth rate

\begin{equation}
    \lambda = \beta  S^{*} - \gamma,
\end{equation}

which is directly proportional to the size of the susceptibility pool. In the context of epidemiological models, $\lambda$ can be thought of as an indicator of the susceptible population that is easily interpretable: if $\lambda > 1$, then $I$ increases exponentially. Moreover, $\lambda$ can be estimated by linear regression, as the coefficient of a 1-lag autoregressive model with no intercept $ I_t = \lambda I_{t-1}$. In this work, we utilize a 3-week time window to estimate the value of $\lambda$.

\textbf{Implementing $\lambda$ in practice:} For each week $t$, we calculated the value of $\lambda_t$ using the most recent data available. Retrospectively, a proxy event was defined as a period where the value $\lambda_t>1$ for at least two weeks. If two time periods satisfied such conditions but were only separated at most four weeks apart, they were merged into a single, bigger period.

\subsubsection*{Implementing the Multiple Source Method}

Our Early Warning system $\mathbf{M}$  was designed to track COVID-19 outbreak events by iteratively identifying proxy signals $\mathbf{x}_i$ that have experienced events preceding a target ($\mathbf{y}$) events, and combining them into a single output signal $M_t$, where  $0\leq M_t <1$ for all $t>0$. This output signal $M_t$ can be interpreted as an indicator for  $\mathbf{y}$ experiencing an outbreak in the near future (up to 6 weeks). Given $\mathbf{y}$ and $\mathbf{X}$ up to week $t$ for an specific location,  we implement our method as follows:

\textbf{Data Preparation:} We begin by identifying all the outbreak events for $\mathbf{y}$ and the proxy events in our dataset $\mathbf{X}$. These events are used to establish a performance ranking for each proxy $\mathbf{x_i}$ via the following labeling rules (applied individually to each proxy):

\begin{itemize}
    \item \textbf{True Positive (TP):} If a proxy event $t_{e,\mathbf{x}_i}$ precedes an outbreak onset event  $t_{e,\mathbf{y}}$ by at most $6$ weeks, we label $t_{e,\mathbf{x}_i}$ as true positive.

    \item \textbf{False Positive (FP):} If a proxy event $t_{e,\mathbf{x}_i}$ occurs but there is no outbreak onset event happening in the next $6$ weeks, then we label that proxy's event as a false positive, also referred as false alarm.
    \item \textbf{False Negative (FN):} If an outbreak onset event $t_{e,\mathbf{y}}$ occurs but no proxy event occurs at most $6$ weeks retrospectively, then we identify that as a false negative.
\end{itemize}

The number of TP, FP and FN is then calculated for each proxy $\mathbf{x_i}$. With the aim of focusing specifically in the detection of outbreak events and avoiding a highly imbalanced dataset, we do not optimize for True Negatives events, since every week where the activity was not an outbreak onset or within the outbreak period (as marked by $R_t$) would be a potential true negative.

\textbf{Feature Selection:} A performance ranking is created based on the TP, FP and FN values of each proxy. Given our main objective is to identify outbreak events for $t_{e,\mathbf{y}}$ earlier in time, we prioritize signals maximizing true positives  while minimizing false positives. A subset of six proxies $\chi = \{\mathbf{x_1}, \mathbf{x_2}, \cdots, \mathbf{x_6}  \}$ with highest TP and lowest FP is then selected  to be used as the input for our early warning system. In the case more than six are fit to be used within our early warning system, those six proxies are then selected randomly.

\textbf{Combining Selected Proxies into an Early Warning System:} The value $M_t$ of our early warning system is given by the expression

\begin{equation*}
    M_t =  2S(n_{e,t})-1 
\end{equation*}

where

\begin{equation*}
    n_{e,t} =  \sum_{\mathbf{x_i}\in \chi}g(\mathbf{x_i},t), \\
\end{equation*}

and 

\begin{equation*}
 g(\mathbf{x_i},t) = \begin{cases} 
      1  & \  \text{if} \ t-k\leq t_{e,\mathbf{x_i}} \leq t  \\
      0 & \text{otherwise}
   \end{cases}    
\end{equation*}

is the number of proxy events that have occurred in the past $k=3$ weeks, and


\begin{equation*}
    S(x) = \frac{1}{1+e^{-x}}
\end{equation*}
 
is the well known sigmoid function. Intuitively, the quantity $M_t$ changes between 0 and 0.9951 ($\sim 1$) depending on the number of proxies events (from zero to six) within the past $k$ weeks before the week $t$.
 
\textbf{Thresholding:} Although each proxy $\mathbf{x}_i$  has been selected based on the premise that its events have successfully tracked our target $\mathbf{y}$ during training (and thus, even a low value $M_t$ may convey some relevant information about an incoming event), there may be some instances when not all proxies in the set $\chi$ have activated prior to an outbreak onset event. Similarly, as we compute the value of $M_t$ every week, there may be some instances when a proxy (for example, Google search activity spiking due to non-COVID related events) triggers an alarm, and thus increasing the value of $M_t$. Based on these possibilities, and with the purpose of having a more practical way of interpreting $M_t$, we define a decision threshold $\tau$ which we use to map  $M_t$ into a "yes/no" methodology. If $M_t>\tau$, then we interpret it as our early warning system is expecting an outbreak onset event happening in the near future. In practice, we find this threshold by computing the performance of our early warning system as a function of the threshold $\tau$ (similar to an ROC curve) and selecting the threshold that maximizes a metric of interest (precision, for example).

Given that $M_t$ is calculated every week using a k-week moving window, our early warning system events consist in a subset of weeks in which $M_t > \tau$ (this is to be contrasted to a $\mathbf{x_i}$ event, which consists only of a single week). Based on this behavior, we define a different set of event labels which we use to compute the performance of our early warning system. We label our events in the following way:

\begin{itemize}
    \item \textbf{True Positive (TP):} If $M_t > \tau$ prior $t_{e,\mathbf{y}}$ within a retrospective window of $w$ weeks.
    
    \item \textbf{Strict False Positive (FP):} If $M_t > \tau$ but no outbreak onset event $t_{e,\mathbf{y}}$ is observed in the following $w$ weeks.
    
    \item \textbf{Relaxed False Positive (RFP):} Given a set of $m$ subsequent weeks where $M_t > \tau$, we count all dates as a single misfire (in comparison to a strict misfire, which would $m$ misfires instead of 1).
    
    \item \textbf{False Negative (FN):} If $t_{e,\mathbf{y}}$ occurs but $M_t > \tau$ is not observed retrospectively within a $w$ week window.
\end{itemize}

We also define

\begin{align}
    m_1 = \frac{TP}{TP+FP+FN}
    \label{m1}
\end{align}
and 
\begin{align}
    m_2 = \frac{TP}{TP+FN+RFP}.
    \label{m2}
\end{align}

Given that the $M_t$ values are inherently connected overtime (if $M_t > \tau$ then it may take some weeks to deactivate), optimizing for $m_1$ usually lead to high threshold values as a way to avoid a high number of strict false positives (i.e. keeping $M_t > \tau$ a high number of weeks). Although this is desirable in practice (having an system which only activates when it is certain that an event is going to happen, and below the threshold otherwise), $m_1$ may cause our early warning system to overfit given a) The very low number of events scenario may not convey enough information to find an adequate threshold and b) If $M_t > 0$ for the first occasion at week $t$, then it takes at least $k$ weeks to change its value. On the other hand, optimizing for $m_2$ encourages the selection of lower threshold values as it does not penalize $M_t > \tau$ being true for a long period. Nonetheless, $m_2$ may go as low as a threshold of $\tau=0$ if there are no False Negatives  or if the number of Relaxed False Positives is the same below a certain threshold value. We thus opt for optimizing the averaged sum of $m_1$ and $m_2$. Precisely, our optimal threshold $\tau_{opt}$ is given by $$\tau_{opt} = \text{argmax}_{\tau \in _{[0,1]}} \left( \frac{m_1 + m_2}{2}\right),$$ where both $m_1$ and $m_2$ given by Eqs. \ref{m1} and \ref{m2} depend on $\tau$.

\textbf{Out of sample experiment description:}  For a given location with outbreak onset events $\mathbf{E} = \left \{ e_1, e_2,,..., e_{n} \right \}$, we utilized $e_1$ as the first event for training, with the aim to predict $e_2$. As a next step, we incorporate $e_2$ into the training dataset, and thus use both $e_1$ and $e_2$ to train our early warning system, and predict the following event $e_3$. We repeated this procedure until the last event $e_{n}$. We also defined the out-of-sample period as the time interval between the week when the last outbreak event in the training dataset ended and the week when the out-of-sample outbreak onset occurred. 

We counted the number of times when a method triggered an alarm based on the following labels:

\begin{itemize}
    \item \textbf{Early warnings:} When the early warning system triggered an alarm preceding the outbreak onset event with at most six weeks in advance. 
    \item \textbf{Synchronous warnings:} When the early warning system triggered an alarm on the same date as the outbreak onset
    \item \textbf{Late Warnings:} Events where the early warning system triggered an alarm up to two weeks later than the outbreak onset event.
    \item For the Multiple Source method, we considered a \textbf{soft warnings} when the output of the early warning system increased at least  $70\%$ of the decision threshold and the outbreak onset event was successfully observed six weeks after.  
    \item \textbf{Missed outbreaks:} when an outbreak onset event was observed, but no alarm was registered within the observation window. 
    \item In terms of false alarms, we differentiated between two different scenarios: 
    \begin{enumerate}
    \item A \textbf{False Alarm with no increase} occurred when the system triggered an alarm, but no event \emph{and} no increase in COVID-19 cases were observed in the following six weeks.
    \item A \textbf{False Alarm with increase observed} occurred when the early warning system triggered an alarm and was followed by an increase in COVID-19 cases.
    \end{enumerate}

\end{itemize}

\section*{Acknowledgments}
We thank Wolters Kluwer Health’s UpToDate team for providing resource-specific data. We also thank Dr. Bill Hanage and the members of the Center for Communicable Disease Dynamics (CCDD) at the Harvard T.H. Chan School of Public Health for insightful comments that improved this work.
\textbf{Funding:} This project has been funded (in part) by contract 200-2016-91779 with the Centers for Disease Control and Prevention.   
Disclaimer: The findings, conclusions, and views expressed are those of the author(s) and do not necessarily represent the official position of the Centers for Disease Control and Prevention (CDC). Research reported in this publication was also partially supported by the National Institute of General Medical Sciences of the National Institutes of Health under Award Number R01GM130668. The content is solely the responsibility of the authors and does not necessarily represent the official views of the National Institutes of Health. M.S., L.M.S. and L.C. thank the Johnson \& Johnson Foundation for providing institutional research funds and Johnson \& Johnson Global Public Health for their support
\textbf{Author contributions:} M.S, L.M.S, and L.C conceptualized the study, interpreted the results and wrote the first draft of the paper; M.S supervised the study. L.M.S and L.C designed and implemented the methodologies, collected and analyzed the data. C.P and K.V.P contributed to the methods section. B.R provided Twitter data. S.M and A.M provided provided additional subject matter expertise in shaping our methodologies.  All authors contributed to and approved the final version of the manuscript. Competing interests: The authors declare that they have no competing interests. Data and materials availability: All data needed to evaluate the conclusions in the paper are present in the paper and/or the Supplementary Information. Additional data related to this paper may be requested from the authors. Some data sources may require establishing data sharing agreements with the data providers, which have been established with a diverse set of research teams. Code can be made available upon request.

\newpage
\section*{Supplementary Information}

\begin{landscape}
\begin{table}[]
    \centering
    \tiny
    \setlength\tabcolsep{1.5pt}
    \begin{tabular}{lllllllllllll}
\toprule
Model & \multicolumn{3}{l}{Naive} & \multicolumn{3}{l}{Single Source ("How long does covid last?")} & \multicolumn{3}{l}{Single Source ("side effects of vaccine")} & \multicolumn{3}{l}{Multiple Source} \\
Time & Before 2021-12-01 & After 2021-12-01 &       Total &                           Before 2021-12-01 & After 2021-12-01 &       Total &                         Before 2021-12-01 & After 2021-12-01 &       Total & Before 2021-12-01 & After 2021-12-01 &       Total \\
\midrule
Total Events                                &        305 (100\%) &        62 (100\%) &  367 (100\%) &                                  305 (100\%) &        62 (100\%) &  367 (100\%) &                                305 (100\%) &        62 (100\%) &  367 (100\%) &        305 (100\%) &        62 (100\%) &  367 (100\%) \\
Early Warning                               &         305 (100\%) &         62 (100\%) &   367 (100\%) &                                   202 (66\%) &         35 (56\%) &   237 (65\%) &                                 184 (60\%) &         43 (69\%) &   227 (62\%) &         175 (57\%) &         38 (61\%) &   213 (58\%) \\
Sync Warning                                &           0 (0\%) &           0 (0\%) &     0 (0\%) &                                     23 (8\%) &           2 (3\%) &     25 (7\%) &                                   16 (5\%) &           3 (5\%) &     19 (5\%) &          37 (12\%) &           3 (5\%) &    40 (11\%) \\
Late Warning                                &            1 (0\%) &           0 (0\%) &      0 (0\%) &                                     26 (9\%) &           4 (6\%) &     30 (8\%) &                                   14 (5\%) &           1 (2\%) &     15 (4\%) &          34 (11\%) &           3 (5\%) &    37 (10\%) \\
Soft Warning            &            0 (0\%) &           0 (0\%) &      0 (0\%) &                                      0 (0\%) &           0 (0\%) &      0 (0\%) &                                    0 (0\%) &           0 (0\%) &      0 (0\%) &          45 (15\%) &         16 (26\%) &    61 (17\%) \\
Missed Outbreaks                            &            0 (0\%) &           0 (0\%) &      0 (0\%) &                                    54 (18\%) &         21 (34\%) &    75 (20\%) &                                  91 (30\%) &         15 (24\%) &   106 (29\%) &           14 (5\%) &           2 (3\%) &     16 (4\%) \\
Warning, increase observed &               224 &               28 &         252 &                                         122 &               17 &         139 &                                       148 &               23 &         171 &                34 &                2 &          36 \\
False Alarm                                 &               502 &              110 &         612 &                                         280 &               47 &         327 &                                       374 &              105 &         479 &                95 &               19 &         114 \\
\bottomrule
\end{tabular}

    \caption{\textbf{\textbf{A Summary of the results at the County-level.} Here \emph{EWS} stands for Early Warning System.}}
    \label{tab:table_summary_countylevel}
\end{table}

\begin{table}[]
    \centering
    \tiny
    \setlength\tabcolsep{1.5pt}

    \begin{tabular}{lllllllllllll}
\toprule
Model & \multicolumn{3}{l}{Naive} & \multicolumn{3}{l}{Single Source ("How long does covid last?")} & \multicolumn{3}{l}{Single Source ("chest pain")} & \multicolumn{3}{l}{Multiple Source} \\
Time & Before 2021-12-01 & After 2021-12-01 &       Total &                           Before 2021-12-01 & After 2021-12-01 &       Total &            Before 2021-12-01 & After 2021-12-01 &       Total & Before 2021-12-01 & After 2021-12-01 &       Total \\
\midrule
Total Events                                &        183 (100\%) &        19 (100\%) &  202 (100\%) &                                  183 (100\%) &        19 (100\%) &  202 (100\%) &                   183 (100\%) &        19 (100\%) &  202 (100\%) &        183 (100\%) &        19 (100\%) &  202 (100\%) \\
Early Warning                               &         160 (87\%) &         18 (95\%) &   178 (88\%) &                                   115 (63\%) &         13 (68\%) &   128 (63\%) &                    108 (59\%) &         12 (63\%) &   120 (59\%) &          82 (45\%) &         17 (89\%) &    99 (49\%) \\
Sync Warning                                &          22 (12\%) &           1 (5\%) &    23 (11\%) &                                     11 (6\%) &          2 (11\%) &     13 (6\%) &                      12 (7\%) &           0 (0\%) &     12 (6\%) &          22 (12\%) &           0 (0\%) &    22 (11\%) \\
Late Warning                                &            1 (1\%) &           0 (0\%) &      1 (0\%) &                                     17 (9\%) &           1 (5\%) &     18 (9\%) &                      12 (7\%) &           0 (0\%) &     12 (6\%) &          24 (13\%) &           0 (0\%) &    24 (12\%) \\
Soft Warning            &            0 (0\%) &           0 (0\%) &      0 (0\%) &                                      0 (0\%) &           0 (0\%) &      0 (0\%) &                       0 (0\%) &           0 (0\%) &      0 (0\%) &          39 (21\%) &          2 (11\%) &    41 (20\%) \\
Missed Outbreaks                            &            0 (0\%) &           0 (0\%) &      0 (0\%) &                                    40 (22\%) &          3 (16\%) &    43 (21\%) &                     51 (28\%) &          7 (37\%) &    58 (29\%) &           16 (9\%) &           0 (0\%) &     16 (8\%) \\
Warning, increase observed &                74 &                8 &          82 &                                          57 &                6 &          63 &                           35 &                6 &          41 &                19 &                1 &          20 \\
False Alarm                                 &               171 &               39 &         210 &                                         104 &               15 &         119 &                          112 &                9 &         121 &                30 &                4 &          34 \\
\bottomrule
\end{tabular}
    
    \caption{\textbf{A Summary of the results at the state-level.} Here \emph{EWS} stands for Early Warning System.}
    \label{tab:table_summary_statelevel}
\end{table}
\end{landscape}

\begin{table}[]
    \centering
    
    \scriptsize
    \begin{tabular}{lllllll}
\toprule
N = 367 & Early Warning & Sync Warning & Late Warning & Missed Outbreaks & Warning, activity increases & False Alarm \\
\midrule
gt\_howLongDoesCovidLast\_rt  &     237 (65\%) &      25 (7\%) &      30 (8\%) &         74 (20\%) &                         139 &         327 \\
gt\_sideEffectsOfVaccine\_rt  &     227 (62\%) &      19 (5\%) &      15 (4\%) &        105 (29\%) &                         171 &         479 \\
gt\_chestPain\_rt             &     201 (55\%) &      19 (5\%) &     38 (10\%) &        108 (29\%) &                         128 &         301 \\
gt\_covid\_19Who              &     169 (46\%) &      25 (7\%) &      20 (5\%) &        152 (41\%) &                          87 &         269 \\
JHU\_deaths\_rt               &     147 (40\%) &      16 (4\%) &      31 (8\%) &        172 (47\%) &                          93 &         249 \\
gt\_covidSymptoms\_rt         &     145 (40\%) &      30 (8\%) &     50 (14\%) &        141 (38\%) &                          58 &         122 \\
gt\_afterCovidVaccine\_rt     &     141 (38\%) &      20 (5\%) &      29 (8\%) &        176 (48\%) &                         182 &         463 \\
JHU\_cases\_rt                &     135 (37\%) &     59 (16\%) &     58 (16\%) &        114 (31\%) &                         100 &         204 \\
gt\_effectsOfCovidVaccine\_rt &     107 (29\%) &      14 (4\%) &      20 (5\%) &        225 (61\%) &                         110 &         374 \\
gt\_fever\_rt                 &      85 (23\%) &      24 (7\%) &      21 (6\%) &        236 (64\%) &                          40 &         107 \\
gt\_covid\_19                 &      79 (22\%) &      22 (6\%) &     53 (14\%) &        212 (58\%) &                          52 &         127 \\
gt\_covid\_rt                 &      51 (14\%) &      19 (5\%) &     65 (18\%) &        231 (63\%) &                          10 &          54 \\
twitter\_state\_rt            &      41 (11\%) &      14 (4\%) &     56 (15\%) &        255 (69\%) &                          20 &          40 \\
up2date\_rt\_parag            &      37 (10\%) &       7 (2\%) &      24 (7\%) &        298 (81\%) &                          10 &          32 \\
\bottomrule
\end{tabular}
    \caption{\textbf{County-Level results for the  Single Source Method} }
    \label{tab:table_singlesource_countylevel}
\end{table}

\begin{table}[]
    \centering
    \scriptsize
    \begin{tabular}{lllllll}
\toprule
N=202 & Early Warning & Sync Warning & Late Warning & Missed Outbreaks & Warning, activity increases & False Alarm \\
\midrule
gt\_howLongDoesCovidLast\_rt  &     128 (63\%) &      13 (6\%) &      18 (9\%) &         42 (21\%) &                          63 &         119 \\
gt\_chestPain\_rt             &     120 (59\%) &      12 (6\%) &      12 (6\%) &         57 (28\%) &                          41 &         121 \\
gt\_covid\_19Who              &     112 (55\%) &      11 (5\%) &      19 (9\%) &         59 (29\%) &                          51 &         104 \\
gt\_sideEffectsOfVaccine\_rt  &      96 (48\%) &      12 (6\%) &      18 (9\%) &         75 (37\%) &                          56 &         165 \\
gt\_covidSymptoms\_rt         &      88 (44\%) &       8 (4\%) &      15 (7\%) &         90 (45\%) &                          28 &          45 \\
JHU\_cases\_rt                &      78 (39\%) &     22 (11\%) &     33 (16\%) &         68 (34\%) &                          29 &          97 \\
gt\_afterCovidVaccine\_rt     &      68 (34\%) &       8 (4\%) &       8 (4\%) &        117 (58\%) &                          67 &         149 \\
JHU\_deaths\_rt               &      65 (32\%) &      13 (6\%) &      13 (6\%) &        110 (54\%) &                          28 &          72 \\
gt\_effectsOfCovidVaccine\_rt &      63 (31\%) &       6 (3\%) &       7 (3\%) &        125 (62\%) &                          30 &          90 \\
gt\_covid\_19                 &      46 (23\%) &       9 (4\%) &     30 (15\%) &        116 (57\%) &                          21 &          54 \\
gt\_fever\_rt                 &      45 (22\%) &       5 (2\%) &      15 (7\%) &        136 (67\%) &                          24 &          42 \\
twitter\_state\_rt            &      27 (13\%) &       8 (4\%) &     28 (14\%) &        138 (68\%) &                          10 &          11 \\
gt\_covid\_rt                 &      26 (13\%) &       5 (2\%) &     21 (10\%) &        149 (74\%) &                          10 &          14 \\
up2date\_rt\_parag            &       18 (9\%) &       5 (2\%) &      14 (7\%) &        164 (81\%) &                           8 &           9 \\
\bottomrule
\end{tabular}
    \caption{\textbf{State-Level results for the  Single Source Method} }
    \label{tab:table_singlesource_statelevel}
\end{table}

\begin{table}[]
    \centering
            	\begin{adjustwidth}{-0.5in}{0in} 
    \tiny \begin{tabular}{lllllllll}
\toprule
{} & Early Warning & Sync Warning & Late Warning & Soft Warning & Missed Outbreaks & Warning & False Alarm & Non-deployed \\
{} &  &  &  & &  & increase observed & False Alarm & Non-deployed \\
Data Source(s) Used                                        &               &              &              &                                  &                  &                                             &             &              \\
\midrule
Local Epi, Google Searches, Neighbor Data          &      99 (49\%) &     22 (11\%) &     24 (12\%) &                         41 (20\%) &          16 (8\%) &                                          20 &          34 &       0 (0\%) \\
Local Epi, Google Searches, Neighbor Data, UpTo... &      99 (49\%) &     22 (11\%) &     24 (12\%) &                         41 (20\%) &          16 (8\%) &                                        20.0 &        34.0 &       0 (0\%) \\
Local Epi, Google Searches, Neighbor Data, Twitter &      99 (49\%) &     22 (11\%) &     24 (12\%) &                         41 (20\%) &          16 (8\%) &                                        20.0 &        34.0 &       0 (0\%) \\
Local Epi, Google Searches, Neighbor Data, Twit... &      99 (49\%) &     22 (11\%) &     24 (12\%) &                         41 (20\%) &          16 (8\%) &                                        20.0 &        34.0 &       0 (0\%) \\
Local Epi, Google Searches, Neighbor Data, Appl... &      99 (49\%) &     21 (10\%) &     24 (12\%) &                         42 (21\%) &          16 (8\%) &                                        20.0 &        36.0 &       0 (0\%) \\
Local Epi, Google Searches, Neighbor Data, Appl... &      99 (49\%) &     21 (10\%) &     24 (12\%) &                         42 (21\%) &          16 (8\%) &                                        20.0 &        36.0 &       0 (0\%) \\
Local Epi, Google Searches, Neighbor Data, Twit... &      99 (49\%) &     21 (10\%) &     24 (12\%) &                         42 (21\%) &          16 (8\%) &                                        20.0 &        36.0 &       0 (0\%) \\
Local Epi, Google Searches, Neighbor Data, Twit... &      99 (49\%) &     21 (10\%) &     24 (12\%) &                         42 (21\%) &          16 (8\%) &                                        20.0 &        36.0 &       0 (0\%) \\
Google Searches, Neighbor Data                     &      97 (48\%) &     23 (11\%) &     25 (12\%) &                         41 (20\%) &          16 (8\%) &                                        20.0 &        37.0 &       0 (0\%) \\
Google Searches, Neighbor Data, UpToDate           &      97 (48\%) &     23 (11\%) &     25 (12\%) &                         41 (20\%) &          16 (8\%) &                                        20.0 &        37.0 &       0 (0\%) \\
Google Searches, Neighbor Data, Twitter            &      97 (48\%) &     23 (11\%) &     25 (12\%) &                         41 (20\%) &          16 (8\%) &                                        20.0 &        37.0 &       0 (0\%) \\
Google Searches, Neighbor Data, Twitter, UpToDate  &      97 (48\%) &     23 (11\%) &     25 (12\%) &                         41 (20\%) &          16 (8\%) &                                        20.0 &        37.0 &       0 (0\%) \\
Google Searches, Neighbor Data, Apple Mobility     &      97 (48\%) &     22 (11\%) &     25 (12\%) &                         42 (21\%) &          16 (8\%) &                                        20.0 &        39.0 &       0 (0\%) \\
Google Searches, Neighbor Data, Apple Mobility,... &      97 (48\%) &     22 (11\%) &     25 (12\%) &                         42 (21\%) &          16 (8\%) &                                        20.0 &        39.0 &       0 (0\%) \\
Google Searches, Neighbor Data, Twitter, Apple ... &      97 (48\%) &     22 (11\%) &     25 (12\%) &                         42 (21\%) &          16 (8\%) &                                        20.0 &        39.0 &       0 (0\%) \\
Google Searches, Neighbor Data, Twitter, Apple ... &      97 (48\%) &     22 (11\%) &     25 (12\%) &                         42 (21\%) &          16 (8\%) &                                        20.0 &        39.0 &       0 (0\%) \\
Google Searches, Apple Mobility                    &      93 (46\%) &      19 (9\%) &     23 (11\%) &                         50 (25\%) &          17 (8\%) &                                        20.0 &        38.0 &       0 (0\%) \\
Google Searches, Apple Mobility, UpToDate          &      93 (46\%) &      19 (9\%) &     23 (11\%) &                         50 (25\%) &          17 (8\%) &                                        20.0 &        37.0 &       0 (0\%) \\
Google Searches,                                   &      92 (46\%) &     20 (10\%) &     23 (11\%) &                         50 (25\%) &          17 (8\%) &                                        19.0 &        36.0 &       0 (0\%) \\
Google Searches, UpToDate                          &      92 (46\%) &     20 (10\%) &     23 (11\%) &                         50 (25\%) &          17 (8\%) &                                        19.0 &        35.0 &       0 (0\%) \\
Google Searches, Twitter, Apple Mobility           &      92 (46\%) &     20 (10\%) &     24 (12\%) &                         49 (24\%) &          17 (8\%) &                                        20.0 &        38.0 &       0 (0\%) \\
Google Searches, Twitter, Apple Mobility, UpToDate &      92 (46\%) &     20 (10\%) &     24 (12\%) &                         49 (24\%) &          17 (8\%) &                                        20.0 &        37.0 &       0 (0\%) \\
Local Epi, Google Searches, Apple Mobility         &      91 (45\%) &     23 (11\%) &     22 (11\%) &                         49 (24\%) &          17 (8\%) &                                        20.0 &        34.0 &       0 (0\%) \\
Local Epi, Google Searches, Apple Mobility, UpT... &      91 (45\%) &     23 (11\%) &     22 (11\%) &                         49 (24\%) &          17 (8\%) &                                        20.0 &        33.0 &       0 (0\%) \\
Local Epi, Google Searches, Twitter, Apple Mobi... &      91 (45\%) &     23 (11\%) &     23 (11\%) &                         48 (24\%) &          17 (8\%) &                                        20.0 &        34.0 &       0 (0\%) \\
Local Epi, Google Searches, Twitter, Apple Mobi... &      91 (45\%) &     23 (11\%) &     23 (11\%) &                         48 (24\%) &          17 (8\%) &                                        20.0 &        33.0 &       0 (0\%) \\
Google Searches, Twitter                           &      91 (45\%) &     21 (10\%) &     24 (12\%) &                         49 (24\%) &          17 (8\%) &                                        19.0 &        36.0 &       0 (0\%) \\
Google Searches, Twitter, UpToDate                 &      91 (45\%) &     21 (10\%) &     24 (12\%) &                         49 (24\%) &          17 (8\%) &                                        19.0 &        35.0 &       0 (0\%) \\
Local Epi, Google Searches                         &      90 (45\%) &     24 (12\%) &     22 (11\%) &                         50 (25\%) &          16 (8\%) &                                        19.0 &        32.0 &       0 (0\%) \\
Local Epi, Google Searches, UpToDate               &      90 (45\%) &     24 (12\%) &     22 (11\%) &                         50 (25\%) &          16 (8\%) &                                        19.0 &        31.0 &       0 (0\%) \\
Local Epi, Google Searches, Twitter                &      90 (45\%) &     24 (12\%) &     23 (11\%) &                         49 (24\%) &          16 (8\%) &                                        19.0 &        32.0 &       0 (0\%) \\
Local Epi, Google Searches, Twitter, UpToDate      &      90 (45\%) &     24 (12\%) &     23 (11\%) &                         49 (24\%) &          16 (8\%) &                                        19.0 &        31.0 &       0 (0\%) \\
Local Epi, Twitter, Apple Mobility, UpToDate       &      86 (43\%) &     27 (13\%) &      18 (9\%) &                         44 (22\%) &         21 (10\%) &                                        17.0 &        60.0 &       6 (3\%) \\
Local Epi, Apple Mobility                          &      85 (42\%) &     29 (14\%) &       8 (4\%) &                         24 (12\%) &         38 (19\%) &                                        18.0 &        58.0 &      18 (9\%) \\
Local Epi, Twitter, Apple Mobility                 &      84 (42\%) &     28 (14\%) &      17 (8\%) &                         43 (21\%) &         22 (11\%) &                                        17.0 &        61.0 &       8 (4\%) \\
Local Epi, Apple Mobility, UpToDate                &      81 (40\%) &     27 (13\%) &       8 (4\%) &                         38 (19\%) &         36 (18\%) &                                        18.0 &        58.0 &      12 (6\%) \\
Twitter, Apple Mobility                            &      80 (40\%) &      17 (8\%) &      15 (7\%) &                         45 (22\%) &         32 (16\%) &                                        17.0 &        58.0 &      13 (6\%) \\
Twitter, Apple Mobility, UpToDate                  &      80 (40\%) &      17 (8\%) &      16 (8\%) &                         48 (24\%) &         30 (15\%) &                                        19.0 &        55.0 &      11 (5\%) \\
Local Epi, Twitter, UpToDate                       &      79 (39\%) &     39 (19\%) &     27 (13\%) &                           3 (1\%) &         26 (13\%) &                                        10.0 &        20.0 &     28 (14\%) \\
Neighbor Data, Apple Mobility, UpToDate            &      76 (38\%) &     21 (10\%) &      16 (8\%) &                         59 (29\%) &         29 (14\%) &                                        10.0 &        27.0 &       1 (0\%) \\
Local Epi, Neighbor Data, Apple Mobility, UpToDate &      76 (38\%) &     21 (10\%) &     20 (10\%) &                         57 (28\%) &         27 (13\%) &                                        10.0 &        24.0 &       1 (0\%) \\
Neighbor Data, Twitter, Apple Mobility, UpToDate   &      76 (38\%) &     20 (10\%) &      15 (7\%) &                         62 (31\%) &         28 (14\%) &                                         9.0 &        21.0 &       1 (0\%) \\
Apple Mobility,                                    &      75 (37\%) &      14 (7\%) &       5 (2\%) &                         21 (10\%) &         57 (28\%) &                                        16.0 &        56.0 &     30 (15\%) \\
Neighbor Data, Twitter, Apple Mobility             &      74 (37\%) &     20 (10\%) &      14 (7\%) &                         63 (31\%) &         28 (14\%) &                                        11.0 &        22.0 &       3 (1\%) \\
Neighbor Data, Apple Mobility                      &      74 (37\%) &      19 (9\%) &      15 (7\%) &                         60 (30\%) &         30 (15\%) &                                        10.0 &        25.0 &       4 (2\%) \\
Local Epi, Neighbor Data, Apple Mobility           &      73 (36\%) &     22 (11\%) &     20 (10\%) &                         57 (28\%) &         26 (13\%) &                                         9.0 &        23.0 &       4 (2\%) \\
Local Epi, Neighbor Data, Twitter, Apple Mobili... &      73 (36\%) &     21 (10\%) &     20 (10\%) &                         62 (31\%) &         25 (12\%) &                                         8.0 &        20.0 &       1 (0\%) \\
Apple Mobility, UpToDate                           &      73 (36\%) &      14 (7\%) &       7 (3\%) &                         32 (16\%) &         53 (26\%) &                                        18.0 &        58.0 &     23 (11\%) \\
Local Epi, Twitter                                 &      72 (36\%) &     40 (20\%) &     28 (14\%) &                           0 (0\%) &         23 (11\%) &                                         8.0 &        22.0 &     39 (19\%) \\
Local Epi, Neighbor Data, Twitter, Apple Mobility  &      70 (35\%) &     23 (11\%) &     20 (10\%) &                         60 (30\%) &         26 (13\%) &                                         9.0 &        21.0 &       3 (1\%) \\
Twitter, UpToDate                                  &      67 (33\%) &      18 (9\%) &     37 (18\%) &                           0 (0\%) &         36 (18\%) &                                         9.0 &        21.0 &     44 (22\%) \\
Neighbor Data, Twitter, UpToDate                   &      62 (31\%) &     23 (11\%) &      17 (8\%) &                         55 (27\%) &         38 (19\%) &                                         6.0 &        14.0 &       7 (3\%) \\
Neighbor Data, UpToDate                            &      59 (29\%) &     25 (12\%) &      19 (9\%) &                         52 (26\%) &         40 (20\%) &                                         7.0 &        15.0 &       7 (3\%) \\
Local Epi, Neighbor Data, Twitter, UpToDate        &      59 (29\%) &     23 (11\%) &     22 (11\%) &                         59 (29\%) &         32 (16\%) &                                         5.0 &        13.0 &       7 (3\%) \\
Local Epi, Neighbor Data, UpToDate                 &      58 (29\%) &     25 (12\%) &     23 (11\%) &                         55 (27\%) &         34 (17\%) &                                         7.0 &        13.0 &       7 (3\%) \\
Neighbor Data, Twitter                             &      56 (28\%) &     27 (13\%) &      16 (8\%) &                         54 (27\%) &         39 (19\%) &                                         6.0 &        13.0 &      10 (5\%) \\
Twitter,                                           &      56 (28\%) &      17 (8\%) &     39 (19\%) &                           0 (0\%) &         32 (16\%) &                                         7.0 &        19.0 &     58 (29\%) \\
Neighbor Data,                                     &      55 (27\%) &     23 (11\%) &      17 (8\%) &                         53 (26\%) &         43 (21\%) &                                         6.0 &        13.0 &      11 (5\%) \\
Local Epi, Neighbor Data                           &      53 (26\%) &     24 (12\%) &     22 (11\%) &                         56 (28\%) &         36 (18\%) &                                         6.0 &        11.0 &      11 (5\%) \\
Local Epi, Neighbor Data, Twitter                  &      51 (25\%) &     28 (14\%) &     22 (11\%) &                         57 (28\%) &         34 (17\%) &                                         5.0 &        11.0 &      10 (5\%) \\
Local Epi, UpToDate                                &      37 (18\%) &     48 (24\%) &      13 (6\%) &                           0 (0\%) &         43 (21\%) &                                         8.0 &         9.0 &     61 (30\%) \\
Local Epi,                                         &      23 (11\%) &     43 (21\%) &      10 (5\%) &                           0 (0\%) &         21 (10\%) &                                         4.0 &         6.0 &    105 (52\%) \\
UpToDate,                                          &       17 (8\%) &       6 (3\%) &       6 (3\%) &                           0 (0\%) &         49 (24\%) &                                         5.0 &         3.0 &    124 (61\%) \\
\bottomrule
\end{tabular}
  \end{adjustwidth}
    \caption{\textbf{State-level performance of the Multiple Source method for different data sources used for the training step.} An additional column, 'Non-deployed,' indicates the number of outbreak events where our Multiple Source method did not find a reasonable set of proxies to fit a model. Thus, an early warning system was not deployed to forecast the out-of-sample period.}
    \label{tab:my_label}
  
\end{table}

\begin{table}[]
    \centering
            	\begin{adjustwidth}{-0.5in}{0in} 
    \tiny \begin{tabular}{lllllllll}
\toprule
{} & Early Warning & Sync Warning & Late Warning & Soft Warning & Missed Outbreaks & Warning & False Alarm & Non-deployed \\
{} &  &  &  & &  & increase observed & False Alarm & Non-deployed \\
Data Source(s) Used                                        &               &              &              &                                  &                  &                                             &             &              \\
\midrule
Local Epi, Google Searches                         &     224 (61\%) &      27 (7\%) &     36 (10\%) &                         64 (17\%) &          16 (4\%) &                                          43 &         117 &       0 (0\%) \\
Local Epi, Google Searches, UpToDate               &     224 (61\%) &      27 (7\%) &     36 (10\%) &                         64 (17\%) &          16 (4\%) &                                        43.0 &       117.0 &       0 (0\%) \\
Local Epi, Google Searches, Apple Mobility         &     224 (61\%) &      27 (7\%) &     36 (10\%) &                         64 (17\%) &          16 (4\%) &                                        43.0 &       117.0 &       0 (0\%) \\
Local Epi, Google Searches, Apple Mobility, UpT... &     224 (61\%) &      27 (7\%) &     36 (10\%) &                         64 (17\%) &          16 (4\%) &                                        43.0 &       117.0 &       0 (0\%) \\
Local Epi, Google Searches, Twitter                &     224 (61\%) &      27 (7\%) &     36 (10\%) &                         64 (17\%) &          16 (4\%) &                                        43.0 &       117.0 &       0 (0\%) \\
Local Epi, Google Searches, Twitter, UpToDate      &     224 (61\%) &      27 (7\%) &     36 (10\%) &                         64 (17\%) &          16 (4\%) &                                        43.0 &       117.0 &       0 (0\%) \\
Local Epi, Google Searches, Twitter, Apple Mobi... &     224 (61\%) &      27 (7\%) &     36 (10\%) &                         64 (17\%) &          16 (4\%) &                                        43.0 &       117.0 &       0 (0\%) \\
Local Epi, Google Searches, Twitter, Apple Mobi... &     224 (61\%) &      27 (7\%) &     36 (10\%) &                         64 (17\%) &          16 (4\%) &                                        43.0 &       117.0 &       0 (0\%) \\
Google Searches,                                   &     220 (60\%) &      23 (6\%) &      32 (9\%) &                         78 (21\%) &          14 (4\%) &                                        42.0 &       114.0 &       0 (0\%) \\
Google Searches, UpToDate                          &     220 (60\%) &      23 (6\%) &      32 (9\%) &                         78 (21\%) &          14 (4\%) &                                        42.0 &       114.0 &       0 (0\%) \\
Google Searches, Apple Mobility                    &     220 (60\%) &      23 (6\%) &      32 (9\%) &                         78 (21\%) &          14 (4\%) &                                        42.0 &       114.0 &       0 (0\%) \\
Google Searches, Apple Mobility, UpToDate          &     220 (60\%) &      23 (6\%) &      32 (9\%) &                         78 (21\%) &          14 (4\%) &                                        42.0 &       114.0 &       0 (0\%) \\
Google Searches, Twitter                           &     220 (60\%) &      23 (6\%) &      32 (9\%) &                         78 (21\%) &          14 (4\%) &                                        42.0 &       114.0 &       0 (0\%) \\
Google Searches, Twitter, UpToDate                 &     220 (60\%) &      23 (6\%) &      32 (9\%) &                         78 (21\%) &          14 (4\%) &                                        42.0 &       114.0 &       0 (0\%) \\
Google Searches, Twitter, Apple Mobility           &     220 (60\%) &      23 (6\%) &      32 (9\%) &                         78 (21\%) &          14 (4\%) &                                        42.0 &       114.0 &       0 (0\%) \\
Google Searches, Twitter, Apple Mobility, UpToDate &     220 (60\%) &      23 (6\%) &      32 (9\%) &                         78 (21\%) &          14 (4\%) &                                        42.0 &       114.0 &       0 (0\%) \\
Google Searches, Neighbor Data                     &     216 (59\%) &      31 (8\%) &     35 (10\%) &                         71 (19\%) &          14 (4\%) &                                        35.0 &       110.0 &       0 (0\%) \\
Google Searches, Neighbor Data, UpToDate           &     216 (59\%) &      31 (8\%) &     35 (10\%) &                         71 (19\%) &          14 (4\%) &                                        35.0 &       110.0 &       0 (0\%) \\
Google Searches, Neighbor Data, Apple Mobility     &     216 (59\%) &      31 (8\%) &     35 (10\%) &                         71 (19\%) &          14 (4\%) &                                        35.0 &       110.0 &       0 (0\%) \\
Google Searches, Neighbor Data, Apple Mobility,... &     216 (59\%) &      31 (8\%) &     35 (10\%) &                         71 (19\%) &          14 (4\%) &                                        35.0 &       110.0 &       0 (0\%) \\
Google Searches, Neighbor Data, Twitter            &     216 (59\%) &      31 (8\%) &     35 (10\%) &                         71 (19\%) &          14 (4\%) &                                        35.0 &       110.0 &       0 (0\%) \\
Google Searches, Neighbor Data, Twitter, UpToDate  &     216 (59\%) &      31 (8\%) &     35 (10\%) &                         71 (19\%) &          14 (4\%) &                                        35.0 &       110.0 &       0 (0\%) \\
Google Searches, Neighbor Data, Twitter, Apple ... &     216 (59\%) &      31 (8\%) &     35 (10\%) &                         71 (19\%) &          14 (4\%) &                                        35.0 &       110.0 &       0 (0\%) \\
Google Searches, Neighbor Data, Twitter, Apple ... &     216 (59\%) &      31 (8\%) &     35 (10\%) &                         71 (19\%) &          14 (4\%) &                                        35.0 &       110.0 &       0 (0\%) \\
Local Epi, Google Searches, Neighbor Data          &     213 (58\%) &     40 (11\%) &     37 (10\%) &                         61 (17\%) &          16 (4\%) &                                        36.0 &       114.0 &       0 (0\%) \\
Local Epi, Google Searches, Neighbor Data, UpTo... &     213 (58\%) &     40 (11\%) &     37 (10\%) &                         61 (17\%) &          16 (4\%) &                                        36.0 &       114.0 &       0 (0\%) \\
Local Epi, Google Searches, Neighbor Data, Appl... &     213 (58\%) &     40 (11\%) &     37 (10\%) &                         61 (17\%) &          16 (4\%) &                                        36.0 &       114.0 &       0 (0\%) \\
Local Epi, Google Searches, Neighbor Data, Appl... &     213 (58\%) &     40 (11\%) &     37 (10\%) &                         61 (17\%) &          16 (4\%) &                                        36.0 &       114.0 &       0 (0\%) \\
Local Epi, Google Searches, Neighbor Data, Twitter &     213 (58\%) &     40 (11\%) &     37 (10\%) &                         61 (17\%) &          16 (4\%) &                                        36.0 &       114.0 &       0 (0\%) \\
Local Epi, Google Searches, Neighbor Data, Twit... &     213 (58\%) &     40 (11\%) &     37 (10\%) &                         61 (17\%) &          16 (4\%) &                                        36.0 &       114.0 &       0 (0\%) \\
Local Epi, Google Searches, Neighbor Data, Twit... &     213 (58\%) &     40 (11\%) &     37 (10\%) &                         61 (17\%) &          16 (4\%) &                                        36.0 &       114.0 &       0 (0\%) \\
Local Epi, Google Searches, Neighbor Data, Twit... &     213 (58\%) &     40 (11\%) &     37 (10\%) &                         61 (17\%) &          16 (4\%) &                                        36.0 &       114.0 &       0 (0\%) \\
Local Epi, Neighbor Data, Twitter, UpToDate        &     135 (37\%) &     59 (16\%) &     41 (11\%) &                         65 (18\%) &         37 (10\%) &                                        29.0 &        60.0 &      30 (8\%) \\
Local Epi, Neighbor Data, Twitter                  &     133 (36\%) &     60 (16\%) &     42 (11\%) &                         65 (18\%) &         35 (10\%) &                                        28.0 &        60.0 &      32 (9\%) \\
Local Epi, Neighbor Data, UpToDate                 &     131 (36\%) &     57 (16\%) &     41 (11\%) &                         66 (18\%) &         41 (11\%) &                                        27.0 &        58.0 &      31 (8\%) \\
Local Epi, Twitter, Apple Mobility, UpToDate       &     130 (35\%) &     40 (11\%) &      16 (4\%) &                         37 (10\%) &         82 (22\%) &                                        17.0 &        50.0 &     62 (17\%) \\
Local Epi, Neighbor Data, Twitter, Apple Mobili... &     129 (35\%) &     60 (16\%) &      33 (9\%) &                         83 (23\%) &         48 (13\%) &                                        29.0 &        61.0 &      14 (4\%) \\
Neighbor Data, Twitter, UpToDate                   &     128 (35\%) &     53 (14\%) &     37 (10\%) &                         70 (19\%) &         39 (11\%) &                                        27.0 &        60.0 &     40 (11\%) \\
Local Epi, Neighbor Data                           &     127 (35\%) &     58 (16\%) &     40 (11\%) &                         63 (17\%) &          34 (9\%) &                                        27.0 &        56.0 &     45 (12\%) \\
Local Epi, Neighbor Data, Twitter, Apple Mobility  &     126 (34\%) &     59 (16\%) &      34 (9\%) &                         87 (24\%) &         46 (13\%) &                                        28.0 &        62.0 &      15 (4\%) \\
Neighbor Data, Twitter                             &     126 (34\%) &     55 (15\%) &     38 (10\%) &                         69 (19\%) &         36 (10\%) &                                        26.0 &        61.0 &     43 (12\%) \\
Local Epi, Neighbor Data, Apple Mobility, UpToDate &     125 (34\%) &     57 (16\%) &      33 (9\%) &                         87 (24\%) &         50 (14\%) &                                        27.0 &        60.0 &      15 (4\%) \\
Local Epi, Twitter, UpToDate                       &     125 (34\%) &     51 (14\%) &      23 (6\%) &                           3 (1\%) &         49 (13\%) &                                        16.0 &        40.0 &    116 (32\%) \\
Local Epi, Twitter, Apple Mobility                 &     125 (34\%) &     41 (11\%) &      19 (5\%) &                          33 (9\%) &         84 (23\%) &                                        22.0 &        61.0 &     65 (18\%) \\
Neighbor Data, UpToDate                            &     124 (34\%) &     52 (14\%) &     37 (10\%) &                         70 (19\%) &         43 (12\%) &                                        25.0 &        58.0 &     41 (11\%) \\
Neighbor Data, Twitter, Apple Mobility, UpToDate   &     123 (34\%) &     53 (14\%) &      29 (8\%) &                         88 (24\%) &         54 (15\%) &                                        30.0 &        61.0 &      20 (5\%) \\
Local Epi, Twitter                                 &     122 (33\%) &     51 (14\%) &      22 (6\%) &                           1 (0\%) &         44 (12\%) &                                        19.0 &        50.0 &    127 (35\%) \\
Local Epi, Neighbor Data, Apple Mobility           &     121 (33\%) &     58 (16\%) &      33 (9\%) &                         82 (22\%) &         53 (14\%) &                                        27.0 &        58.0 &      20 (5\%) \\
Neighbor Data, Twitter, Apple Mobility             &     121 (33\%) &     53 (14\%) &      30 (8\%) &                         90 (25\%) &         51 (14\%) &                                        29.0 &        63.0 &      22 (6\%) \\
Neighbor Data,                                     &     120 (33\%) &     53 (14\%) &     36 (10\%) &                         67 (18\%) &         35 (10\%) &                                        25.0 &        57.0 &     56 (15\%) \\
Neighbor Data, Apple Mobility, UpToDate            &     120 (33\%) &     51 (14\%) &      29 (8\%) &                         90 (25\%) &         56 (15\%) &                                        28.0 &        60.0 &      21 (6\%) \\
Neighbor Data, Apple Mobility                      &     116 (32\%) &     52 (14\%) &      29 (8\%) &                         85 (23\%) &         58 (16\%) &                                        28.0 &        59.0 &      27 (7\%) \\
Local Epi, Apple Mobility, UpToDate                &      92 (25\%) &     44 (12\%) &      21 (6\%) &                         37 (10\%) &         93 (25\%) &                                        17.0 &        44.0 &     80 (22\%) \\
Local Epi, UpToDate                                &      85 (23\%) &     56 (15\%) &      26 (7\%) &                           1 (0\%) &         62 (17\%) &                                        14.0 &        33.0 &    137 (37\%) \\
Twitter, Apple Mobility, UpToDate                  &      84 (23\%) &       9 (2\%) &       2 (1\%) &                          21 (6\%) &        105 (29\%) &                                        14.0 &        36.0 &    146 (40\%) \\
Local Epi, Apple Mobility                          &      82 (22\%) &     46 (13\%) &      21 (6\%) &                          29 (8\%) &         91 (25\%) &                                        16.0 &        46.0 &     98 (27\%) \\
Twitter, Apple Mobility                            &      80 (22\%) &       7 (2\%) &       5 (1\%) &                          19 (5\%) &        104 (28\%) &                                        14.0 &        36.0 &    152 (41\%) \\
Local Epi,                                         &      75 (20\%) &     54 (15\%) &      24 (7\%) &                           0 (0\%) &         42 (11\%) &                                        11.0 &        32.0 &    172 (47\%) \\
Twitter, UpToDate                                  &      69 (19\%) &       6 (2\%) &       7 (2\%) &                           0 (0\%) &          27 (7\%) &                                        11.0 &        19.0 &    258 (70\%) \\
Twitter,                                           &      62 (17\%) &       6 (2\%) &       8 (2\%) &                           0 (0\%) &          10 (3\%) &                                        10.0 &        15.0 &    281 (77\%) \\
Apple Mobility, UpToDate                           &       27 (7\%) &       3 (1\%) &       1 (0\%) &                          24 (7\%) &        125 (34\%) &                                         7.0 &        20.0 &    187 (51\%) \\
Apple Mobility,                                    &       20 (5\%) &       3 (1\%) &       1 (0\%) &                          15 (4\%) &        106 (29\%) &                                         6.0 &        21.0 &    222 (60\%) \\
UpToDate,                                          &        9 (2\%) &       2 (1\%) &       4 (1\%) &                           0 (0\%) &         49 (13\%) &                                         2.0 &         2.0 &    303 (83\%) \\
\bottomrule
\end{tabular}
\end{adjustwidth}
    \caption{\textbf{County-level performance of the Multiple Source with different data sources used for the training step.} An additional column, 'Non-deployed,' indicates the number of outbreak events where our Multiple Source method did not find a reasonable set of proxies to fit a model. Thus, an early warning system was not deployed to forecast the out-of-sample period.}
    \label{tab:my_label}
\end{table}

\begin{figure}
    \centering
    \includegraphics[width=\textwidth,height=\textheight,keepaspectratio]{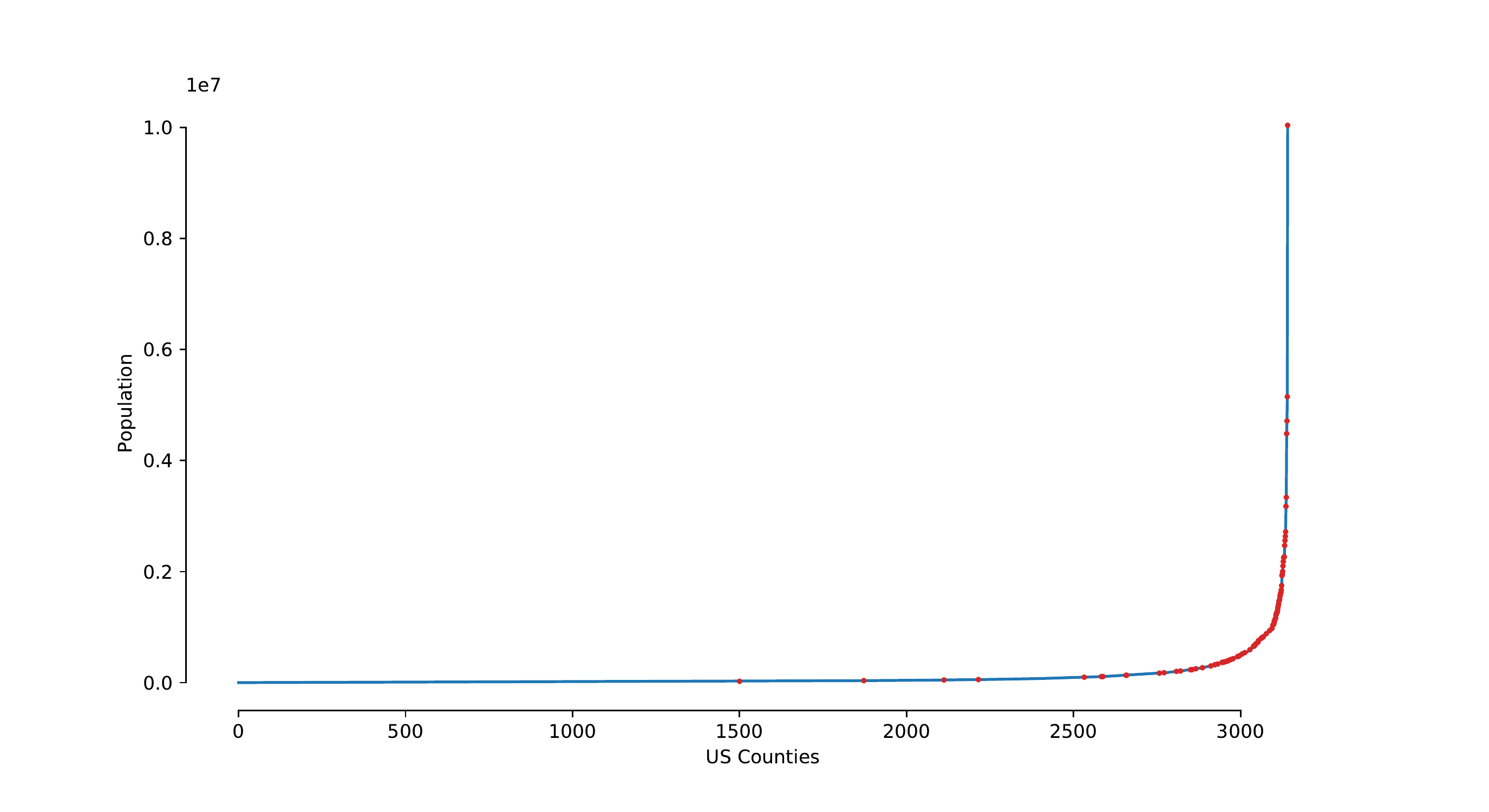} 
    \caption{\textbf{Population sizes for every county within the United States.} Our selected counties are marked with red dots.}
    \label{fig:my_label}
\end{figure}


\newpage

\begin{figure}
    \centering
    \includegraphics[width=0.9\textwidth,height=\textheight,keepaspectratio]{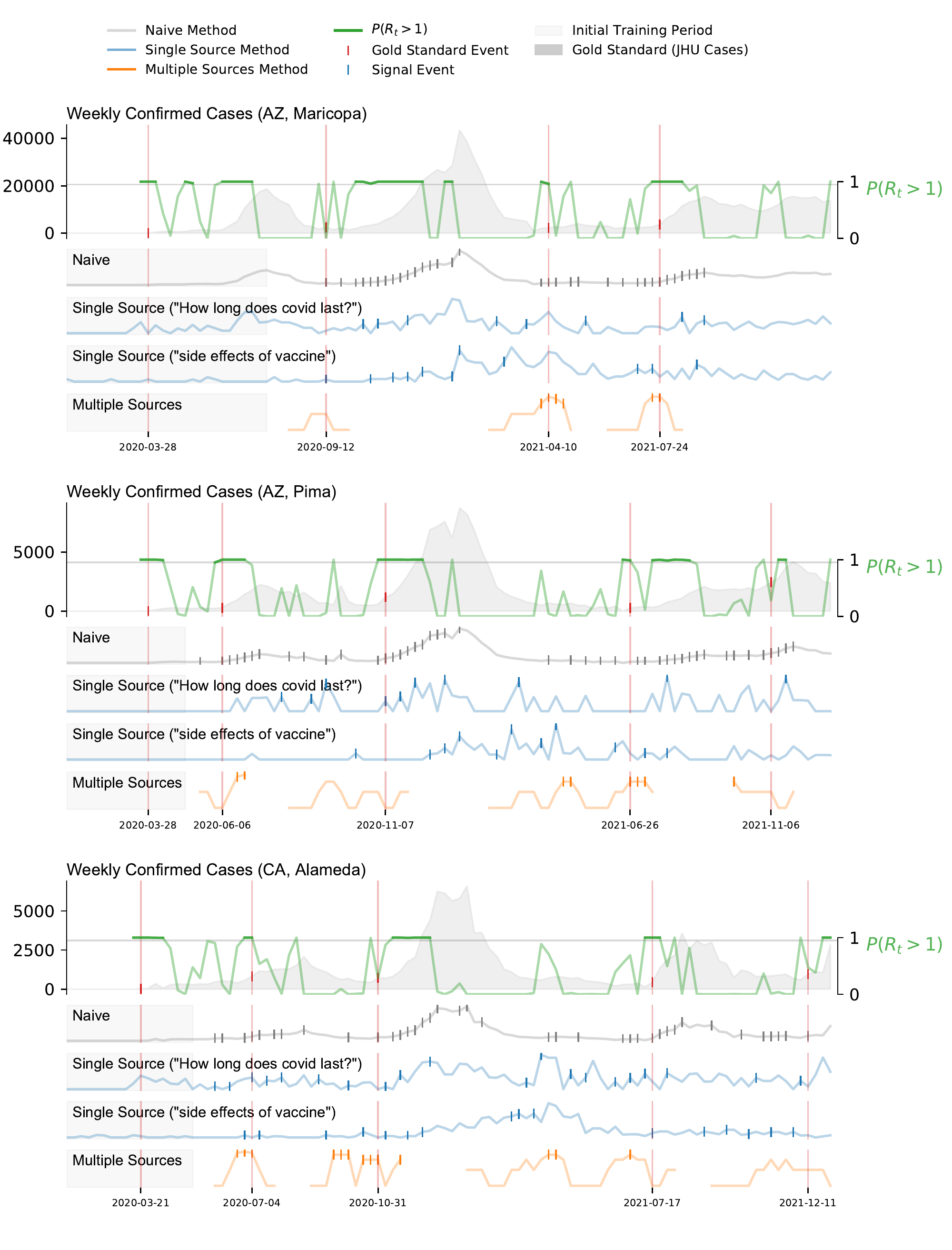}
    \caption{Graphical representation of the reproductive number, $R_t$, along with the weekly confirmed COVID-19 cases (gray-filled curve in the top), and three representative early warning methods (Naive, Single and Multiple Source) at county level.}
    \label{fig:1_county}
\end{figure}

\begin{figure}
    \centering
    \includegraphics[width=0.9\textwidth,height=\textheight,keepaspectratio]{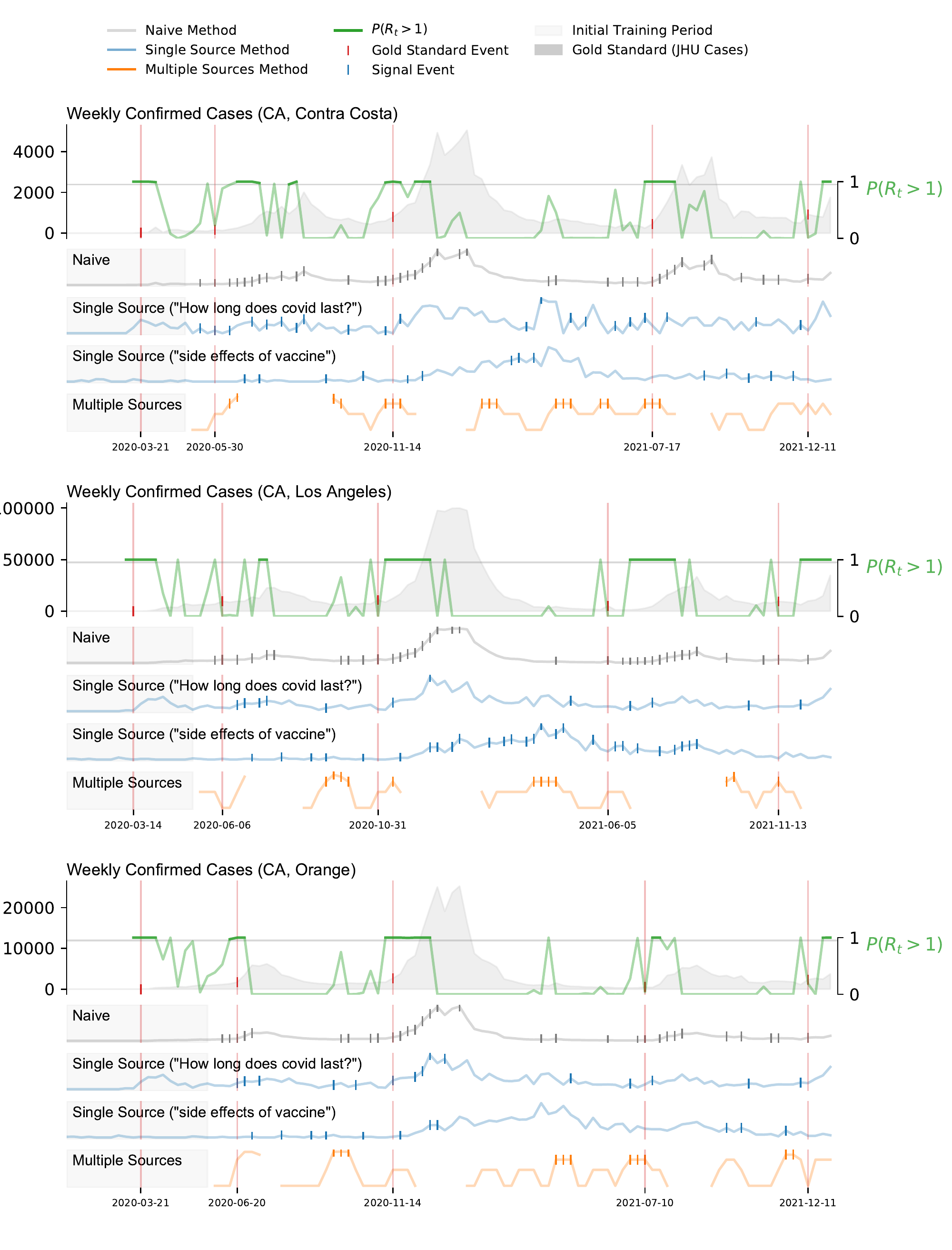}
    \caption{Graphical representation of the reproductive number, $R_t$, along with the weekly confirmed COVID-19 cases (gray-filled curve in the top), and three representative early warning methods (Naive, Single and Multiple Source) at county level.}
    \label{fig:2_county}
\end{figure} 

\begin{figure}
    \centering
    \includegraphics[width=0.9\textwidth,height=\textheight,keepaspectratio]{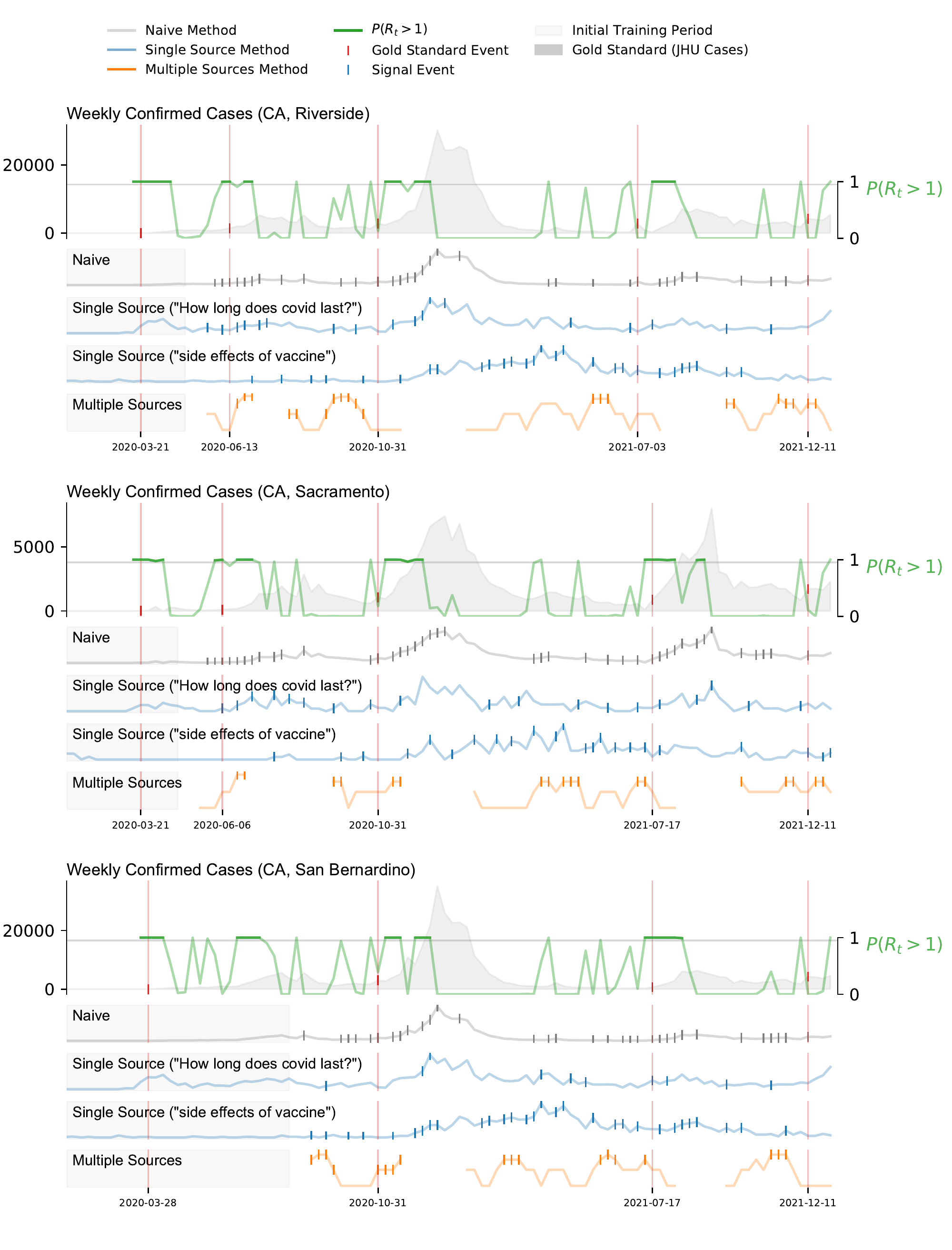}
    \caption{Graphical representation of the reproductive number, $R_t$, along with the weekly confirmed COVID-19 cases (gray-filled curve in the top), and three representative early warning methods (Naive, Single and Multiple Source) at county level.}
    \label{fig:3_county}
\end{figure} 

\begin{figure}
    \centering
    \includegraphics[width=0.9\textwidth,height=\textheight,keepaspectratio]{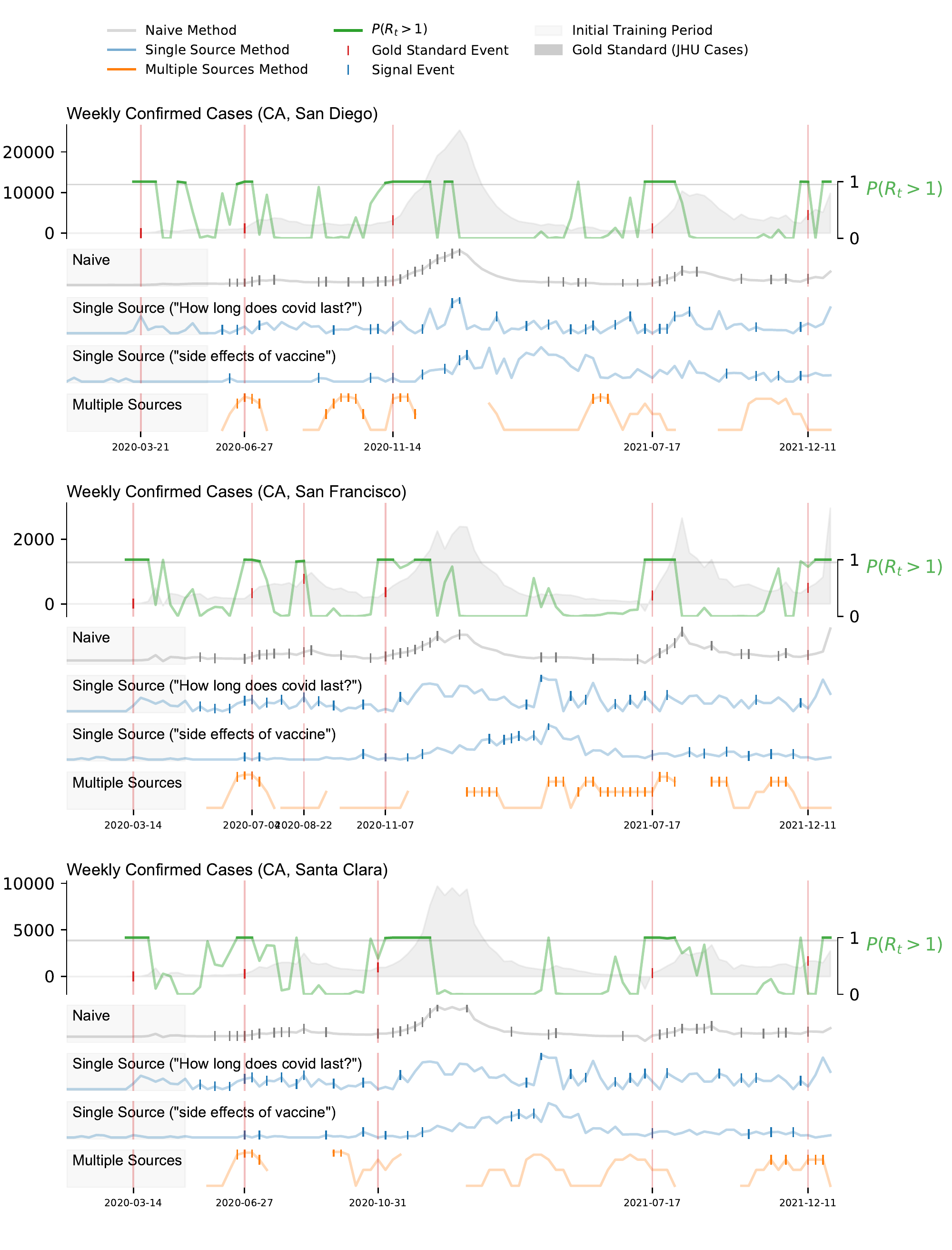}
    \caption{Graphical representation of the reproductive number, $R_t$, along with the weekly confirmed COVID-19 cases (gray-filled curve in the top), and three representative early warning methods (Naive, Single and Multiple Source) at county level.}
    \label{fig:4_county}
\end{figure} 

\begin{figure}
    \centering
    \includegraphics[width=0.9\textwidth,height=\textheight,keepaspectratio]{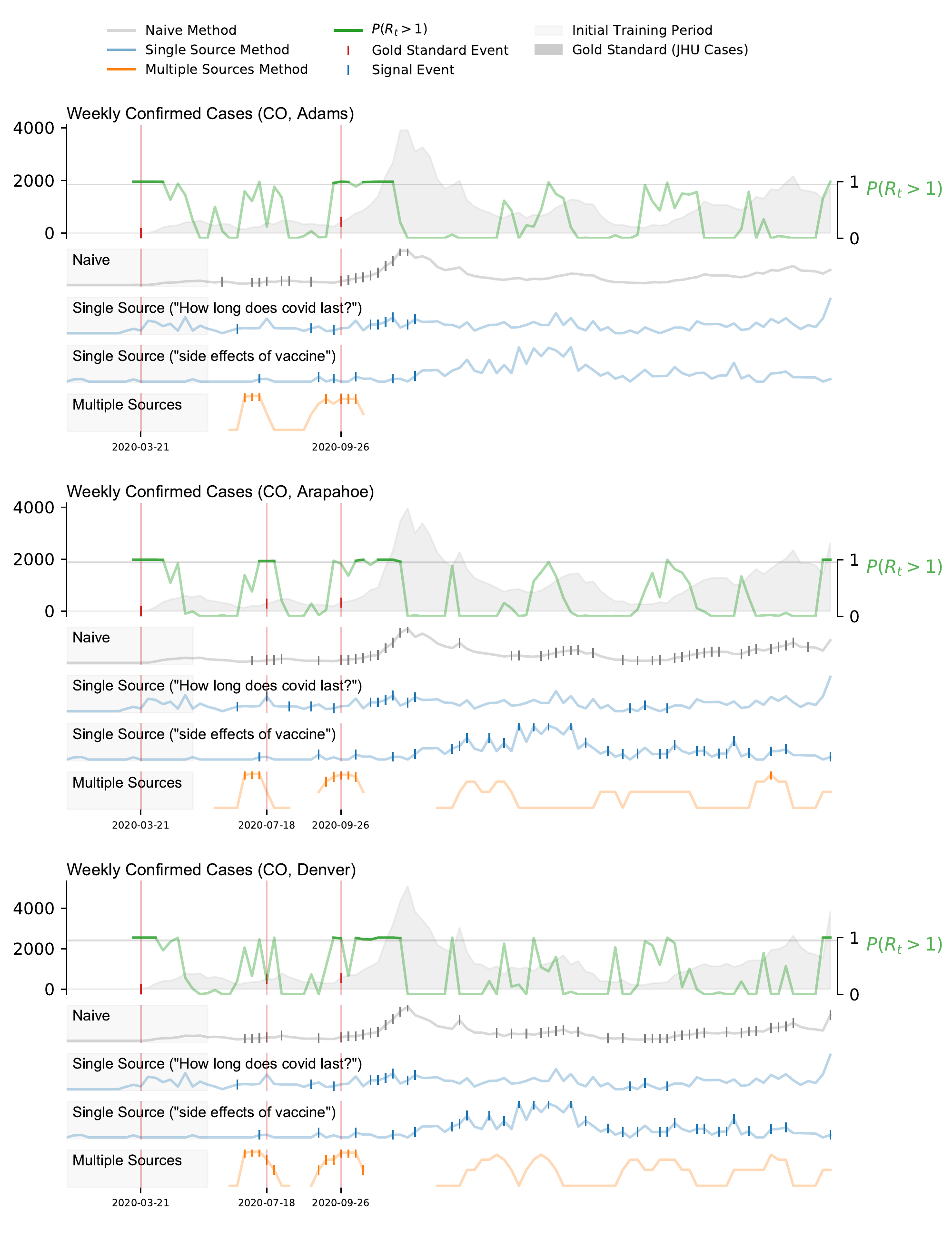}
    \caption{Graphical representation of the reproductive number, $R_t$, along with the weekly confirmed COVID-19 cases (gray-filled curve in the top), and three representative early warning methods (Naive, Single and Multiple Source) at county level.}
    \label{fig:5_county}
\end{figure} 

\begin{figure}
    \centering
    \includegraphics[width=0.9\textwidth,height=\textheight,keepaspectratio]{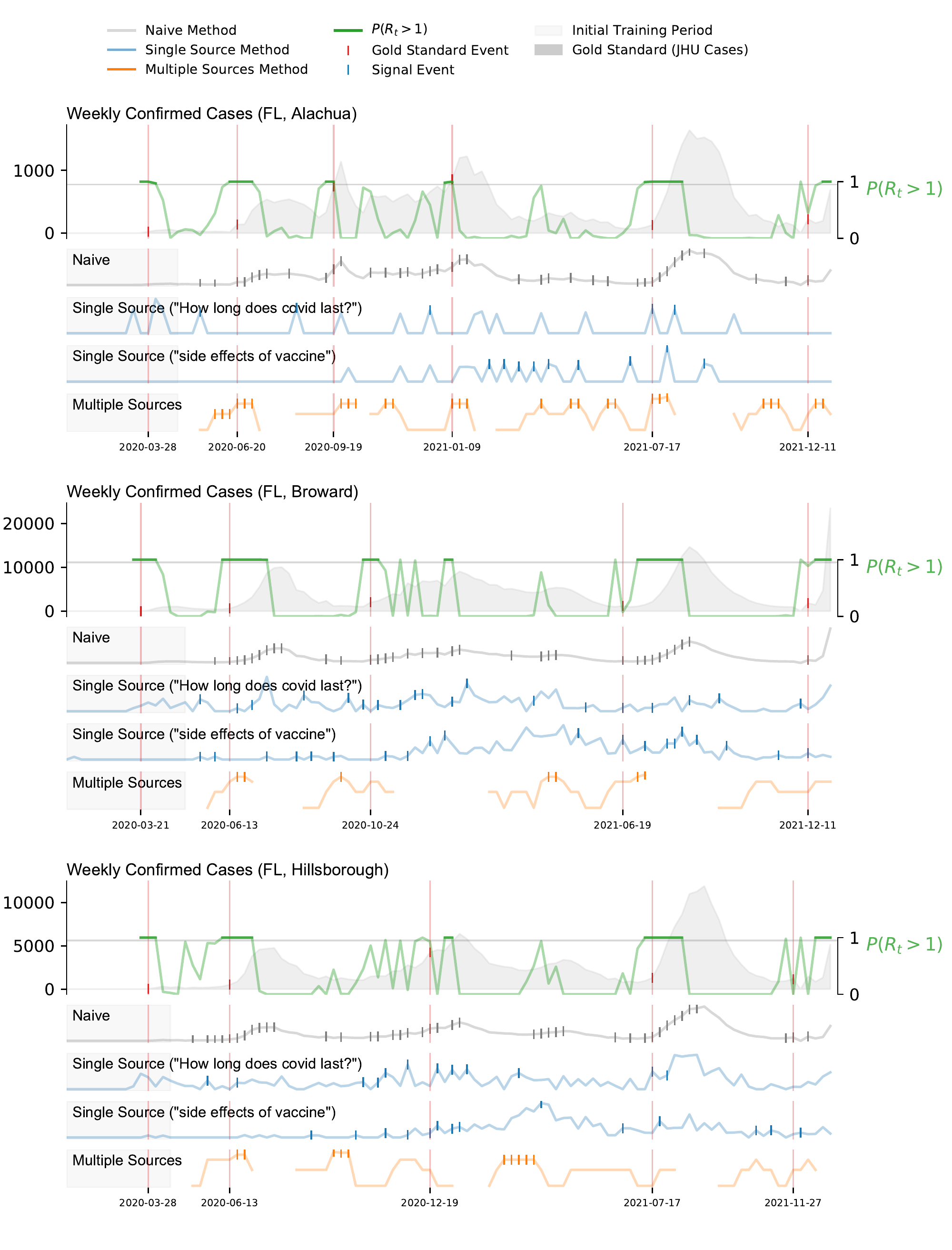}
    \caption{Graphical representation of the reproductive number, $R_t$, along with the weekly confirmed COVID-19 cases (gray-filled curve in the top), and three representative early warning methods (Naive, Single and Multiple Source) at county level.}
    \label{fig:6_county}
\end{figure} 

\begin{figure}
    \centering
    \includegraphics[width=0.9\textwidth,height=\textheight,keepaspectratio]{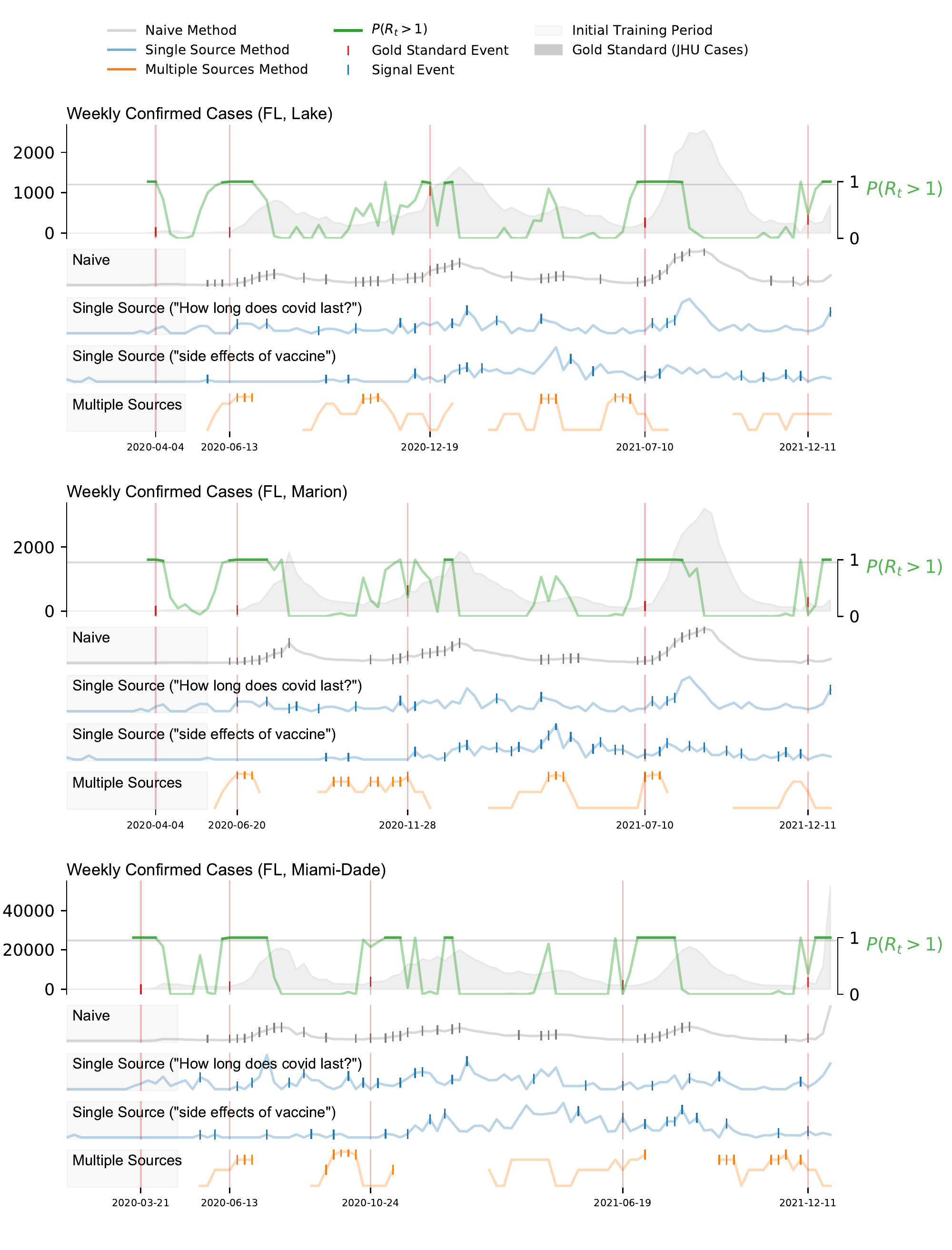}
    \caption{Graphical representation of the reproductive number, $R_t$, along with the weekly confirmed COVID-19 cases (gray-filled curve in the top), and three representative early warning methods (Naive, Single and Multiple Source) at county level.}
    \label{fig:7_county}
\end{figure} 

\begin{figure}
    \centering
    \includegraphics[width=0.9\textwidth,height=\textheight,keepaspectratio]{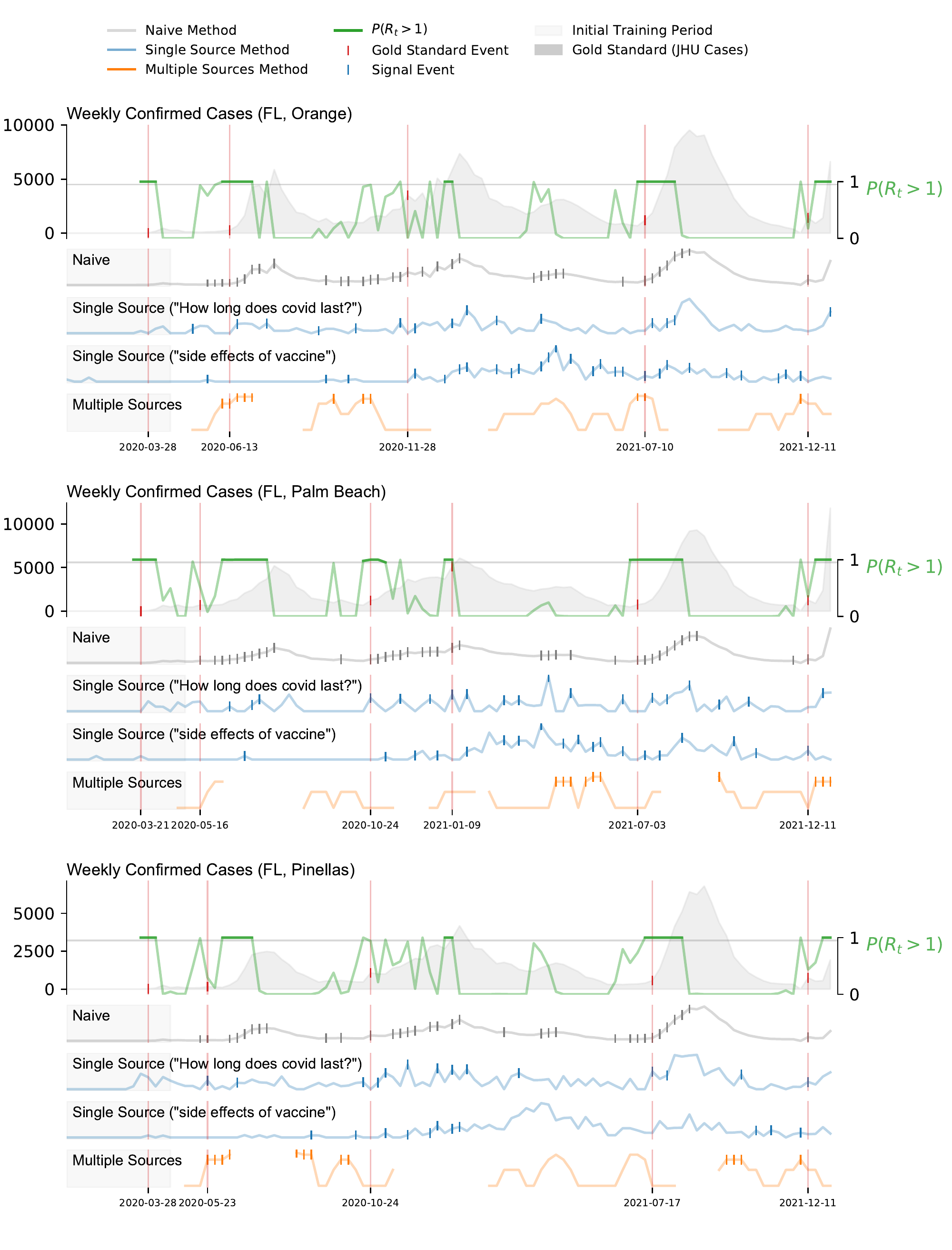}
    \caption{Graphical representation of the probability of resurgence $P(R_t>1)$, along with the weekly confirmed COVID-19 cases (gray-filled curve in the top), and three representative early warning methods (Naive, Single and Multiple Source) at county level.}
    \label{fig:8_county}
\end{figure} 

\begin{figure}
    \centering
    \includegraphics[width=0.9\textwidth,height=\textheight,keepaspectratio]{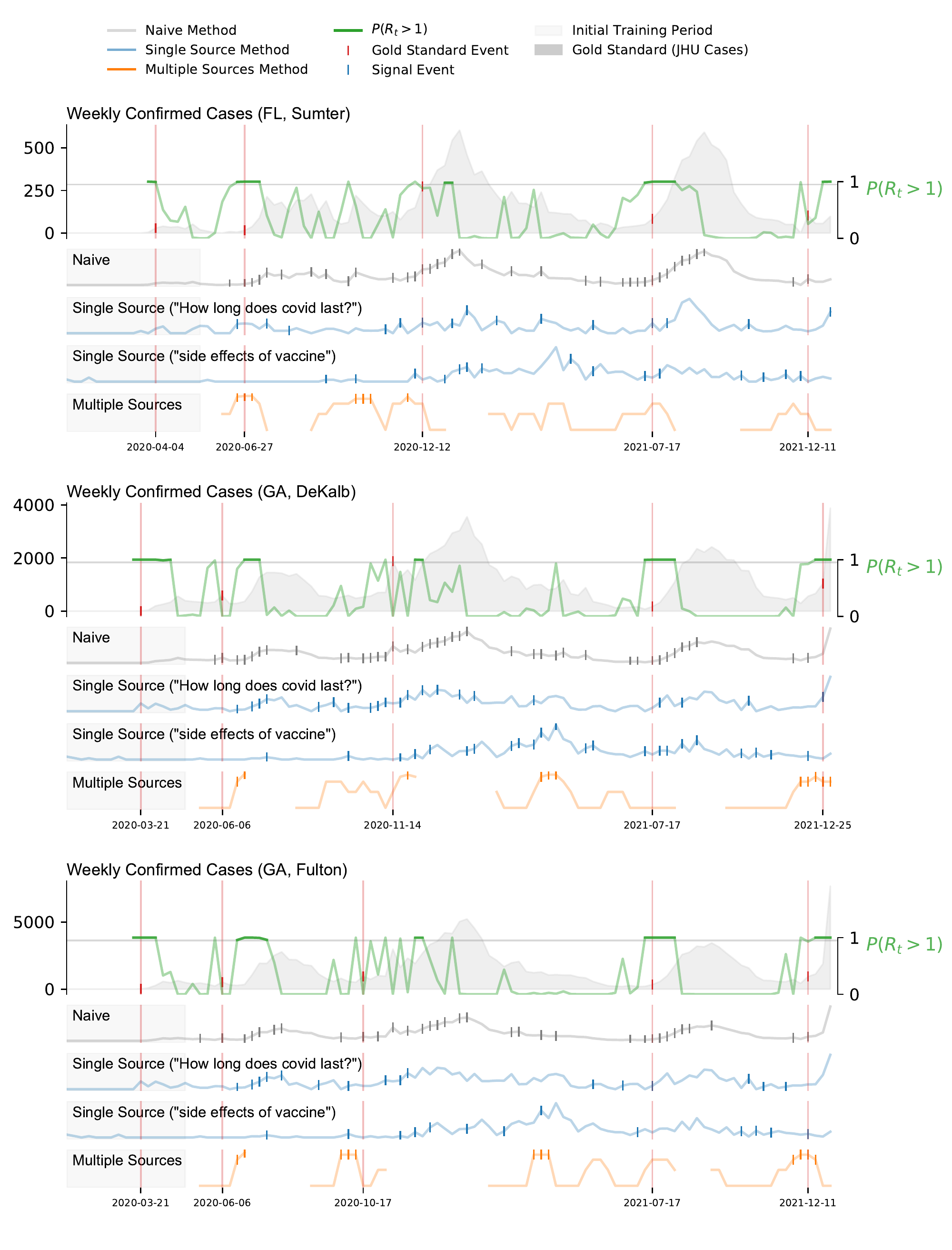}
    \caption{Graphical representation of the reproductive number, $R_t$, along with the weekly confirmed COVID-19 cases (gray-filled curve in the top), and three representative early warning methods (Naive, Single and Multiple Source) at county level.}
    \label{fig:9_county}
\end{figure} 

\begin{figure}
    \centering
    \includegraphics[width=0.9\textwidth,height=\textheight,keepaspectratio]{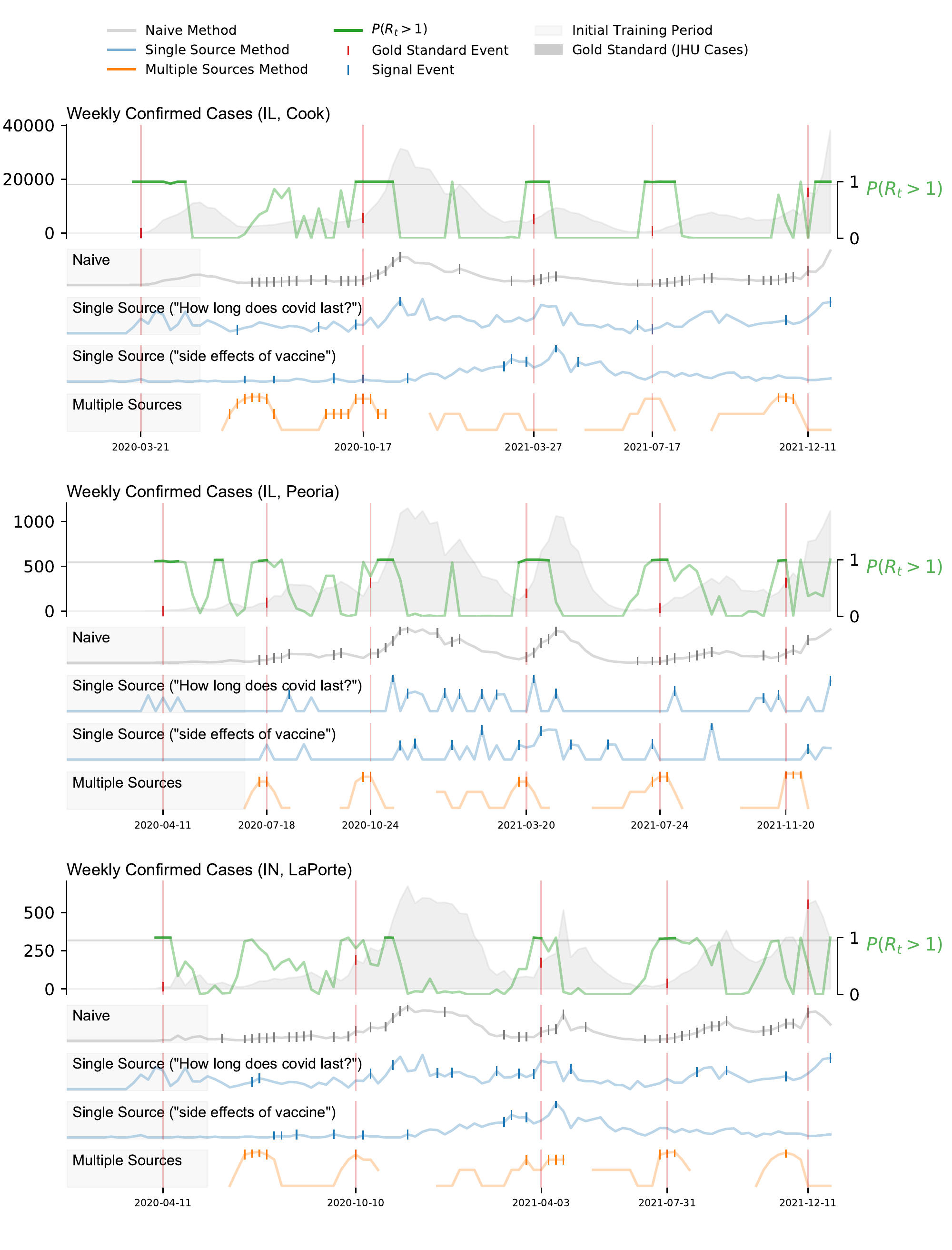}
    \caption{Graphical representation of the reproductive number, $R_t$, along with the weekly confirmed COVID-19 cases (gray-filled curve in the top), and three representative early warning methods (Naive, Single and Multiple Source) at county level.}
    \label{fig:10_county}
\end{figure} 

\begin{figure}
    \centering
    \includegraphics[width=0.9\textwidth,height=\textheight,keepaspectratio]{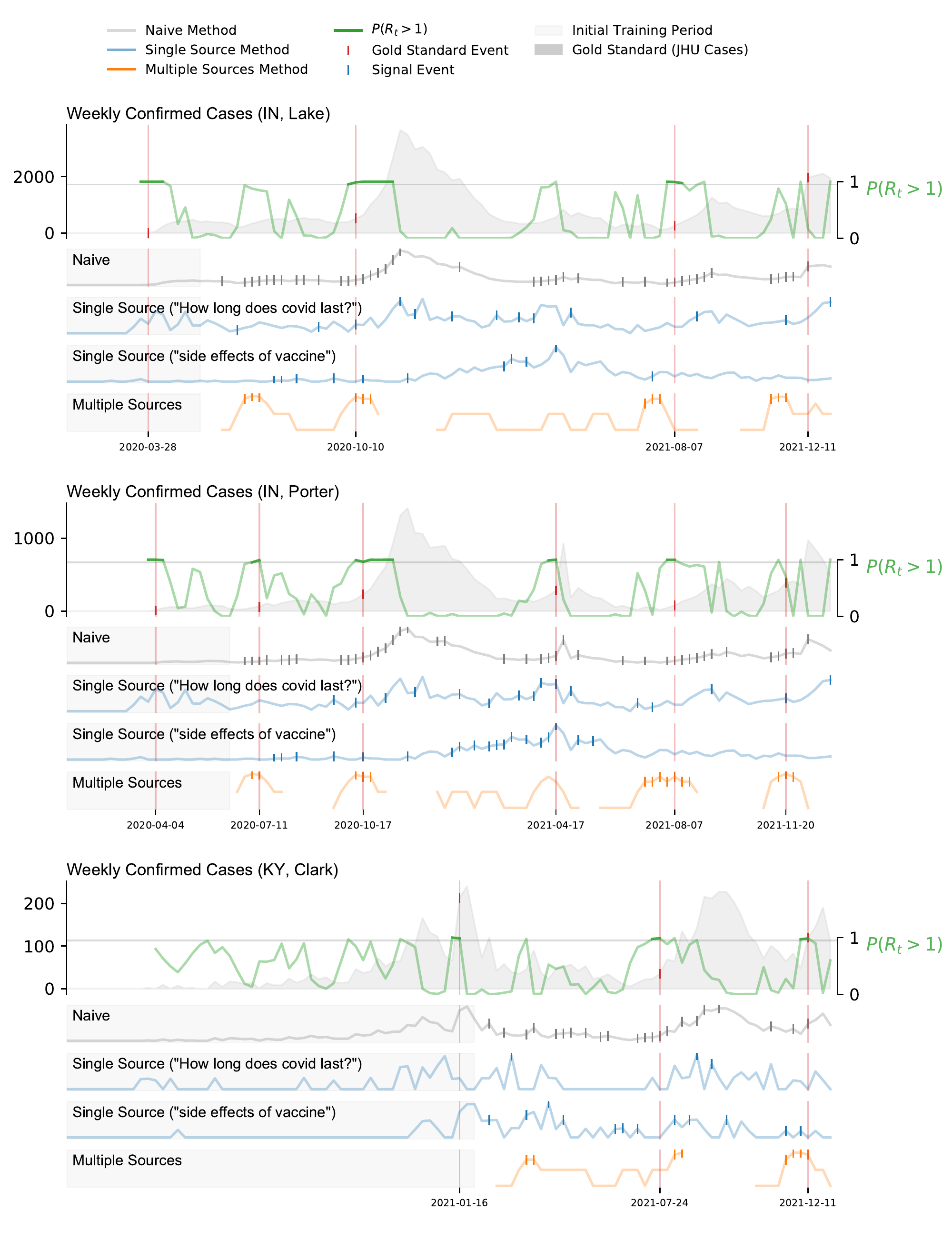}
    \caption{Graphical representation of the reproductive number, $R_t$, along with the weekly confirmed COVID-19 cases (gray-filled curve in the top), and three representative early warning methods (Naive, Single and Multiple Source) at county level.}
    \label{fig:11_county}
\end{figure} 

\begin{figure}
    \centering
    \includegraphics[width=0.9\textwidth,height=\textheight,keepaspectratio]{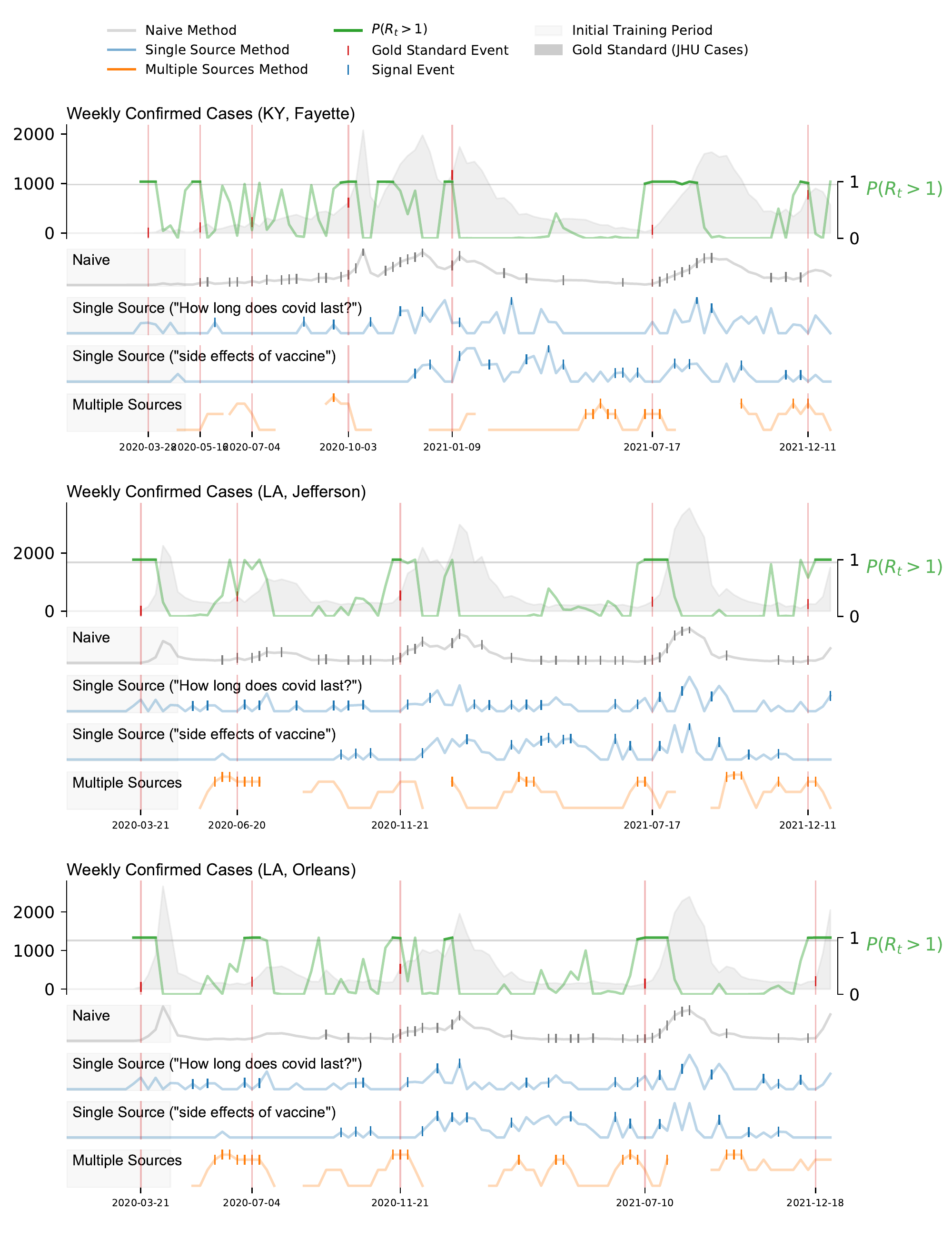}
    \caption{Graphical representation of the reproductive number, $R_t$, along with the weekly confirmed COVID-19 cases (gray-filled curve in the top), and three representative early warning methods (Naive, Single and Multiple Source) at county level.}
    \label{fig:12_county}
\end{figure} 

\begin{figure}
    \centering
    \includegraphics[width=0.9\textwidth,height=\textheight,keepaspectratio]{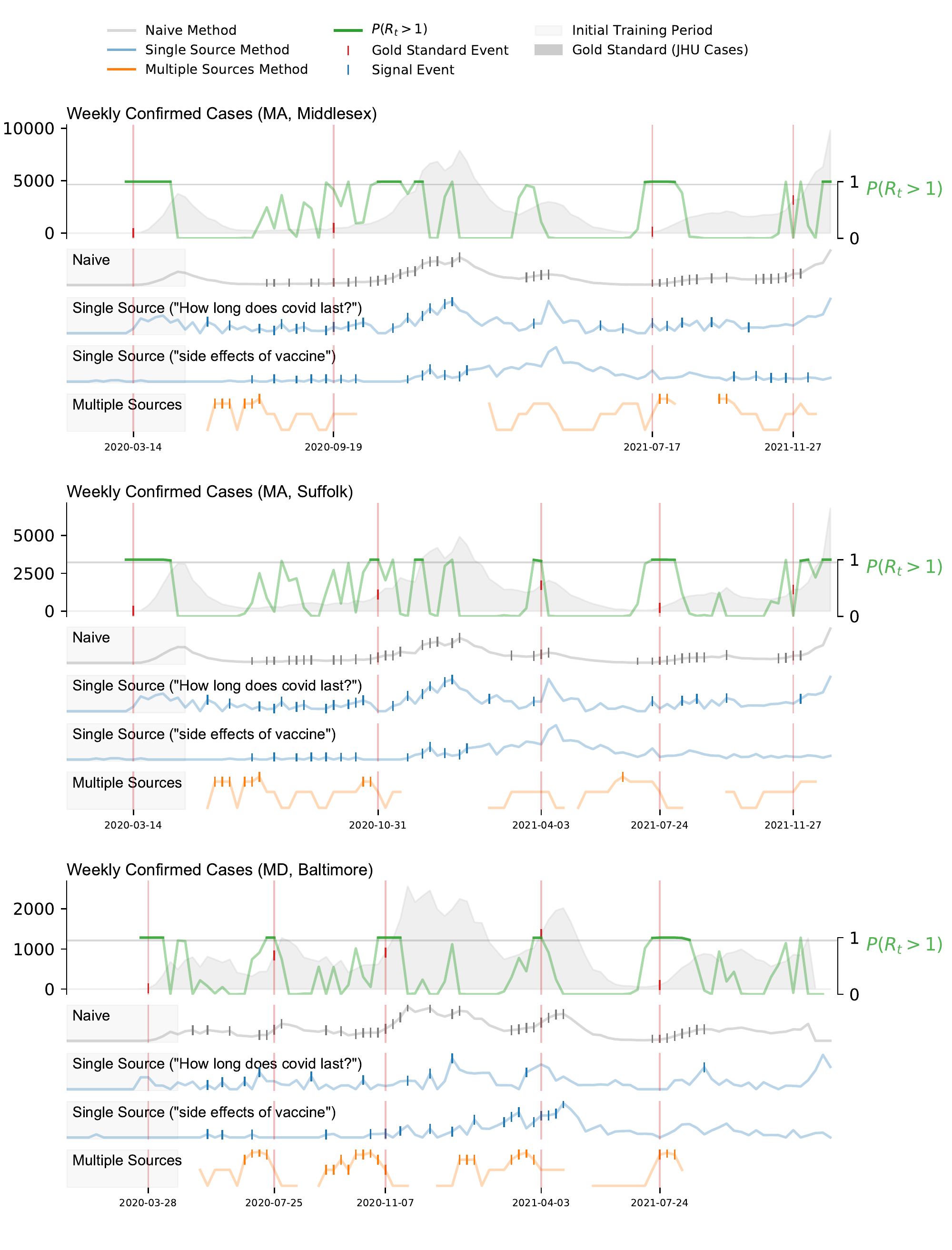}
    \caption{Graphical representation of the reproductive number, $R_t$, along with the weekly confirmed COVID-19 cases (gray-filled curve in the top), and three representative early warning methods (Naive, Single and Multiple Source) at county level.}
    \label{fig:13_county}
\end{figure} 

\begin{figure}
    \centering
    \includegraphics[width=0.9\textwidth,height=\textheight,keepaspectratio]{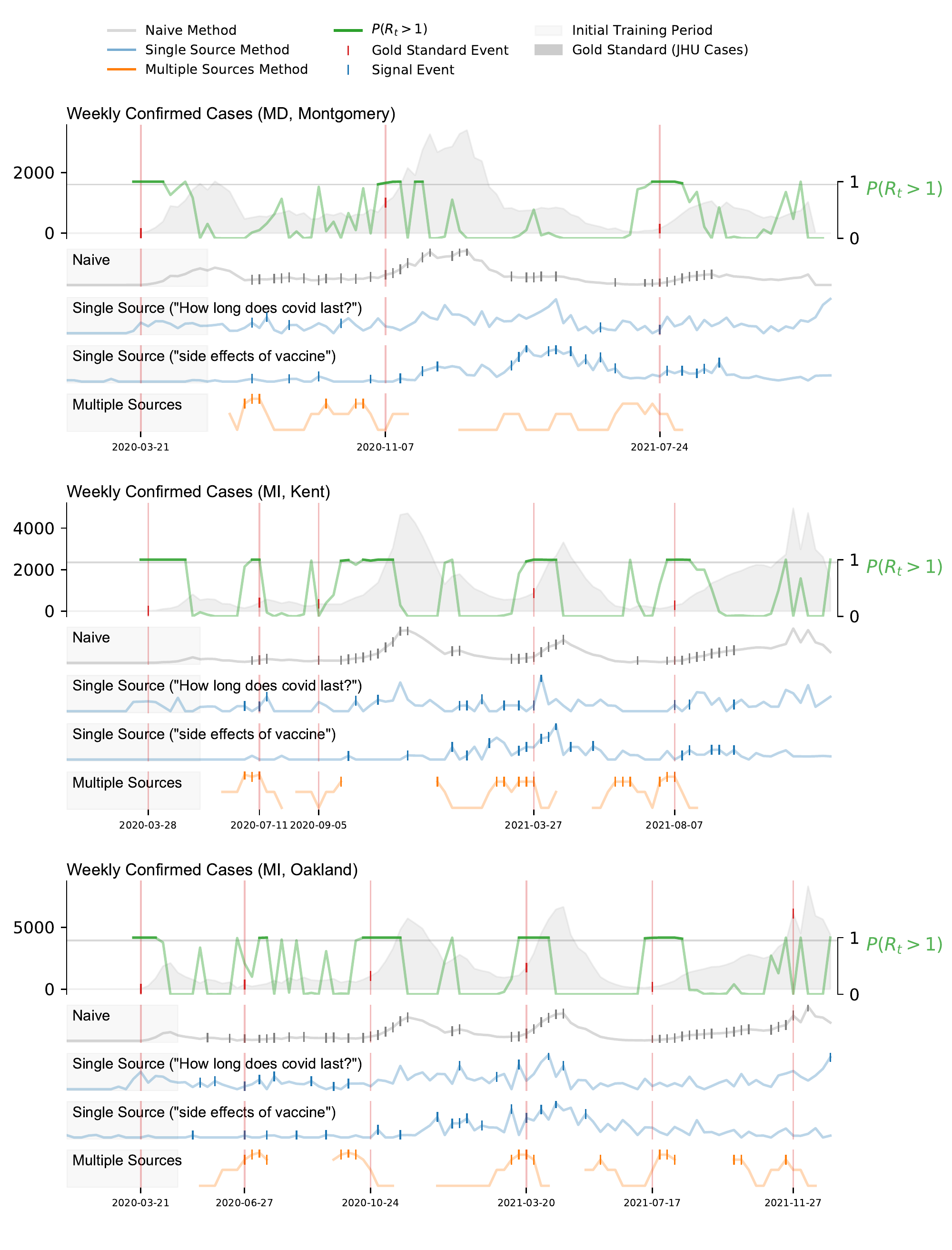}
    \caption{Graphical representation of the reproductive number, $R_t$, along with the weekly confirmed COVID-19 cases (gray-filled curve in the top), and three representative early warning methods (Naive, Single and Multiple Source) at county level.}
    \label{fig:14_county}
\end{figure} 

\begin{figure}
    \centering
    \includegraphics[width=0.9\textwidth,height=\textheight,keepaspectratio]{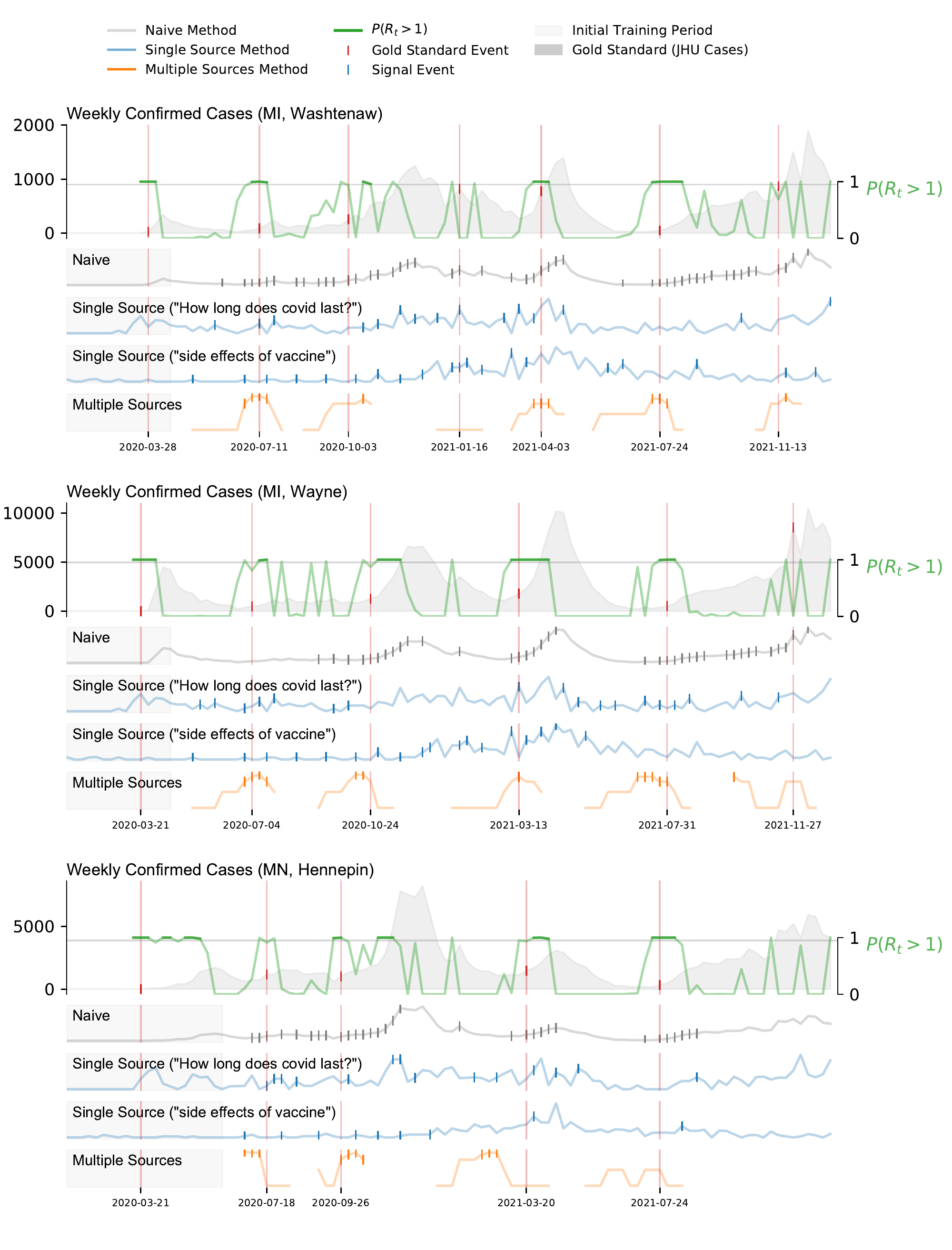}
    \caption{Graphical representation of the reproductive number, $R_t$, along with the weekly confirmed COVID-19 cases (gray-filled curve in the top), and three representative early warning methods (Naive, Single and Multiple Source) at county level.}
    \label{fig:15_county}
\end{figure} 

\begin{figure}
    \centering
    \includegraphics[width=0.9\textwidth,height=\textheight,keepaspectratio]{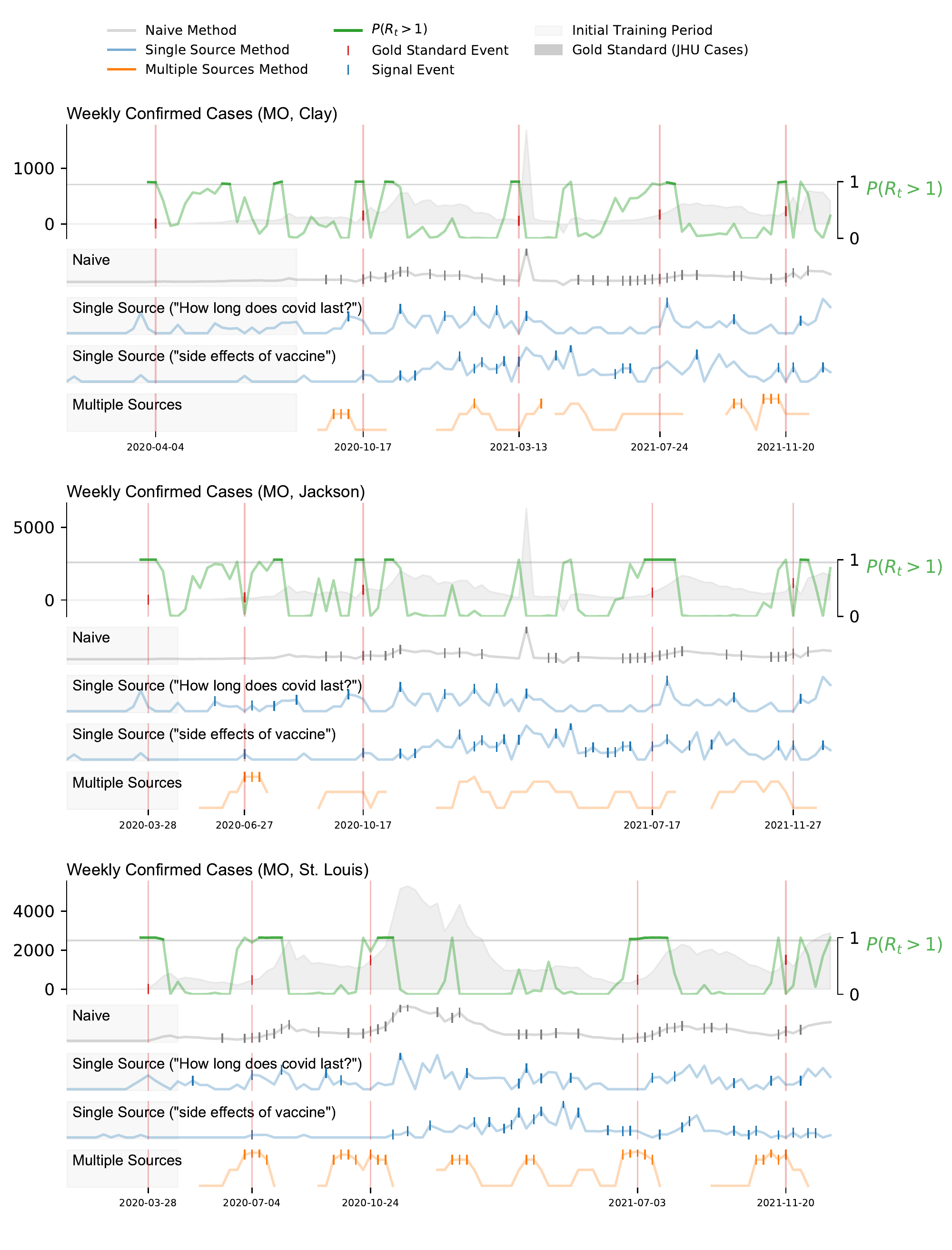}
    \caption{Graphical representation of the reproductive number, $R_t$, along with the weekly confirmed COVID-19 cases (gray-filled curve in the top), and three representative early warning methods (Naive, Single and Multiple Source) at county level.}
    \label{fig:16_county}
\end{figure} 

\begin{figure}
    \centering
    \includegraphics[width=0.9\textwidth,height=\textheight,keepaspectratio]{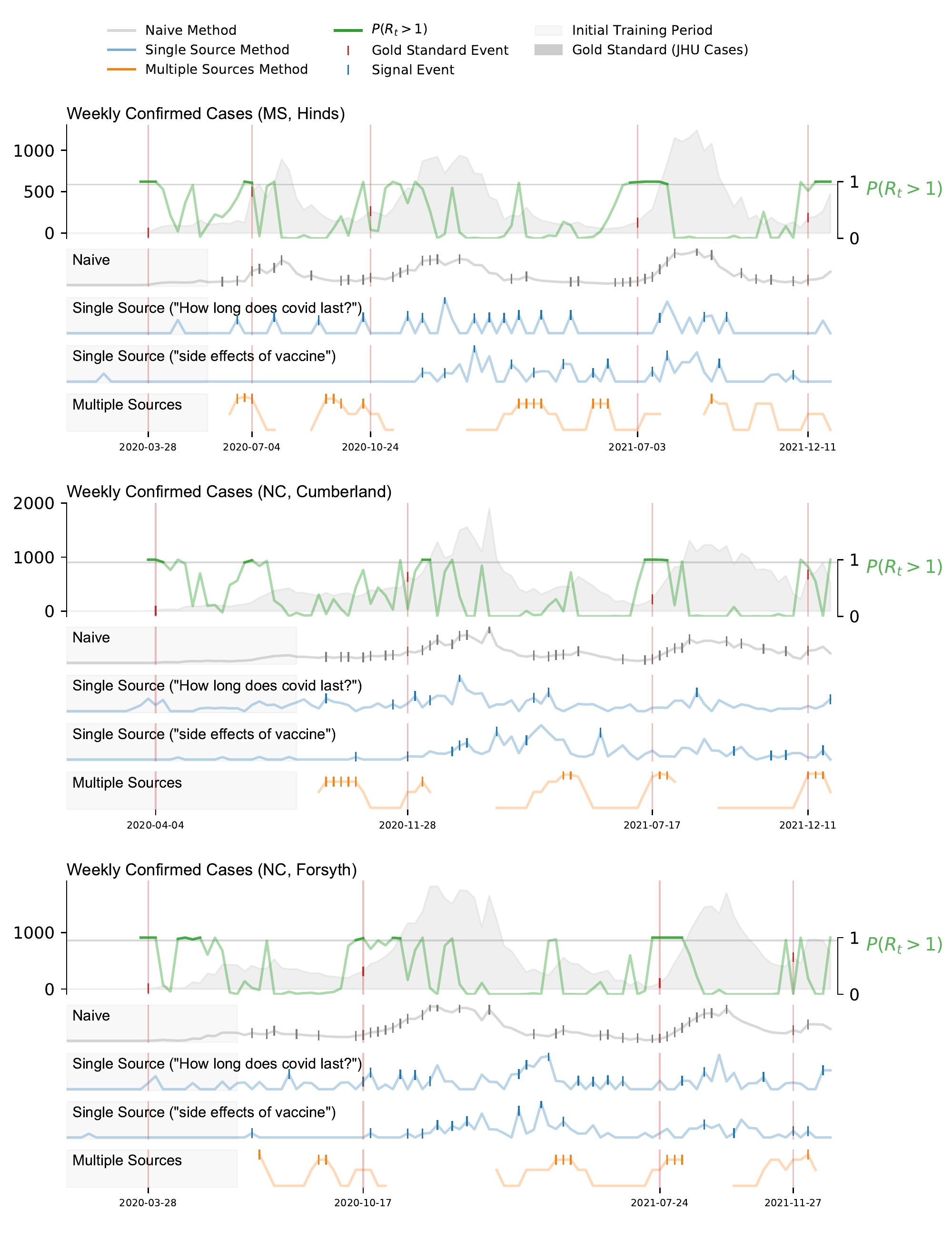}
    \caption{Graphical representation of the reproductive number, $R_t$, along with the weekly confirmed COVID-19 cases (gray-filled curve in the top), and three representative early warning methods (Naive, Single and Multiple Source) at county level.}
    \label{fig:17_county}
\end{figure} 

\begin{figure}
    \centering
    \includegraphics[width=0.9\textwidth,height=\textheight,keepaspectratio]{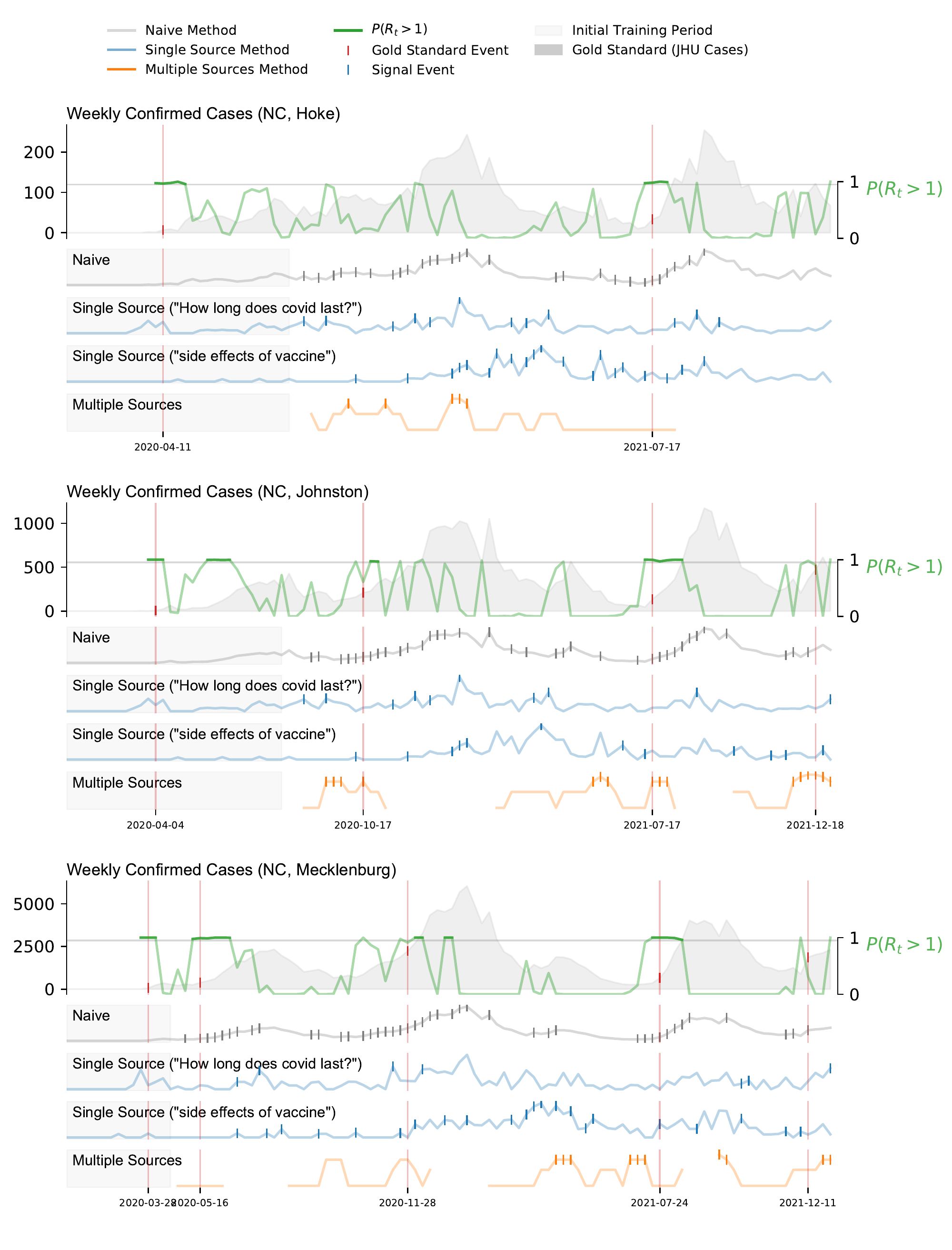}
    \caption{Graphical representation of the reproductive number, $R_t$, along with the weekly confirmed COVID-19 cases (gray-filled curve in the top), and three representative early warning methods (Naive, Single and Multiple Source) at county level.}
    \label{fig:18_county}
\end{figure} 

\begin{figure}
    \centering
    \includegraphics[width=0.9\textwidth,height=\textheight,keepaspectratio]{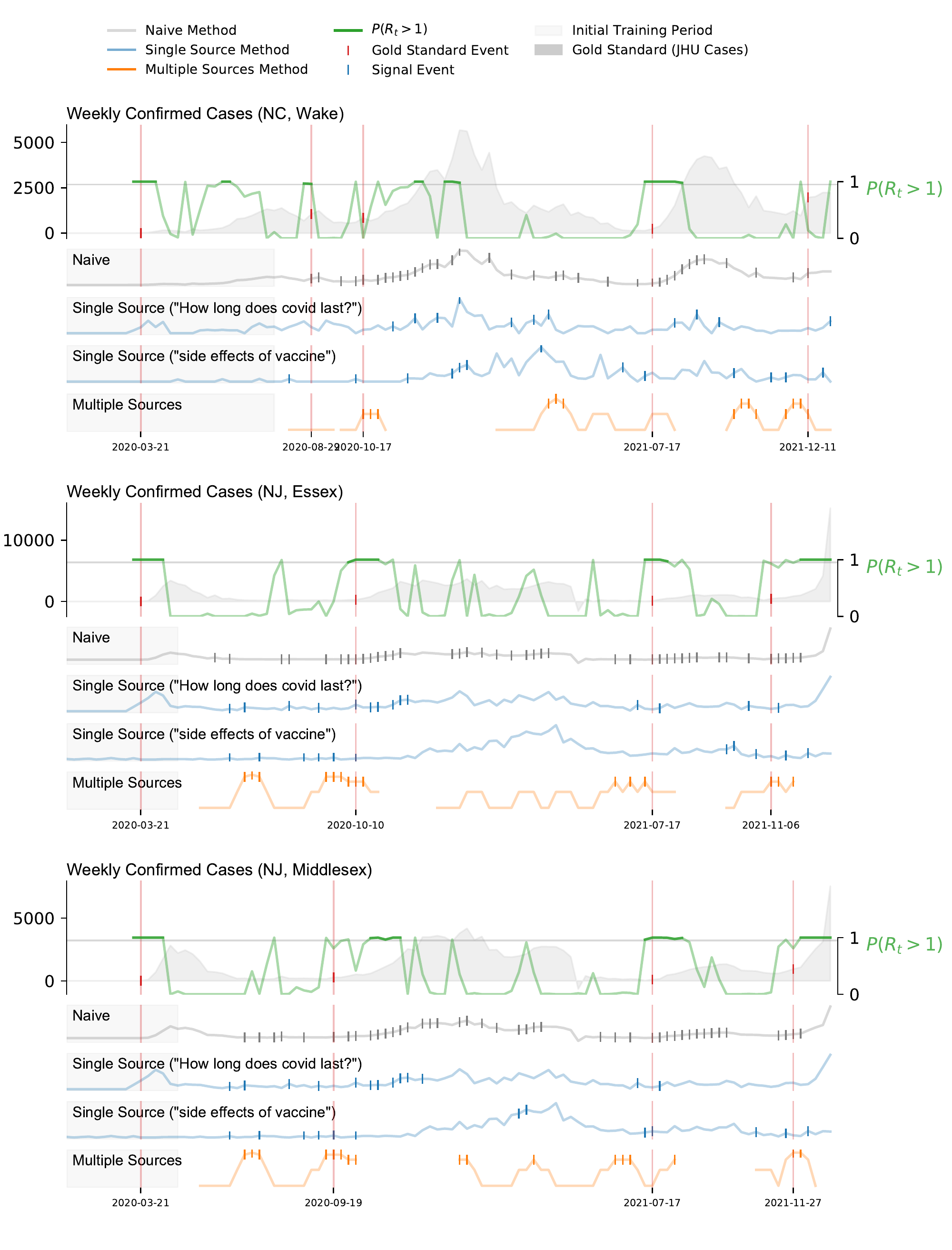}
    \caption{Graphical representation of the reproductive number, $R_t$, along with the weekly confirmed COVID-19 cases (gray-filled curve in the top), and three representative early warning methods (Naive, Single and Multiple Source) at county level.}
    \label{fig:19_county}
\end{figure} 

\begin{figure}
    \centering
    \includegraphics[width=0.9\textwidth,height=\textheight,keepaspectratio]{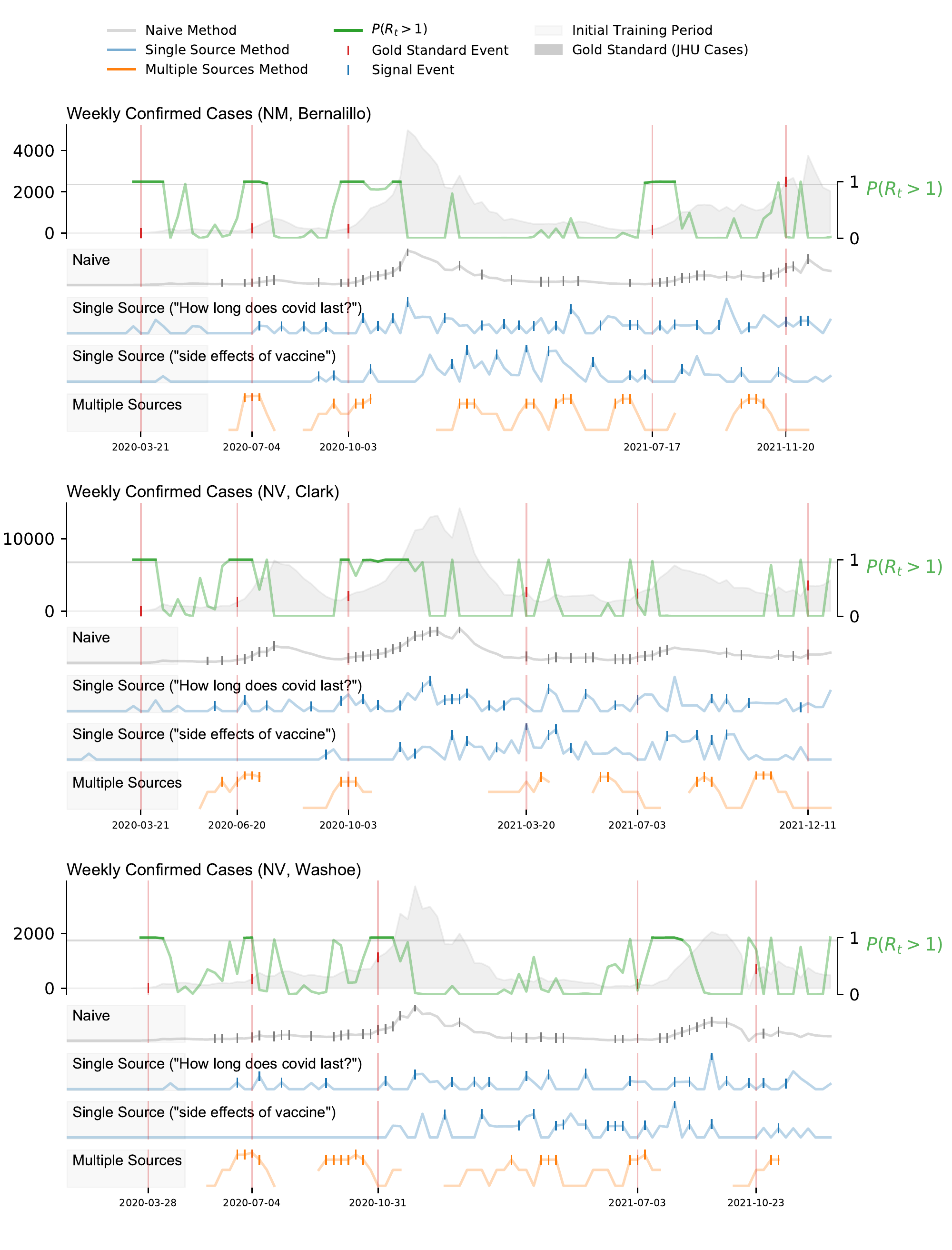}
    \caption{Graphical representation of the reproductive number, $R_t$, along with the weekly confirmed COVID-19 cases (gray-filled curve in the top), and three representative early warning methods (Naive, Single and Multiple Source) at county level.}
    \label{fig:20_county}
\end{figure} 

\begin{figure}
    \centering
    \includegraphics[width=0.9\textwidth,height=\textheight,keepaspectratio]{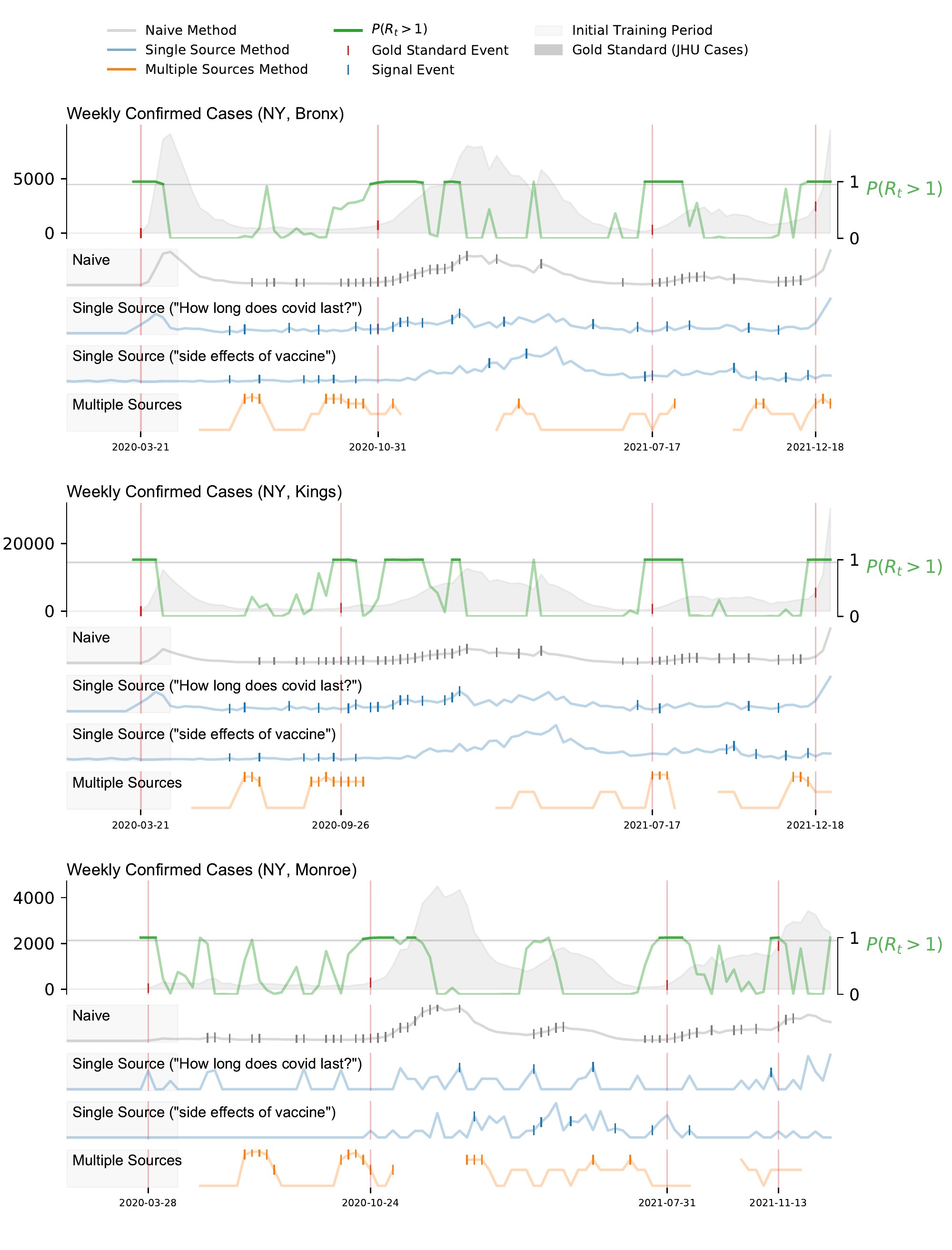}
    \caption{Graphical representation of the reproductive number, $R_t$, along with the weekly confirmed COVID-19 cases (gray-filled curve in the top), and three representative early warning methods (Naive, Single and Multiple Source) at county level.}
    \label{fig:21_county}
\end{figure} 

\begin{figure}
    \centering
    \includegraphics[width=0.9\textwidth,height=\textheight,keepaspectratio]{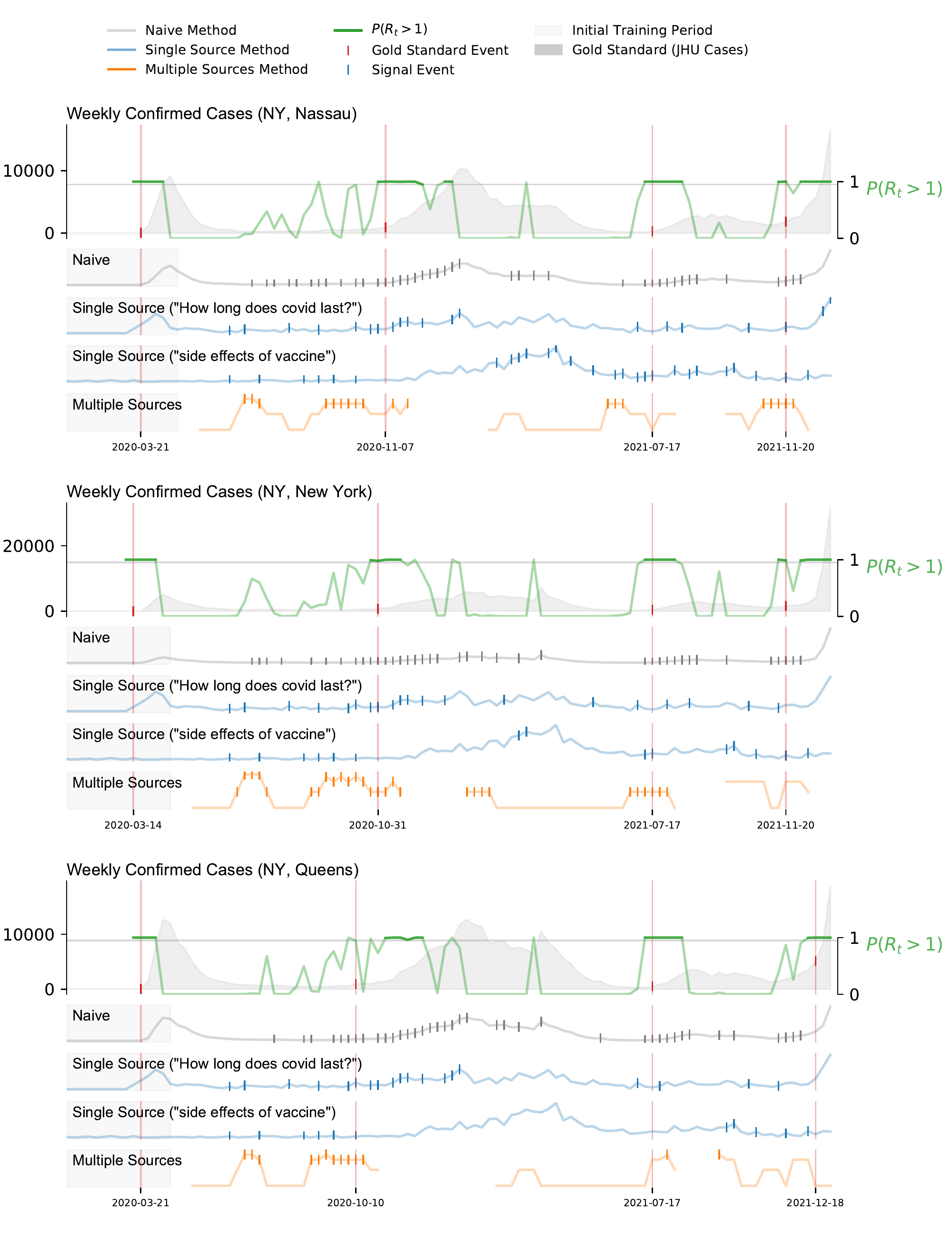}
    \caption{Graphical representation of the reproductive number, $R_t$, along with the weekly confirmed COVID-19 cases (gray-filled curve in the top), and three representative early warning methods (Naive, Single and Multiple Source) at county level.}
    \label{fig:22_county}
\end{figure} 

\begin{figure}
    \centering
    \includegraphics[width=0.9\textwidth,height=\textheight,keepaspectratio]{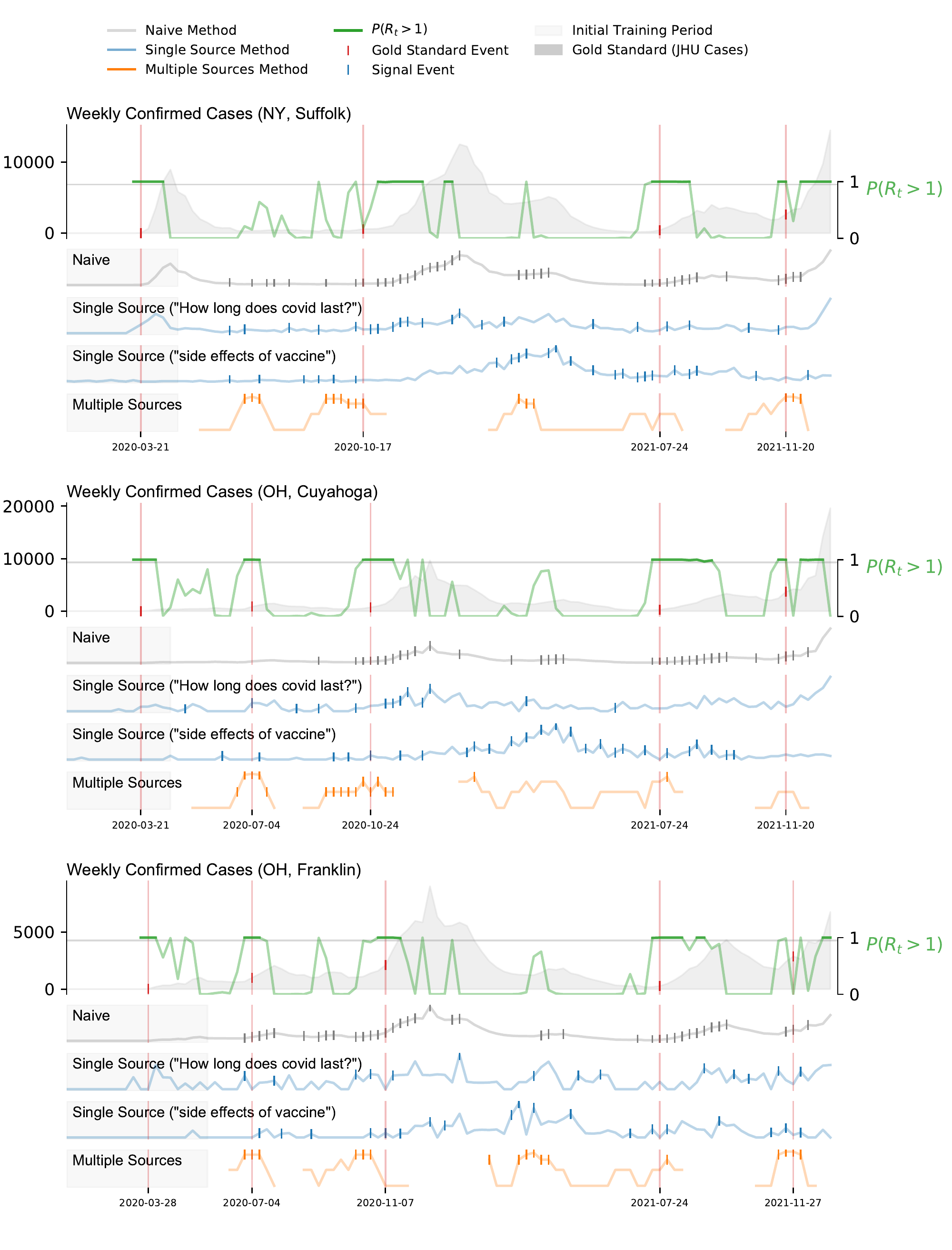}
    \caption{Graphical representation of the reproductive number, $R_t$, along with the weekly confirmed COVID-19 cases (gray-filled curve in the top), and three representative early warning methods (Naive, Single and Multiple Source) at county level.}
    \label{fig:23_county}
\end{figure} 

\begin{figure}
    \centering
    \includegraphics[width=0.9\textwidth,height=\textheight,keepaspectratio]{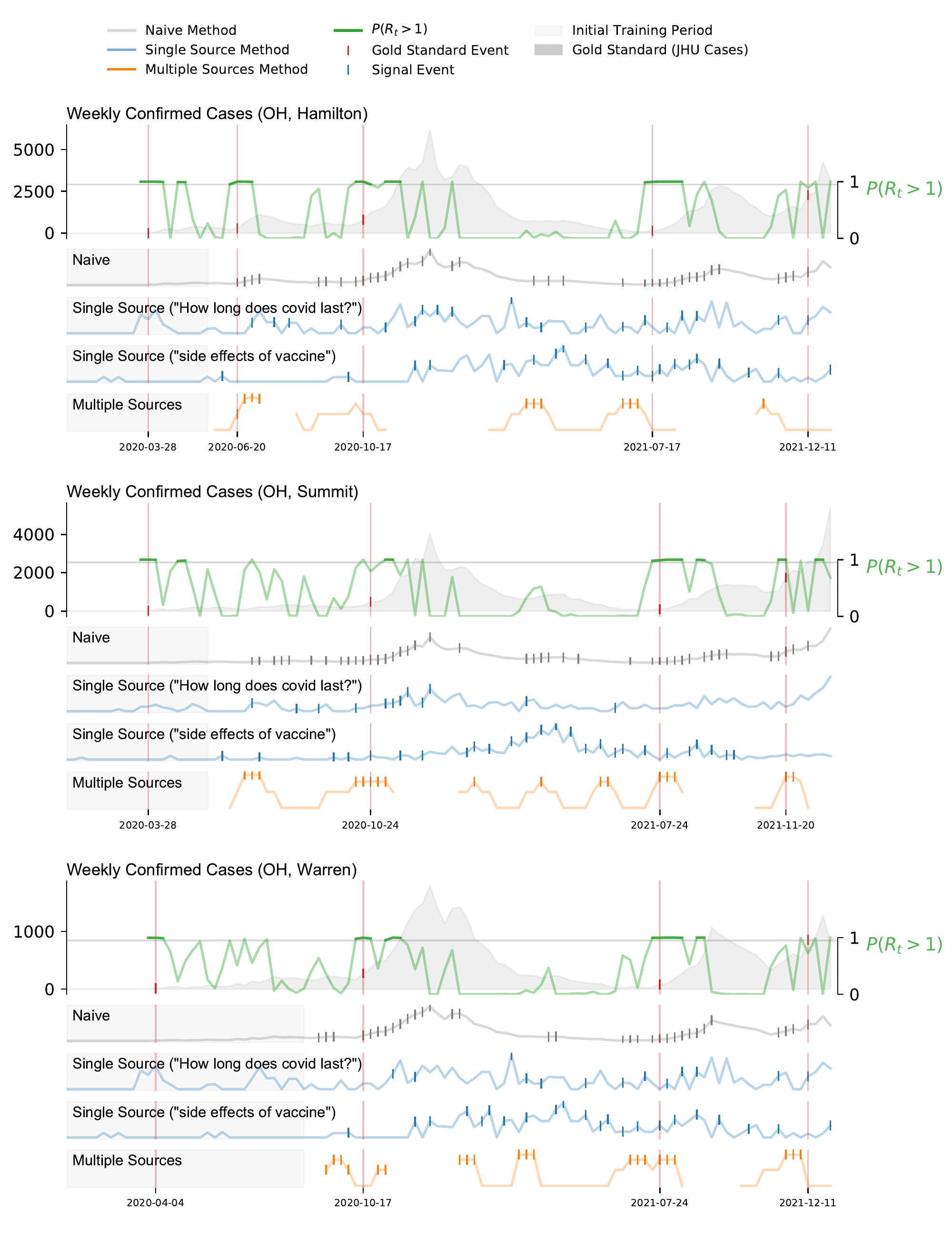}
    \caption{Graphical representation of the reproductive number, $R_t$, along with the weekly confirmed COVID-19 cases (gray-filled curve in the top), and three representative early warning methods (Naive, Single and Multiple Source) at county level.}
    \label{fig:24_county}
\end{figure} 

\begin{figure}
    \centering
    \includegraphics[width=0.9\textwidth,height=\textheight,keepaspectratio]{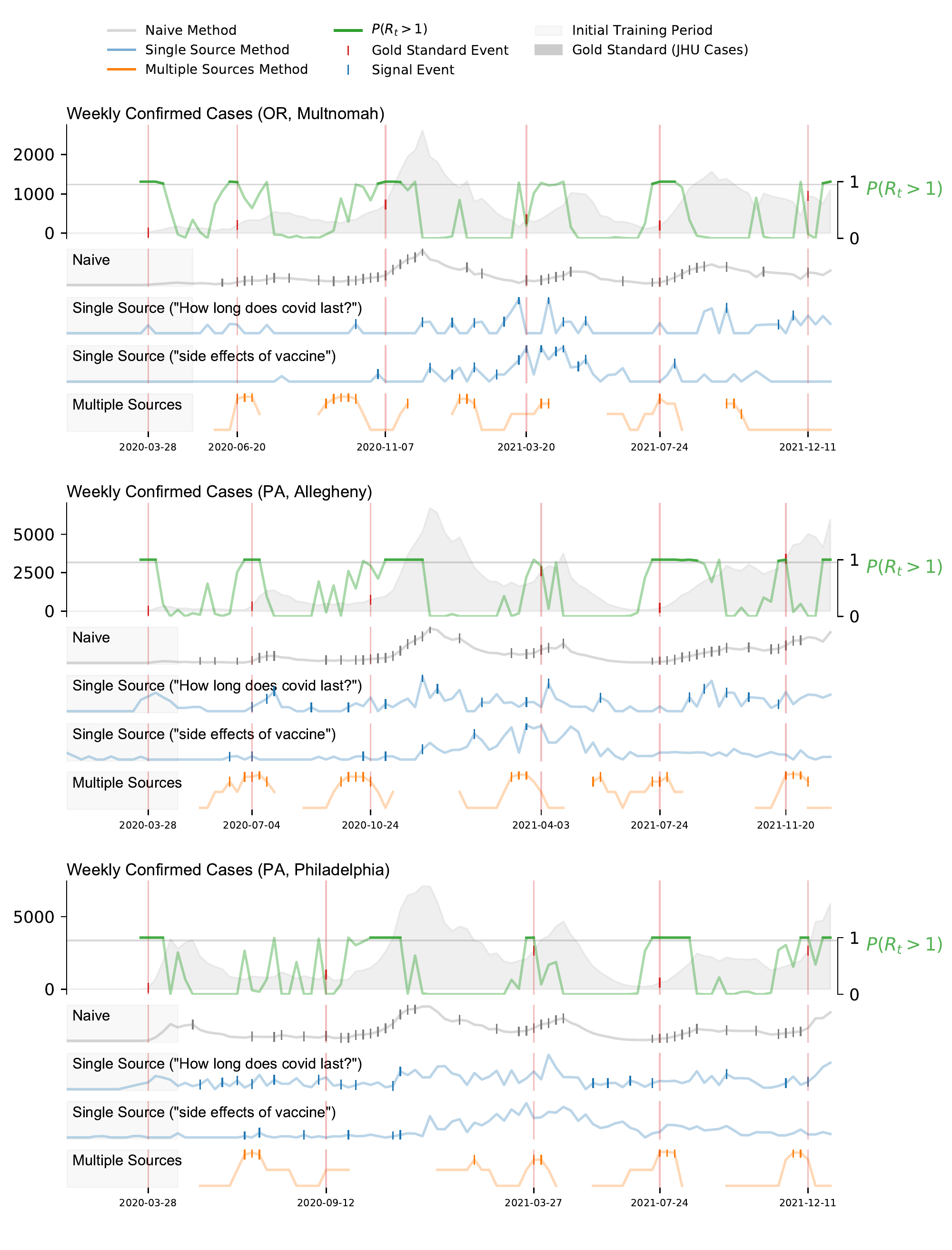}
    \caption{Graphical representation of the reproductive number, $R_t$, along with the weekly confirmed COVID-19 cases (gray-filled curve in the top), and three representative early warning methods (Naive, Single and Multiple Source) at county level.}
    \label{fig:25_county}
\end{figure} 

\begin{figure}
    \centering
    \includegraphics[width=0.9\textwidth,height=\textheight,keepaspectratio]{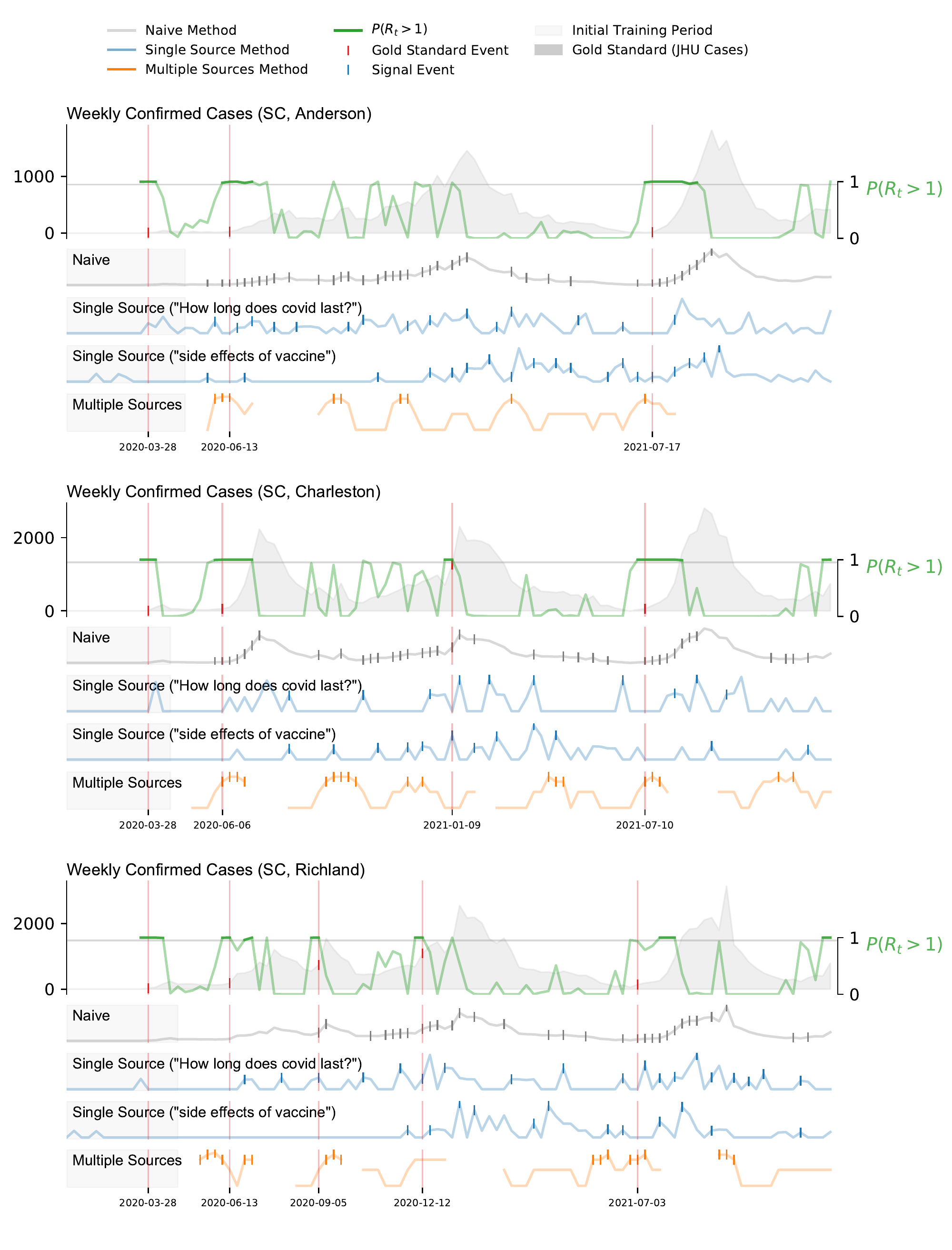}
    \caption{Graphical representation of the reproductive number, $R_t$, along with the weekly confirmed COVID-19 cases (gray-filled curve in the top), and three representative early warning methods (Naive, Single and Multiple Source) at county level.}
    \label{fig:26_county}
\end{figure} 

\begin{figure}
    \centering
    \includegraphics[width=0.9\textwidth,height=\textheight,keepaspectratio]{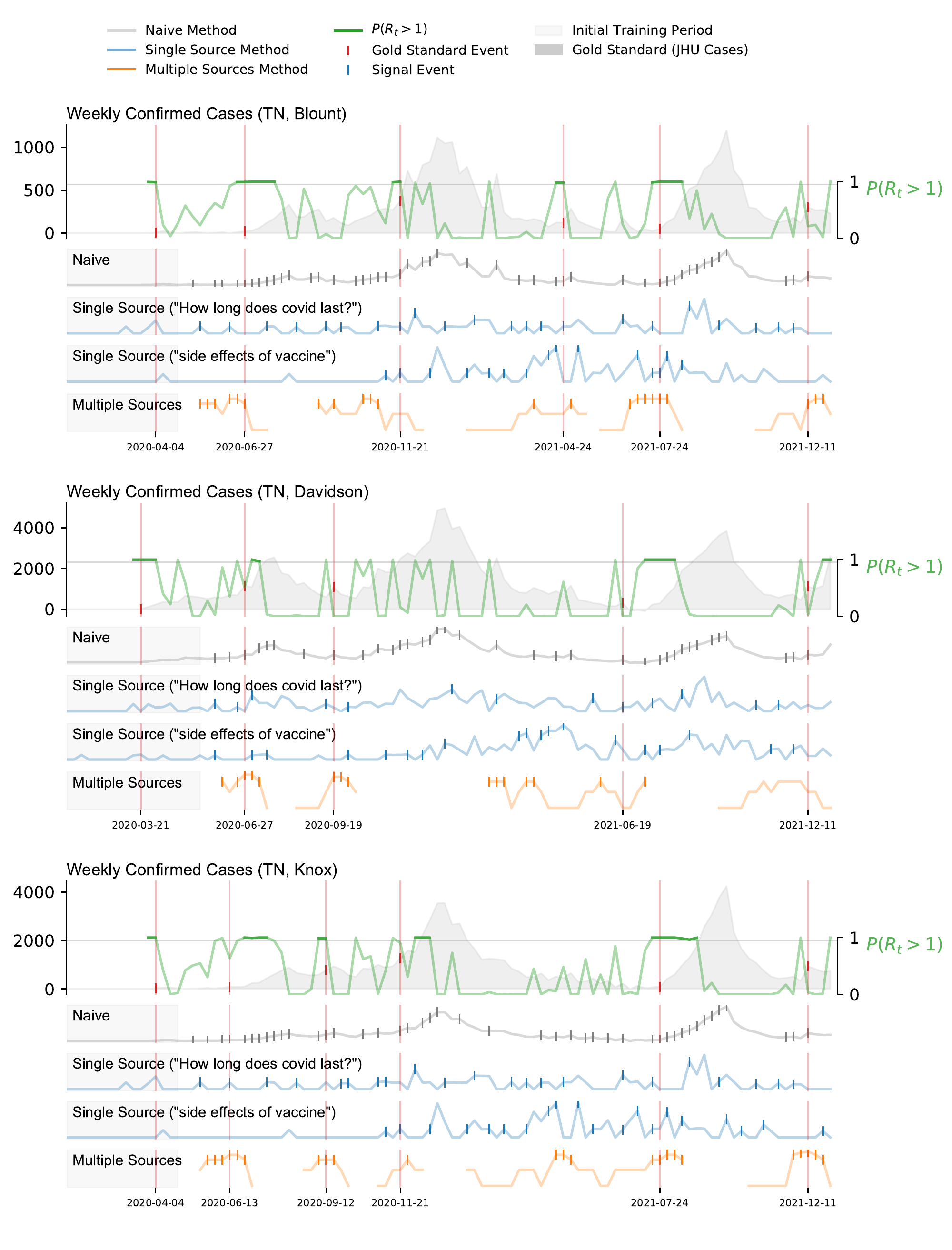}
    \caption{Graphical representation of the reproductive number, $R_t$, along with the weekly confirmed COVID-19 cases (gray-filled curve in the top), and three representative early warning methods (Naive, Single and Multiple Source) at county level.}
    \label{fig:27_county}
\end{figure} 

\begin{figure}
    \centering
    \includegraphics[width=0.9\textwidth,height=\textheight,keepaspectratio]{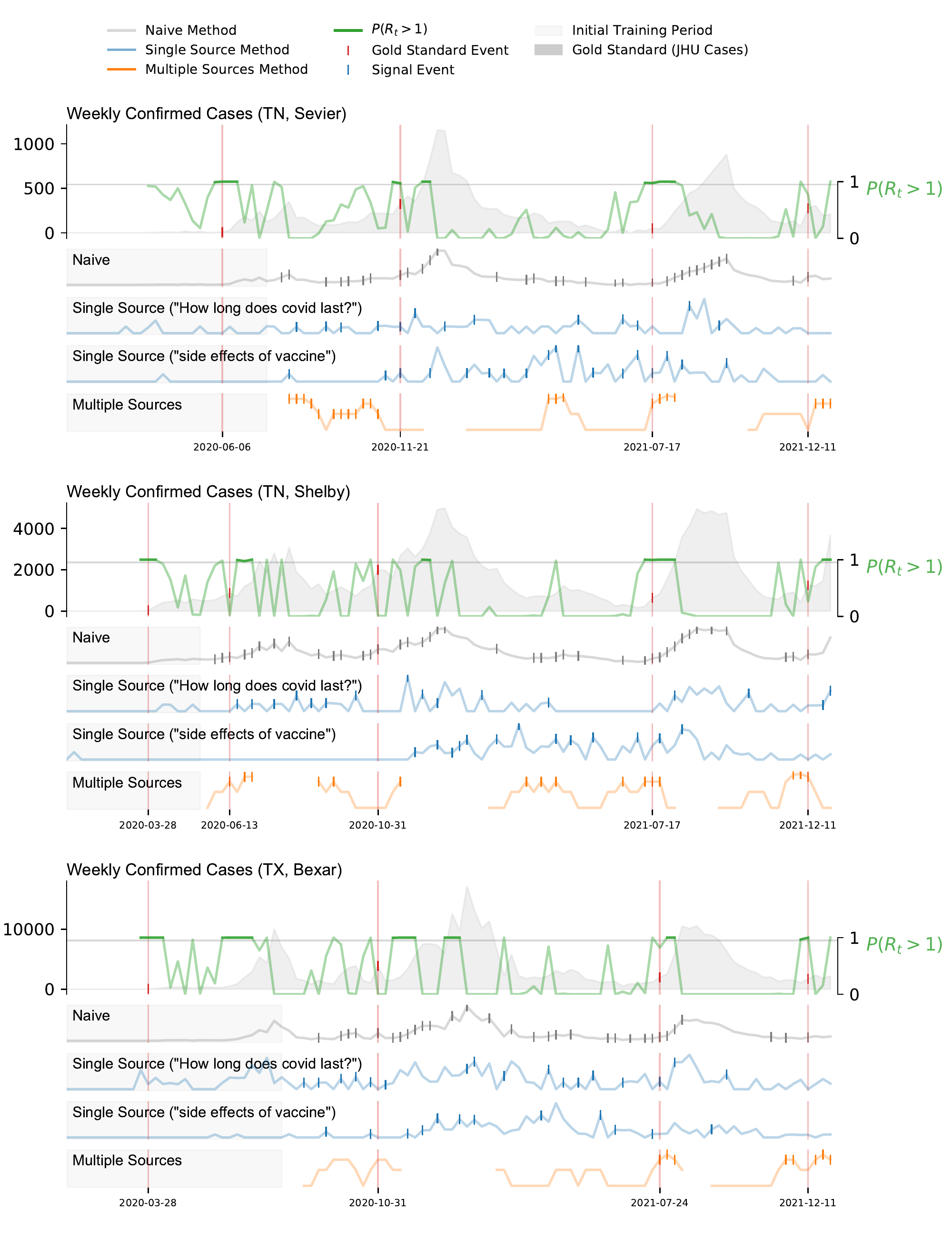}
    \caption{Graphical representation of the reproductive number, $R_t$, along with the weekly confirmed COVID-19 cases (gray-filled curve in the top), and three representative early warning methods (Naive, Single and Multiple Source) at county level.}
    \label{fig:28_county}
\end{figure} 

\begin{figure}
    \centering
    \includegraphics[width=0.9\textwidth,height=\textheight,keepaspectratio]{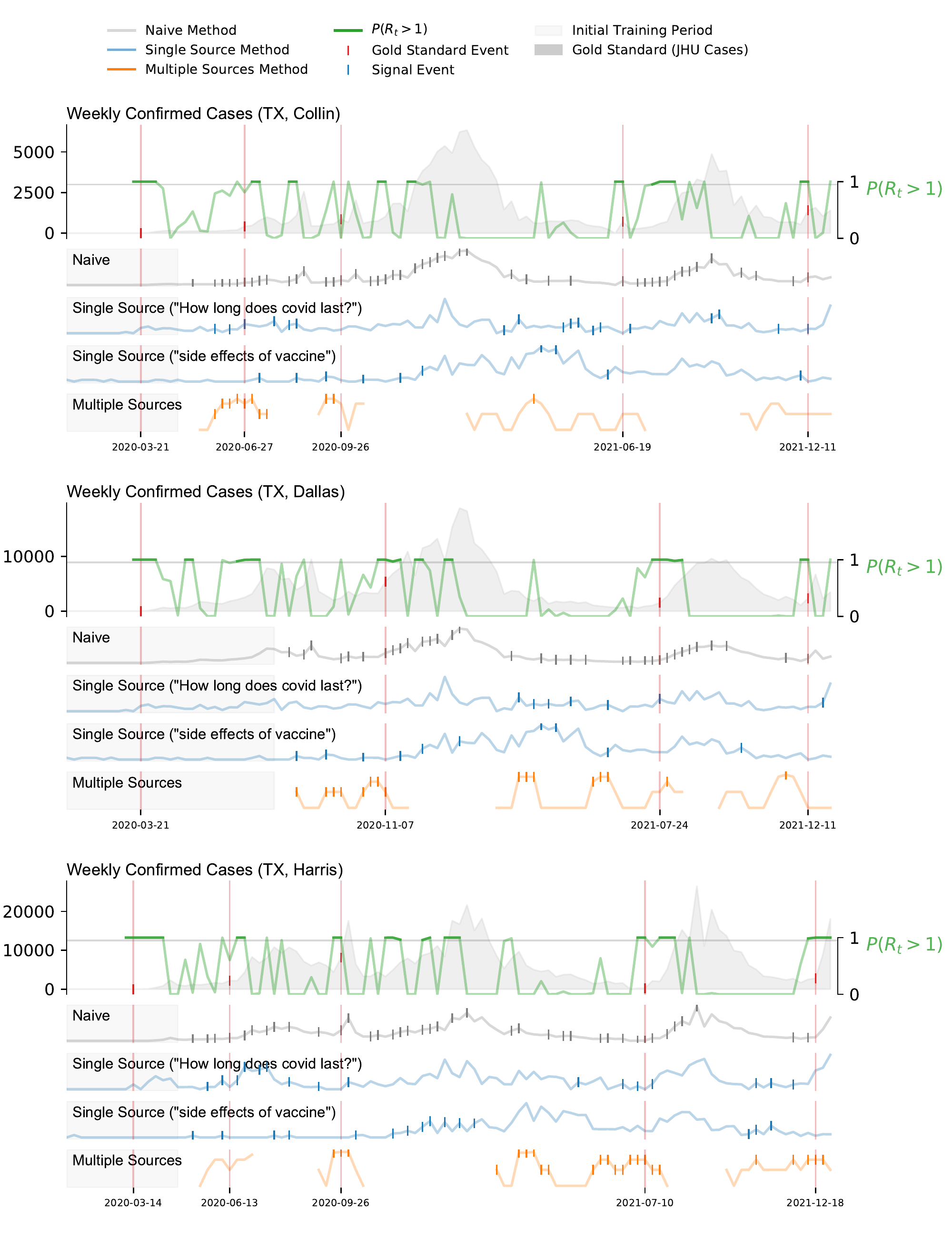}
    \caption{Graphical representation of the reproductive number, $R_t$, along with the weekly confirmed COVID-19 cases (gray-filled curve in the top), and three representative early warning methods (Naive, Single and Multiple Source) at county level.}
    \label{fig:29_county}
\end{figure} 

\begin{figure}
    \centering
    \includegraphics[width=0.9\textwidth,height=\textheight,keepaspectratio]{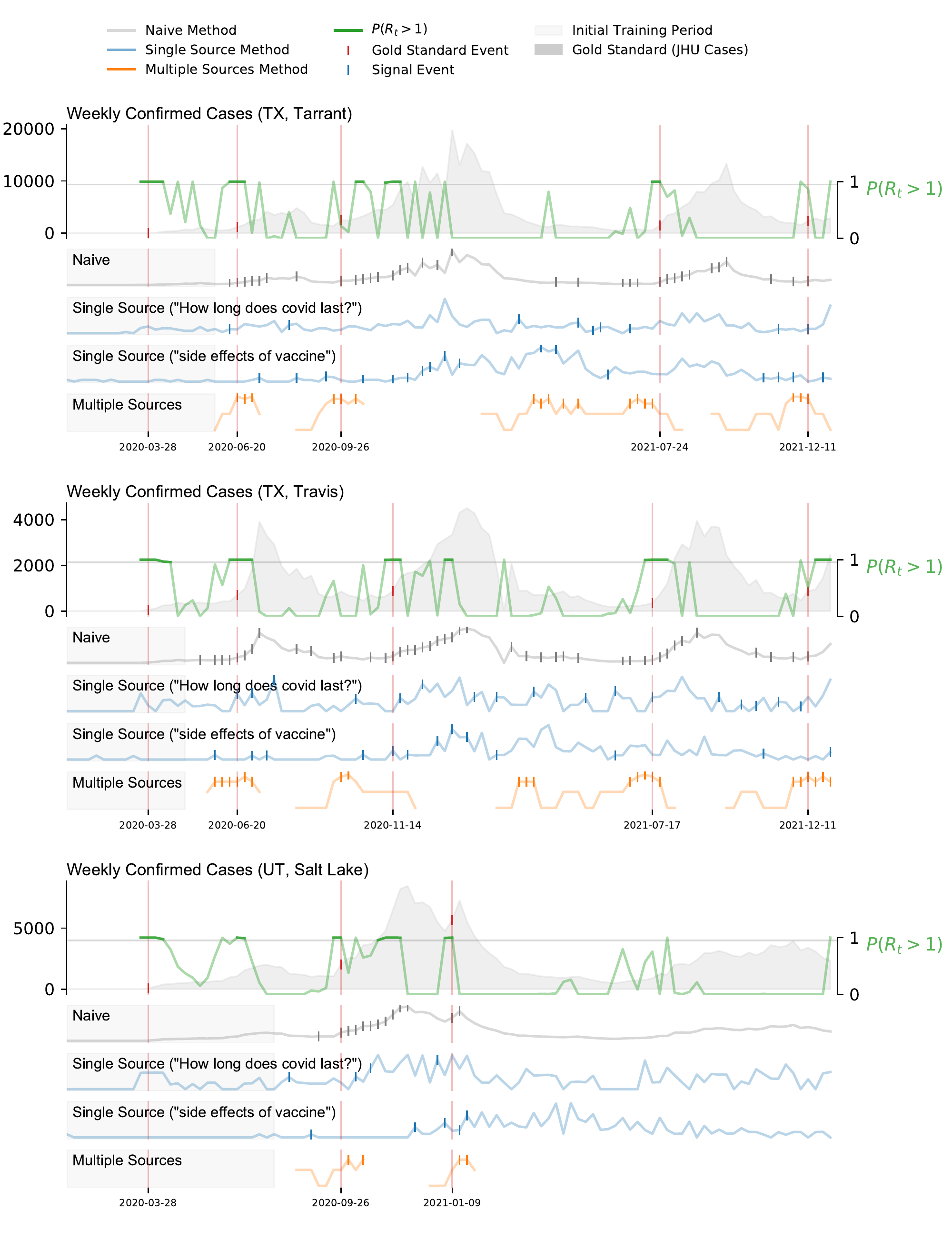}
    \caption{Graphical representation of the reproductive number, $R_t$, along with the weekly confirmed COVID-19 cases (gray-filled curve in the top), and three representative early warning methods (Naive, Single and Multiple Source) at county level.}
    \label{fig:30_county}
\end{figure} 

\begin{figure}
    \centering
    \includegraphics[width=0.9\textwidth,height=\textheight,keepaspectratio]{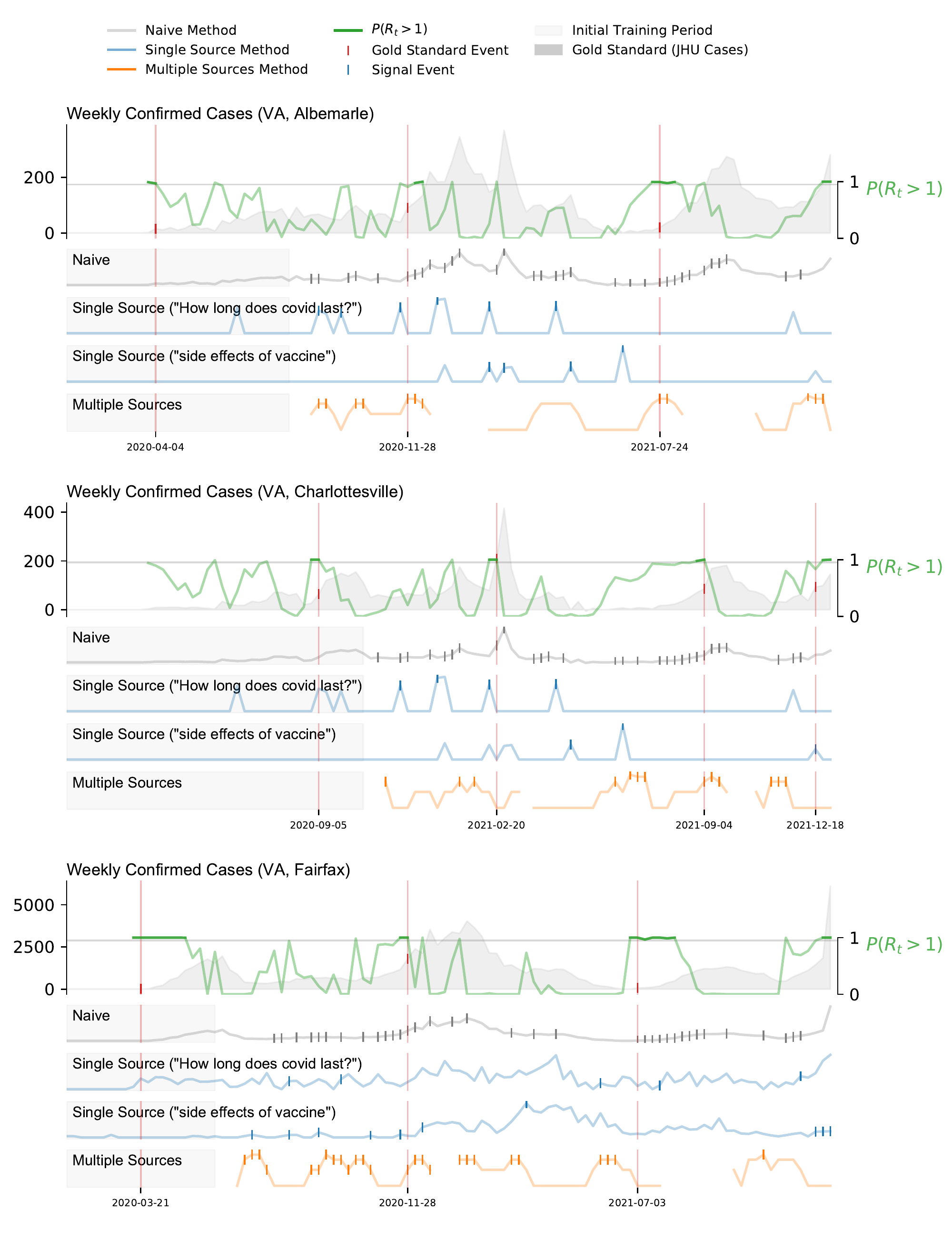}
    \caption{Graphical representation of the reproductive number, $R_t$, along with the weekly confirmed COVID-19 cases (gray-filled curve in the top), and three representative early warning methods (Naive, Single and Multiple Source) at county level.}
    \label{fig:31_county}
\end{figure} 

\begin{figure}
    \centering
    \includegraphics[width=0.9\textwidth,height=\textheight,keepaspectratio]{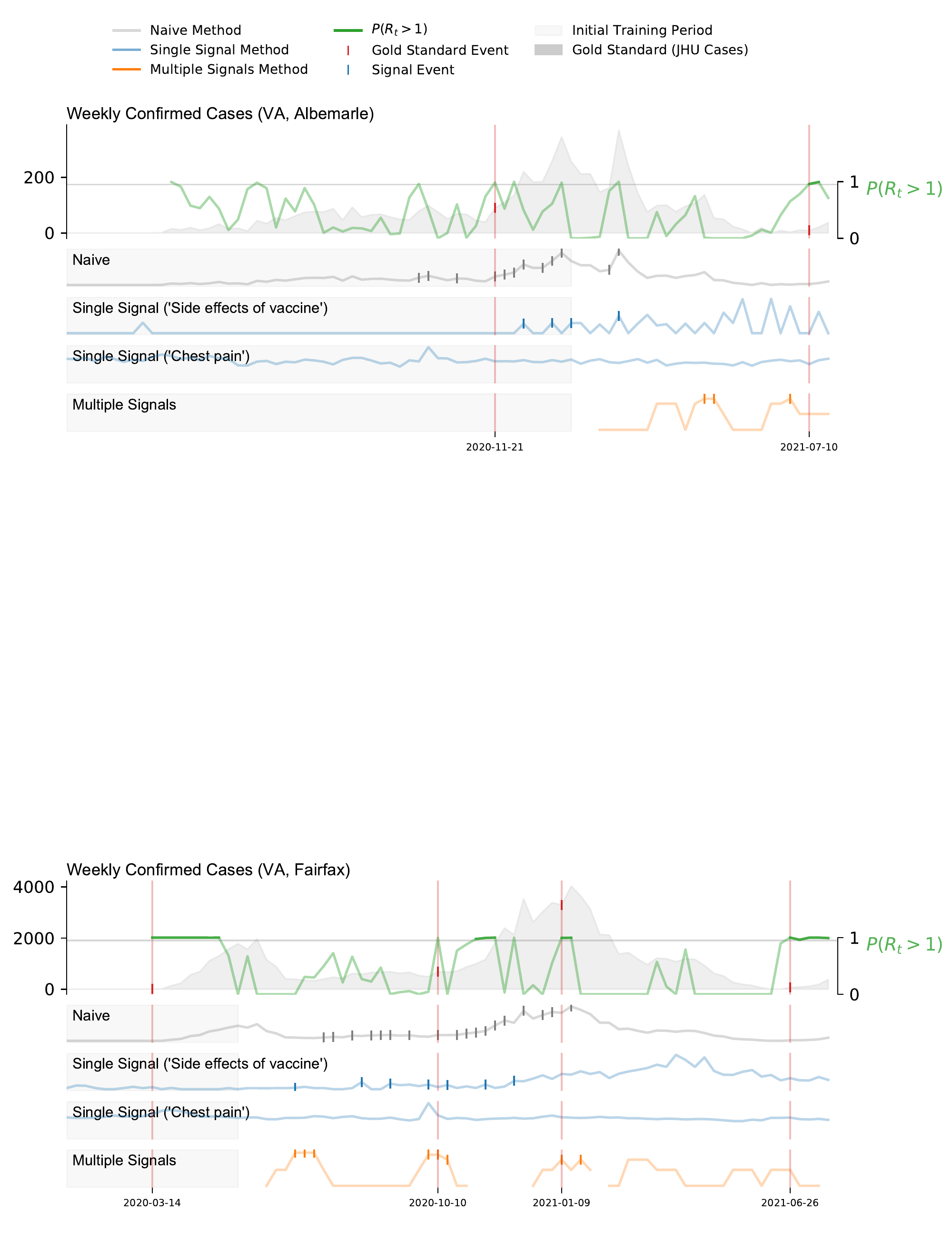}
    \caption{Graphical representation of the reproductive number, $R_t$, along with the weekly confirmed COVID-19 cases (gray-filled curve in the top), and three representative early warning methods (Naive, Single and Multiple Source) at county level.}
    \label{fig:31_county}
\end{figure}

\begin{figure}
    \centering
    \includegraphics[width=0.9\textwidth,height=\textheight,keepaspectratio]{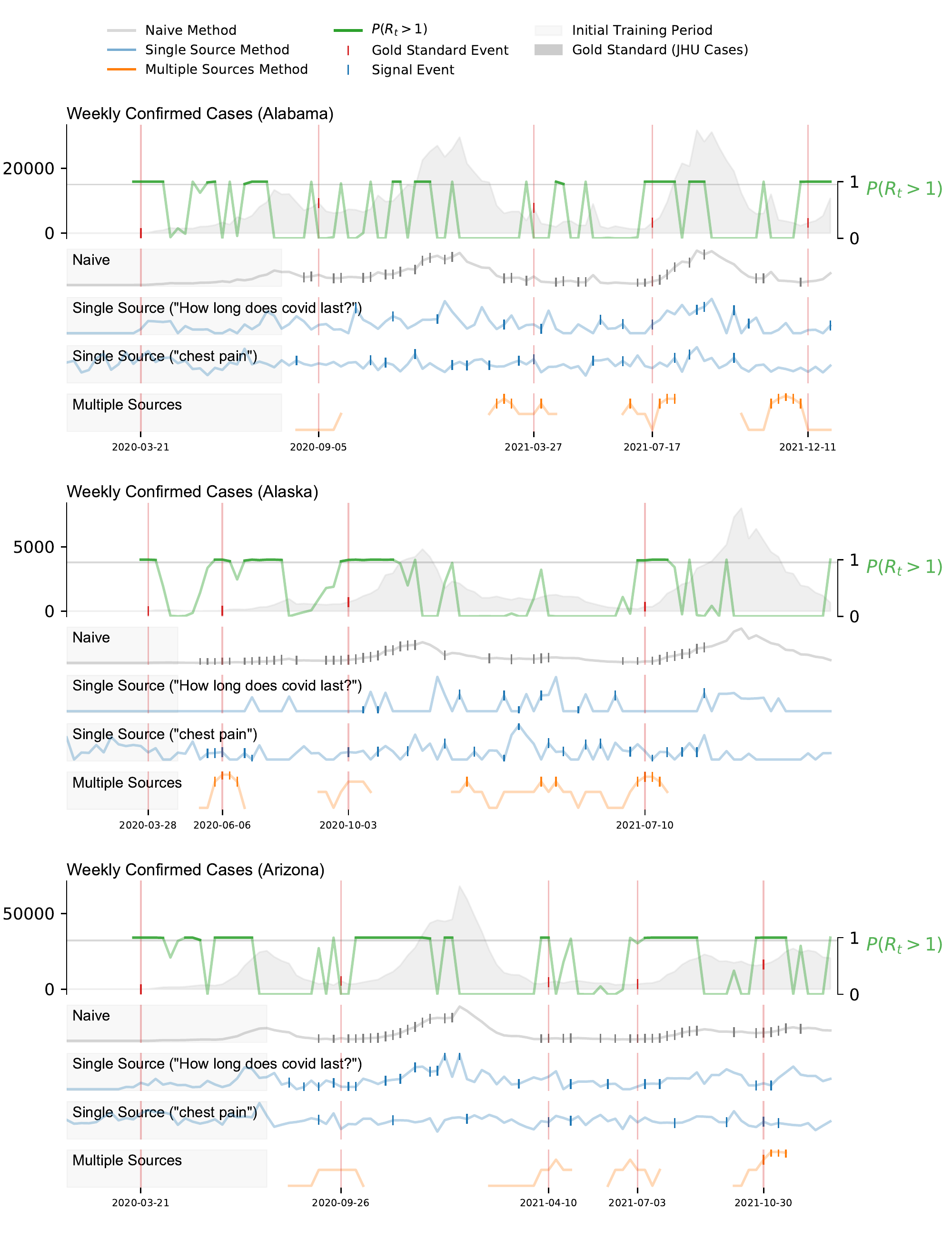}
    \caption{Graphical representation of the reproductive number, $R_t$, along with the weekly confirmed COVID-19 cases (gray-filled curve in the top), and three representative early warning methods (Naive, Single and Multiple Source) at county level.}
    \label{fig:1_state}
\end{figure} 

\begin{figure}
    \centering
    \includegraphics[width=0.9\textwidth,height=\textheight,keepaspectratio]{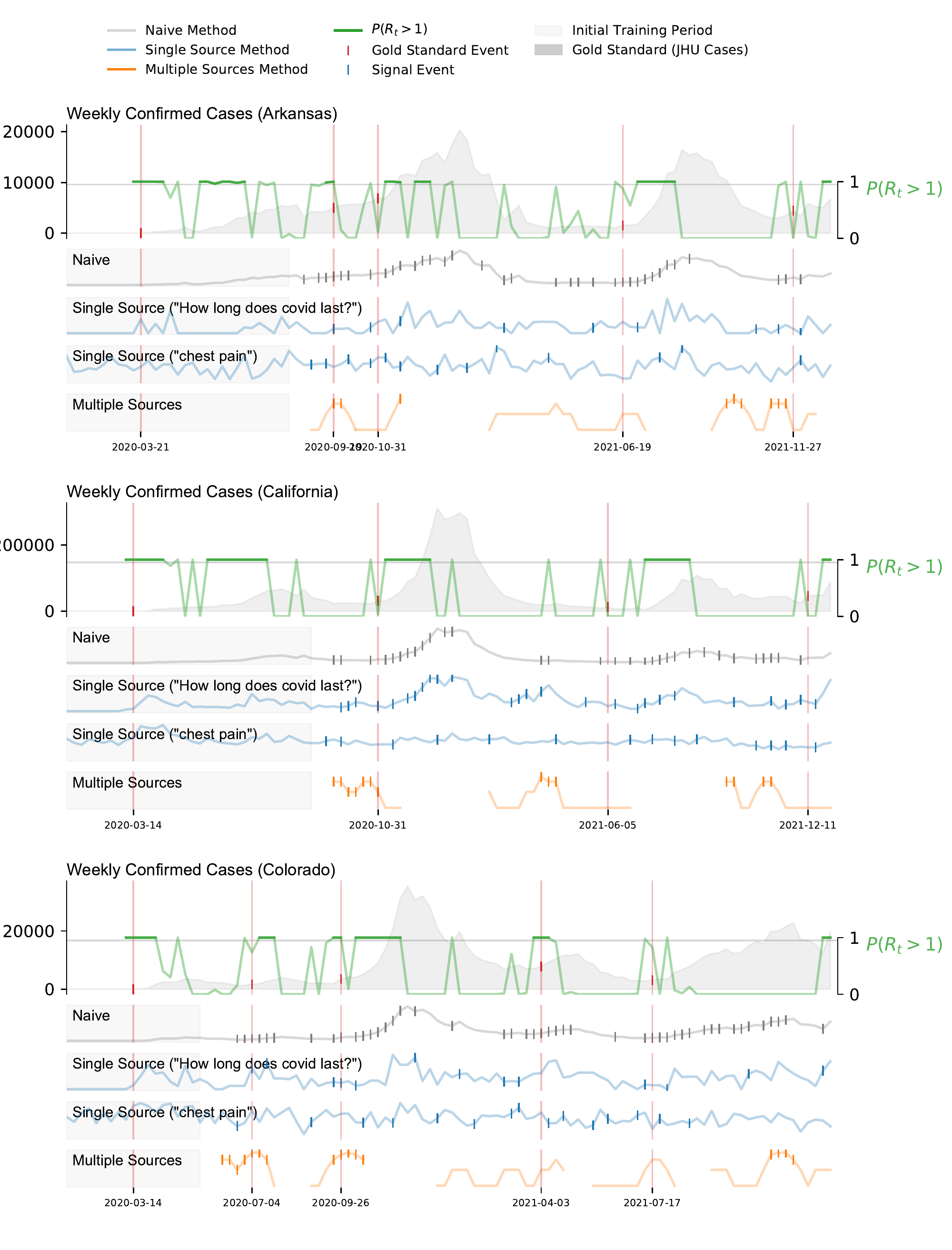}
    \caption{Graphical representation of the reproductive number, $R_t$, along with the weekly confirmed COVID-19 cases (gray-filled curve in the top), and three representative early warning methods (Naive, Single and Multiple Source) at county level.}
    \label{fig:2_state}
\end{figure} 

\begin{figure}
    \centering
    \includegraphics[width=0.9\textwidth,height=\textheight,keepaspectratio]{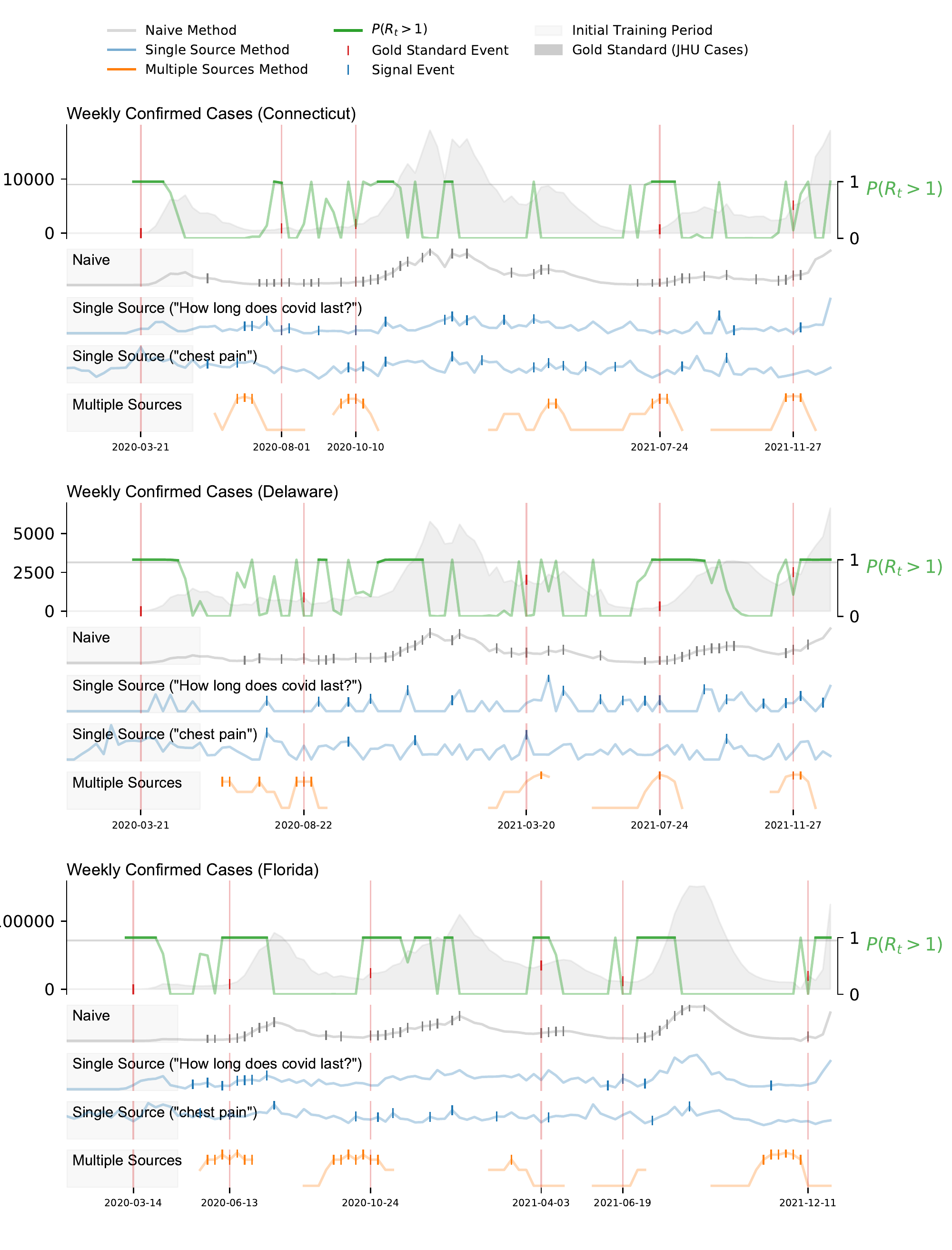}
    \caption{Graphical representation of the reproductive number, $R_t$, along with the weekly confirmed COVID-19 cases (gray-filled curve in the top), and three representative early warning methods (Naive, Single and Multiple Source) at county level.}
    \label{fig:3_state}
\end{figure} 

\begin{figure}
    \centering
    \includegraphics[width=0.9\textwidth,height=\textheight,keepaspectratio]{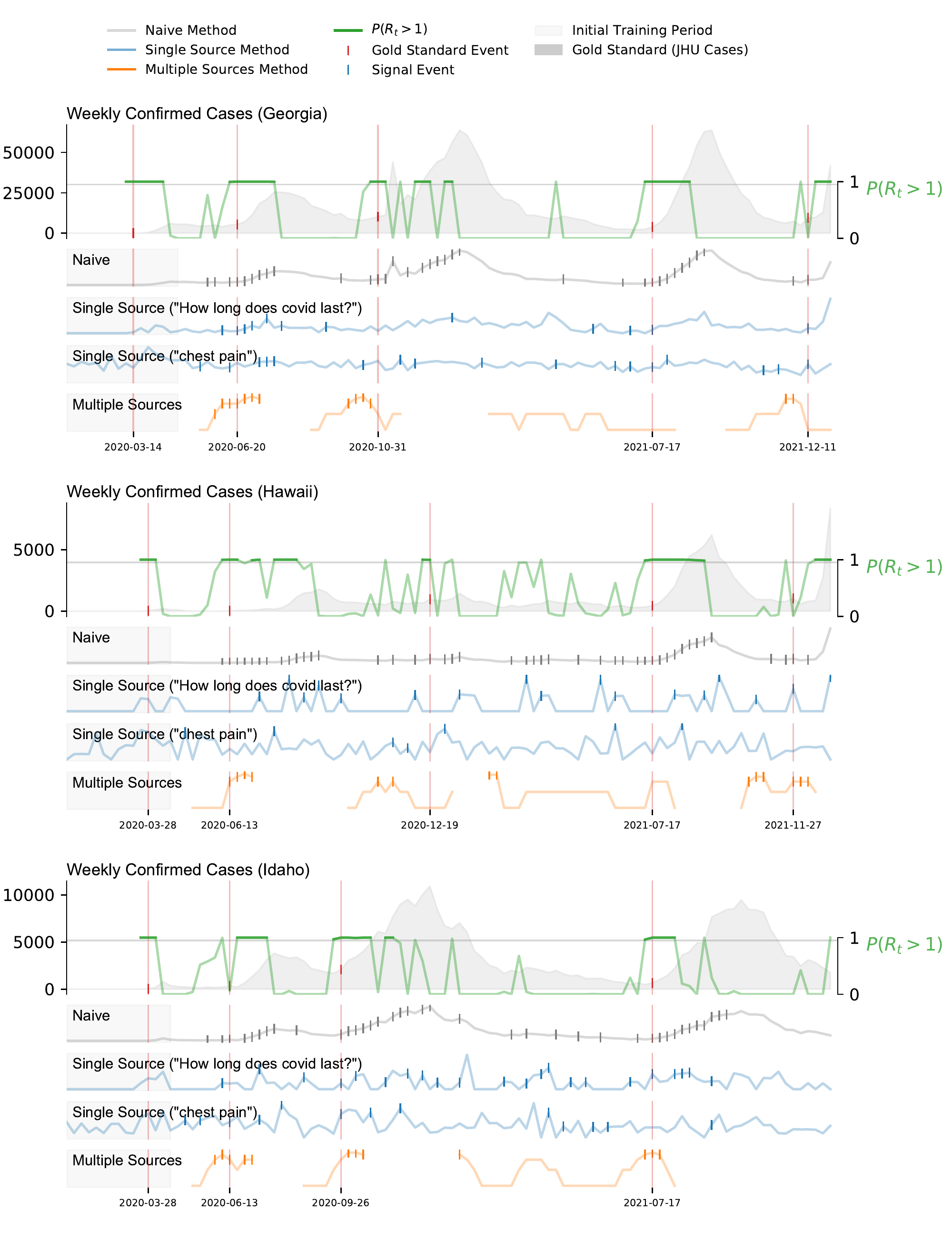}
    \caption{Graphical representation of the reproductive number, $R_t$, along with the weekly confirmed COVID-19 cases (gray-filled curve in the top), and three representative early warning methods (Naive, Single and Multiple Source) at county level.}
    \label{fig:4_state}
\end{figure} 

\begin{figure}
    \centering
    \includegraphics[width=0.9\textwidth,height=\textheight,keepaspectratio]{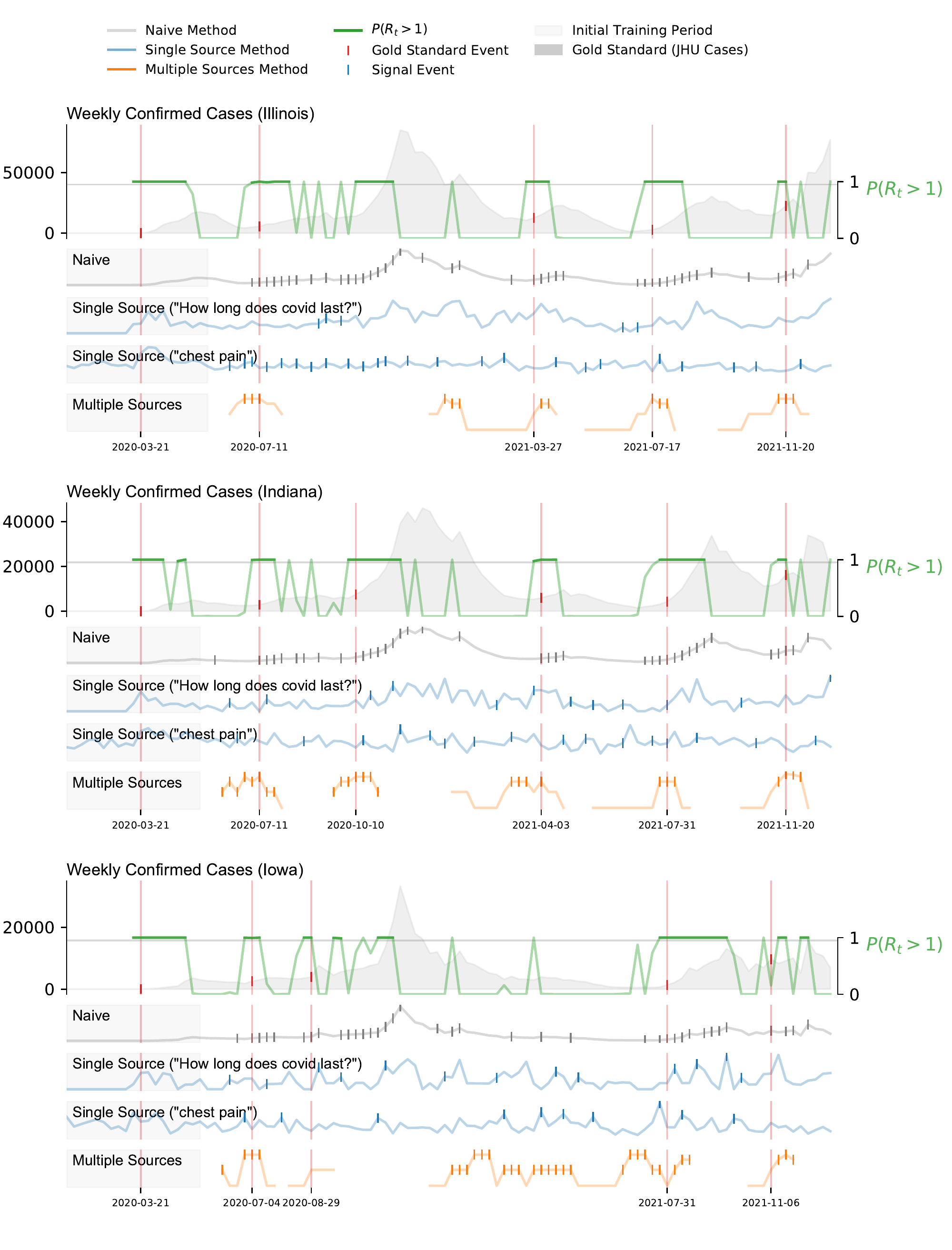}
    \caption{Graphical representation of the reproductive number, $R_t$, along with the weekly confirmed COVID-19 cases (gray-filled curve in the top), and three representative early warning methods (Naive, Single and Multiple Source) at county level.}
    \label{fig:5_state}
\end{figure} 

\begin{figure}
    \centering
    \includegraphics[width=0.9\textwidth,height=\textheight,keepaspectratio]{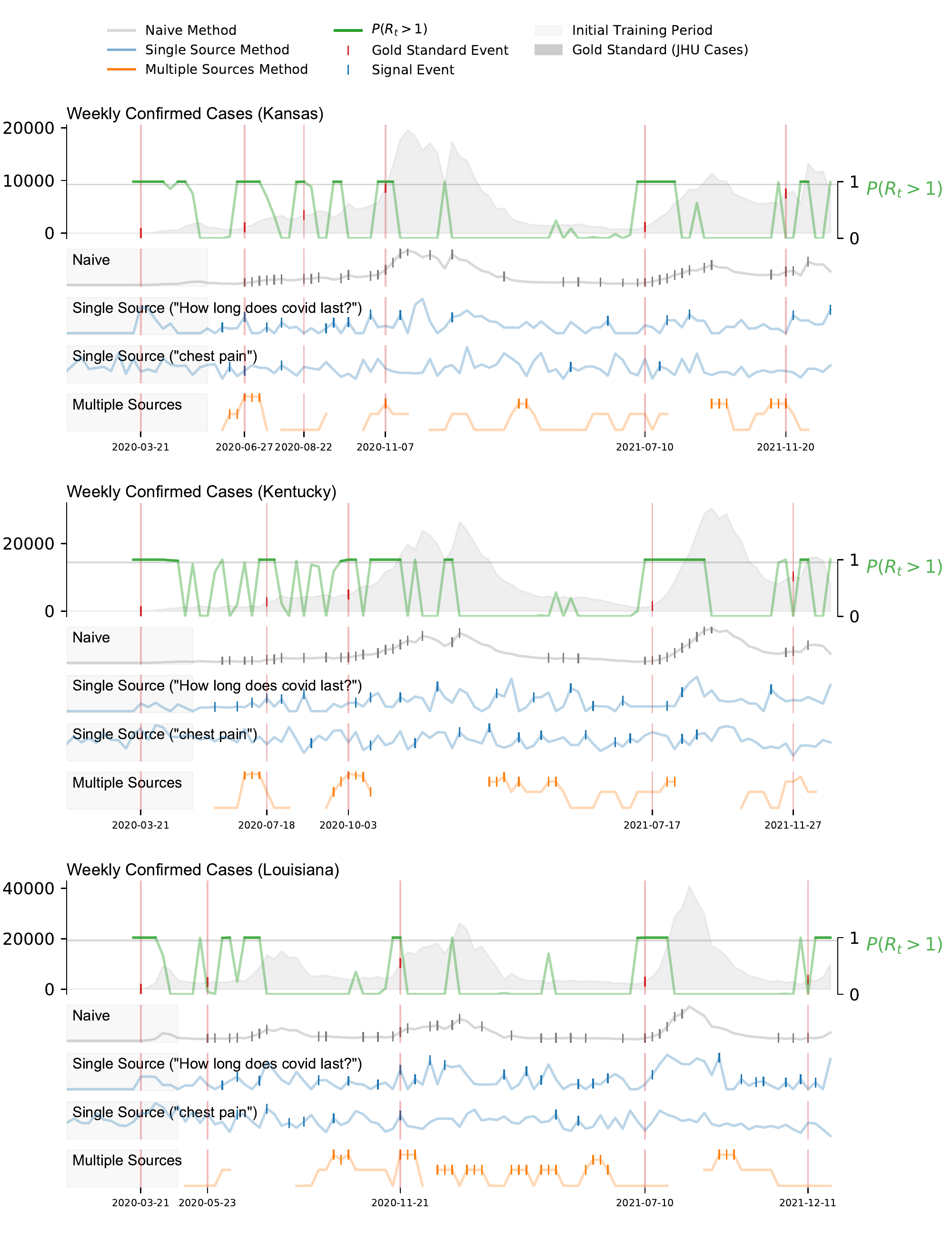}
    \caption{Graphical representation of the reproductive number, $R_t$, along with the weekly confirmed COVID-19 cases (gray-filled curve in the top), and three representative early warning methods (Naive, Single and Multiple Source) at county level.}
    \label{fig:6_state}
\end{figure} 

\begin{figure}
    \centering
    \includegraphics[width=0.9\textwidth,height=\textheight,keepaspectratio]{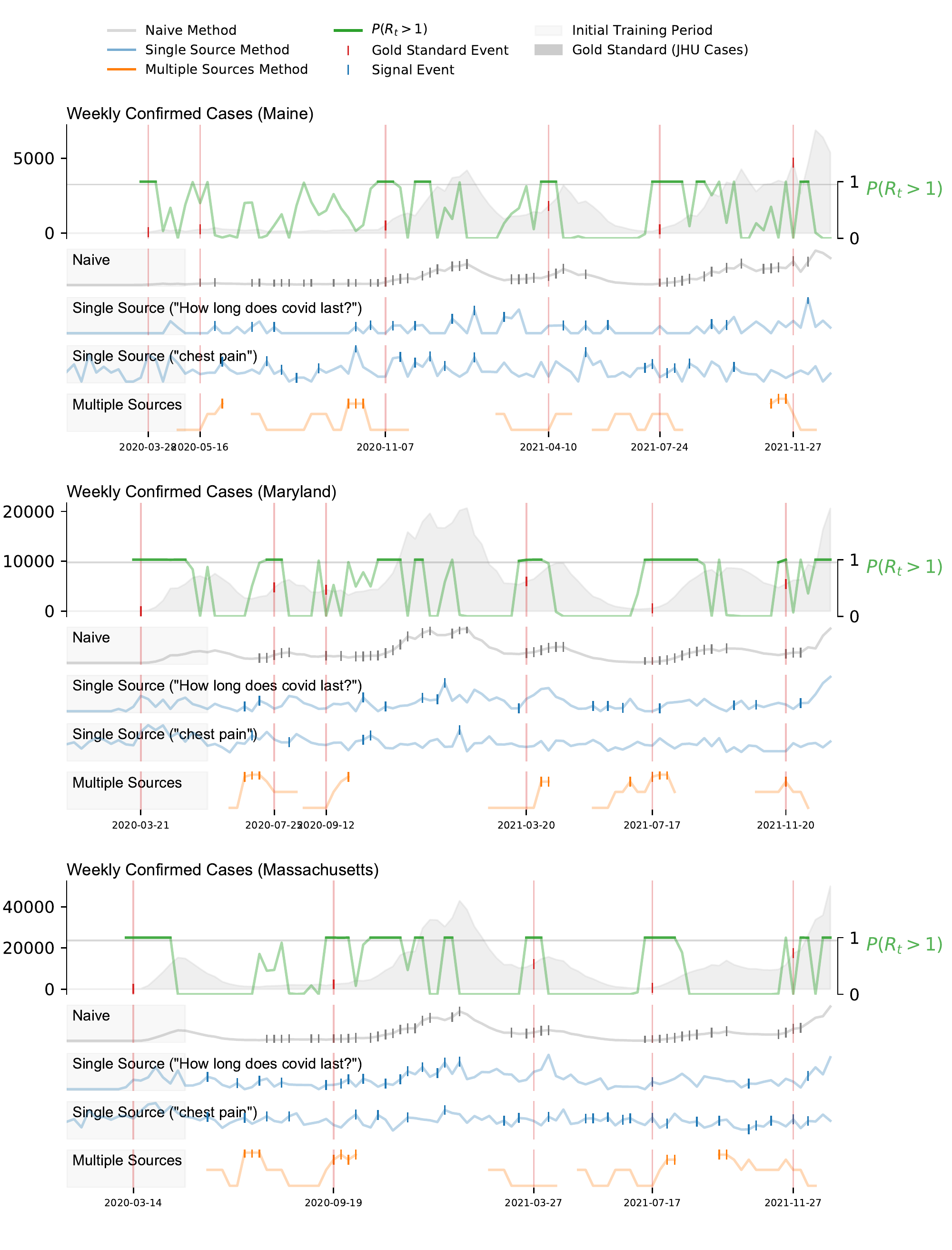}
    \caption{Graphical representation of the reproductive number, $R_t$, along with the weekly confirmed COVID-19 cases (gray-filled curve in the top), and three representative early warning methods (Naive, Single and Multiple Source) at county level.}
    \label{fig:7_state}
\end{figure} 

\begin{figure}
    \centering
    \includegraphics[width=0.9\textwidth,height=\textheight,keepaspectratio]{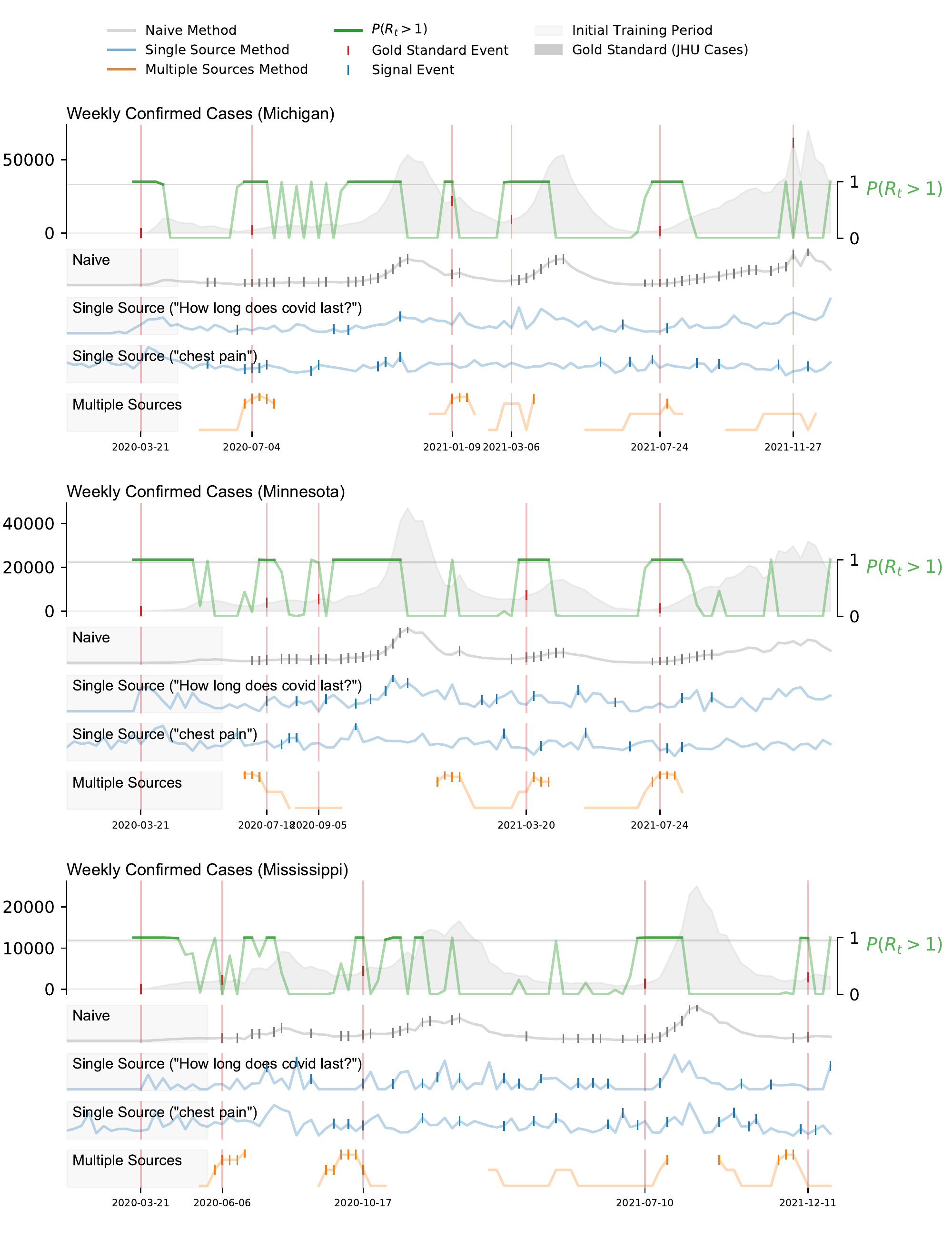}
    \caption{Graphical representation of the reproductive number, $R_t$, along with the weekly confirmed COVID-19 cases (gray-filled curve in the top), and three representative early warning methods (Naive, Single and Multiple Source) at county level.}
    \label{fig:8_state}
\end{figure} 

\begin{figure}
    \centering
    \includegraphics[width=0.9\textwidth,height=\textheight,keepaspectratio]{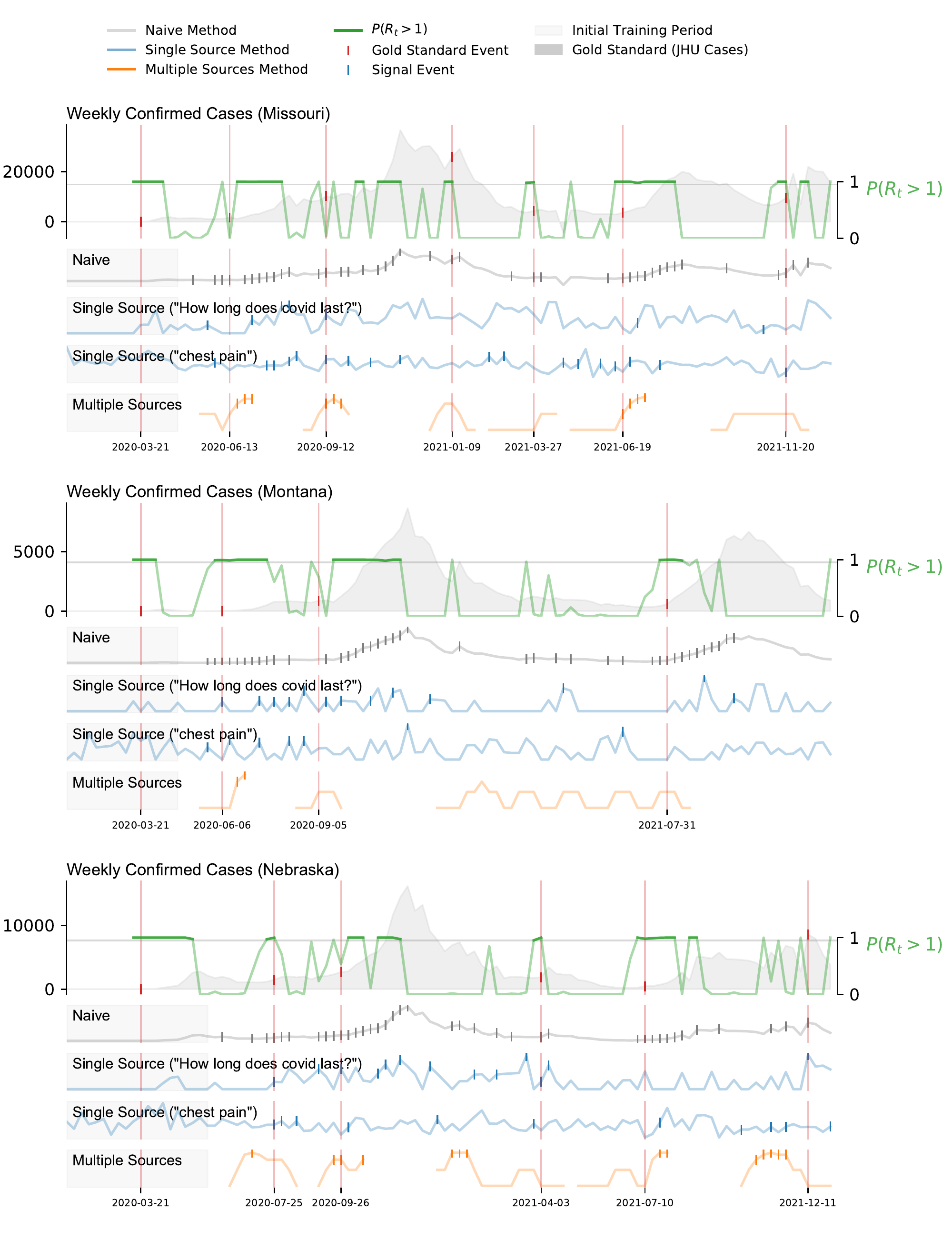}
    \caption{Graphical representation of the reproductive number, $R_t$, along with the weekly confirmed COVID-19 cases (gray-filled curve in the top), and three representative early warning methods (Naive, Single and Multiple Source) at county level.}
    \label{fig:9_state}
\end{figure} 

\begin{figure}
    \centering
    \includegraphics[width=0.9\textwidth,height=\textheight,keepaspectratio]{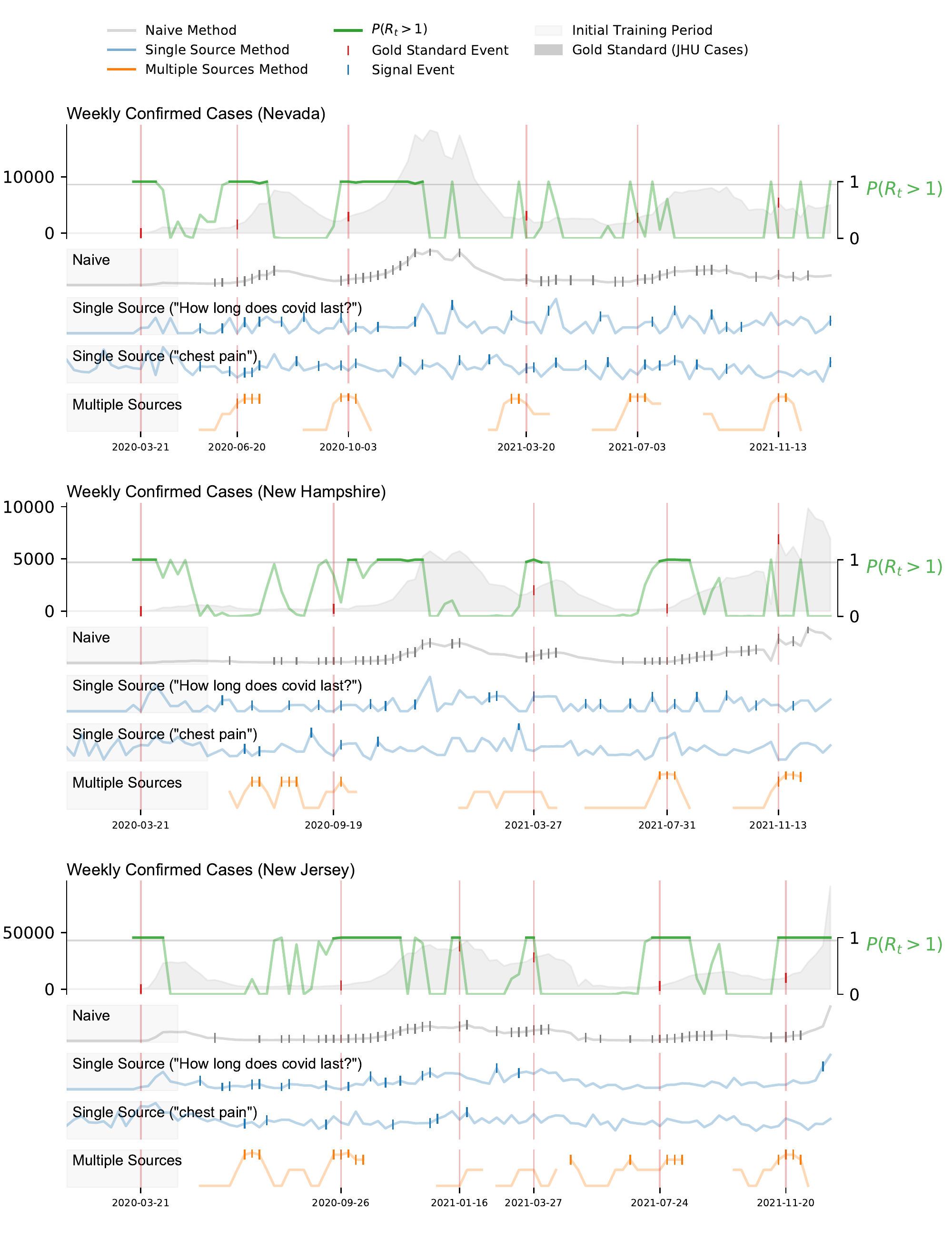}
    \caption{Graphical representation of the reproductive number, $R_t$, along with the weekly confirmed COVID-19 cases (gray-filled curve in the top), and three representative early warning methods (Naive, Single and Multiple Source) at county level.}
    \label{fig:10_state}
\end{figure} 

\begin{figure}
    \centering
    \includegraphics[width=0.9\textwidth,height=\textheight,keepaspectratio]{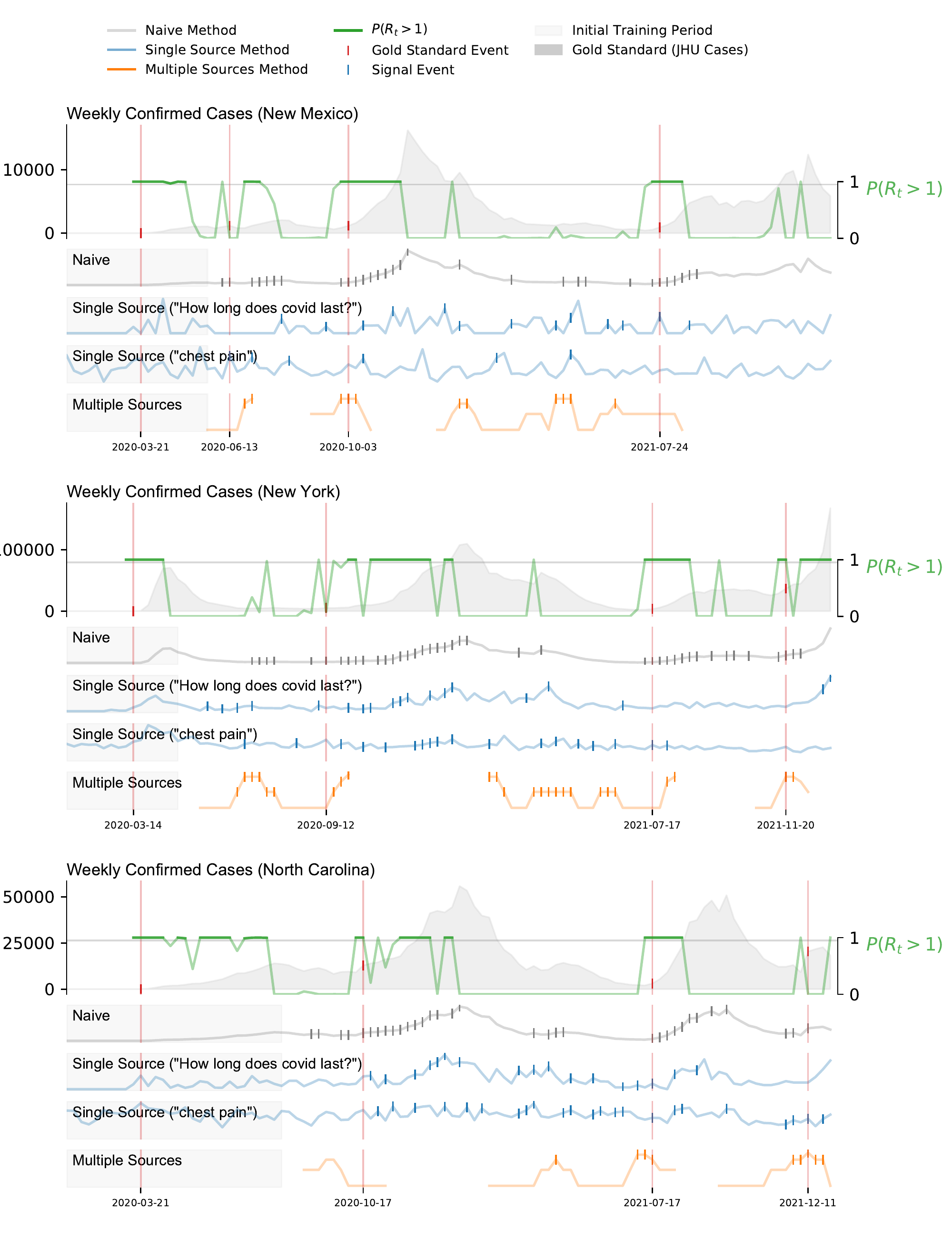}
    \caption{Graphical representation of the reproductive number, $R_t$, along with the weekly confirmed COVID-19 cases (gray-filled curve in the top), and three representative early warning methods (Naive, Single and Multiple Source) at county level.}
    \label{fig:11_state}
\end{figure} 

\begin{figure}
    \centering
    \includegraphics[width=0.9\textwidth,height=\textheight,keepaspectratio]{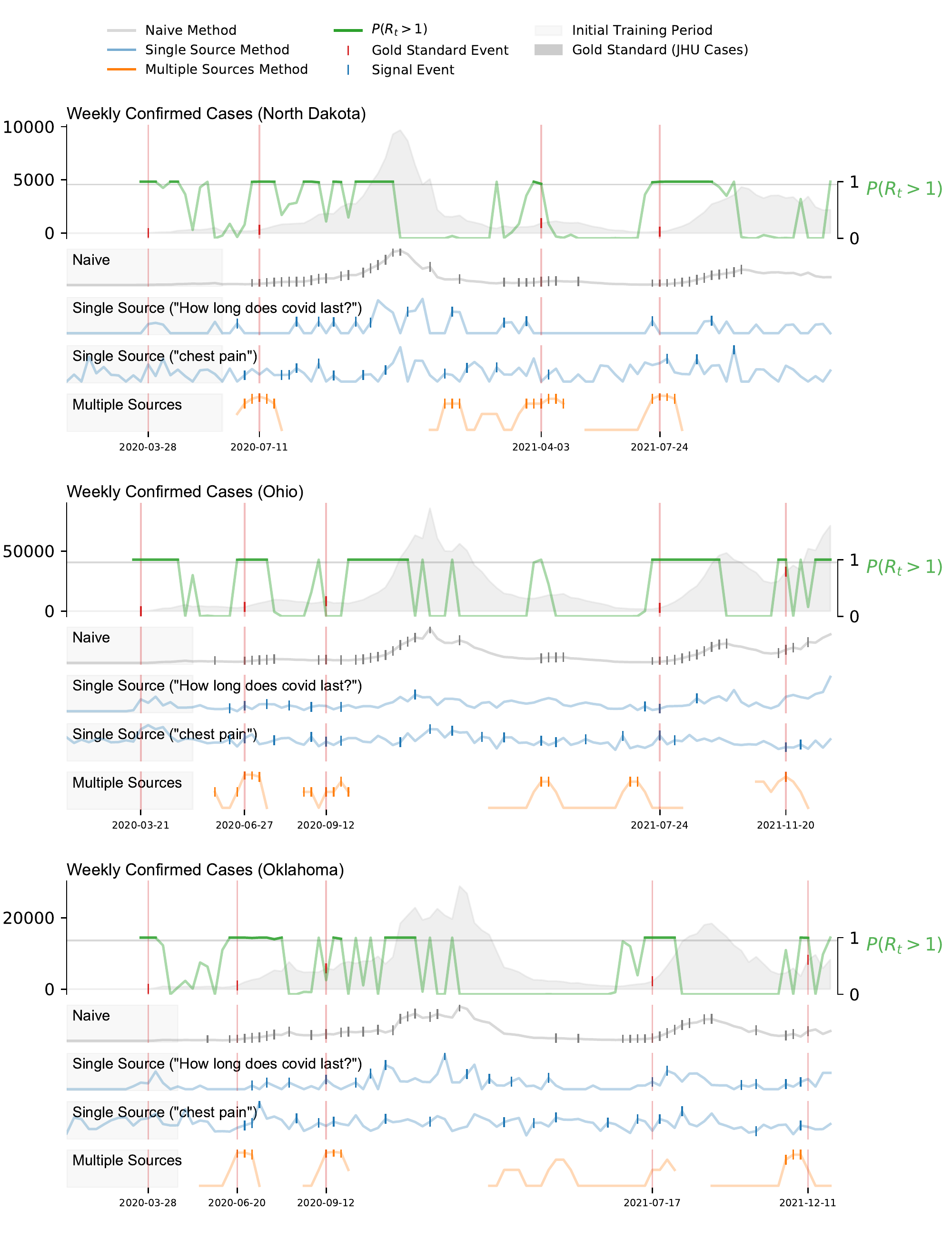}
    \caption{Graphical representation of the reproductive number, $R_t$, along with the weekly confirmed COVID-19 cases (gray-filled curve in the top), and three representative early warning methods (Naive, Single and Multiple Source) at county level.}
    \label{fig:12_state}
\end{figure} 

\begin{figure}
    \centering
    \includegraphics[width=0.9\textwidth,height=\textheight,keepaspectratio]{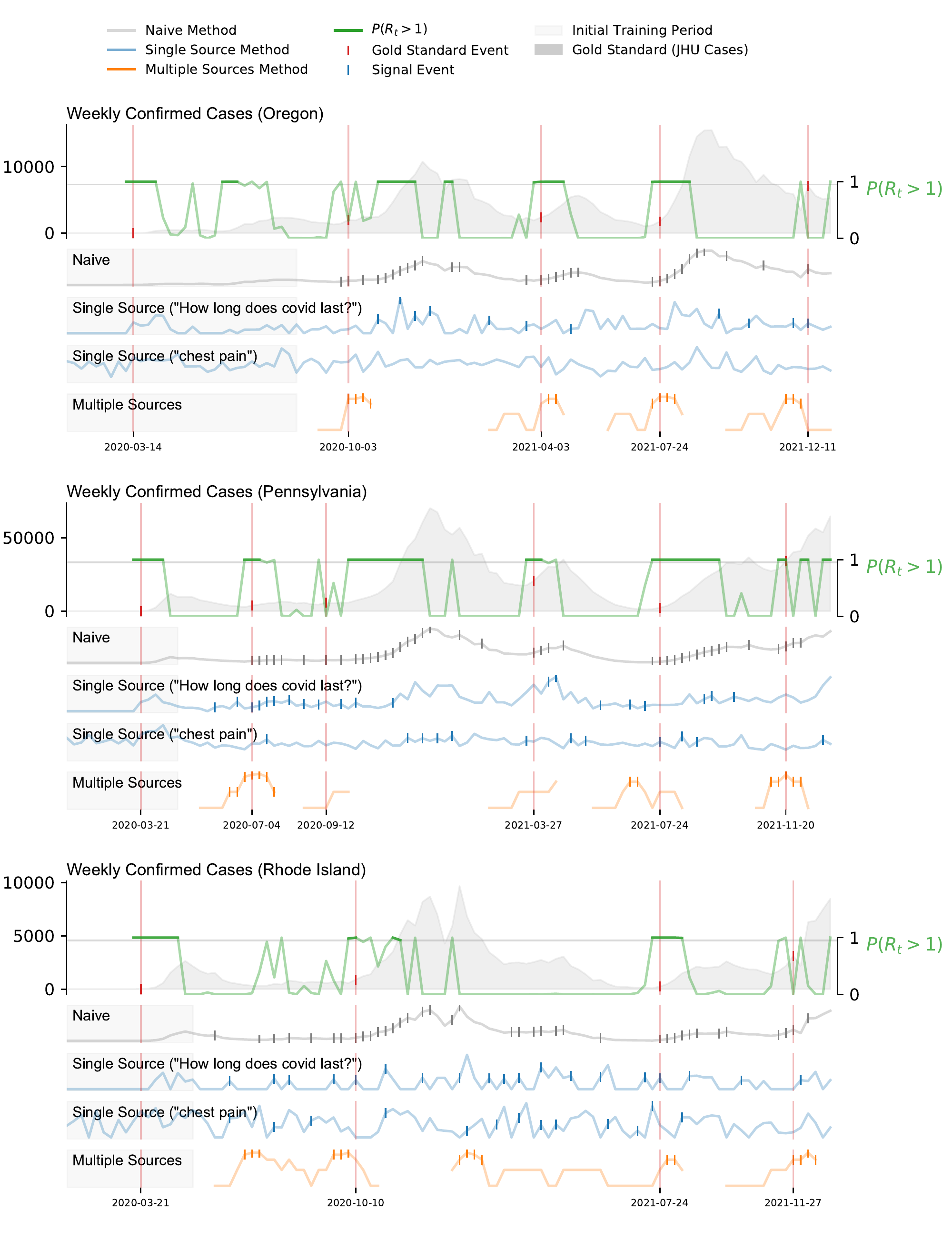}
    \caption{Graphical representation of the reproductive number, $R_t$, along with the weekly confirmed COVID-19 cases (gray-filled curve in the top), and three representative early warning methods (Naive, Single and Multiple Source) at county level.}
    \label{fig:13_state}
\end{figure} 

\begin{figure}
    \centering
    \includegraphics[width=0.9\textwidth,height=\textheight,keepaspectratio]{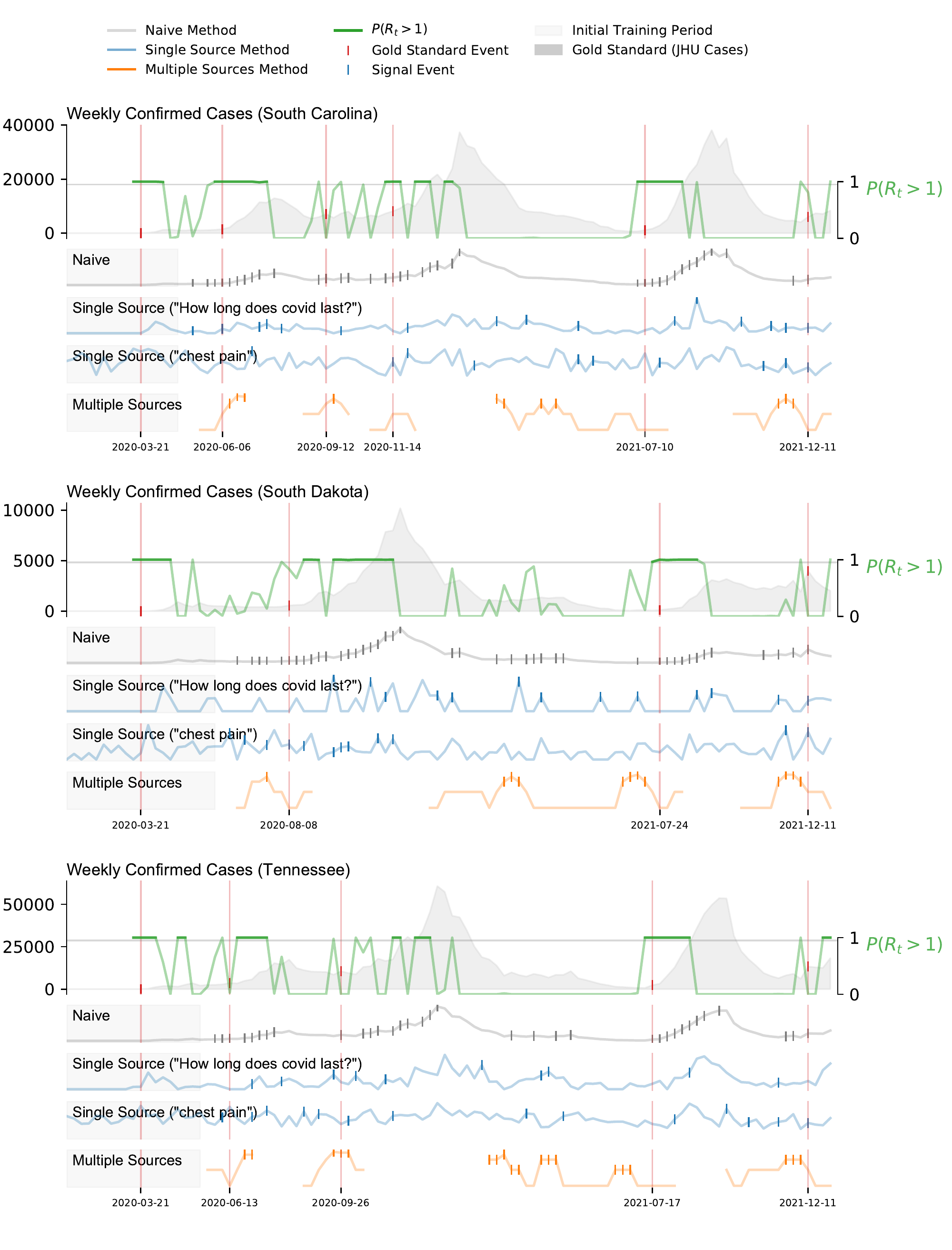}
    \caption{Graphical representation of the reproductive number, $R_t$, along with the weekly confirmed COVID-19 cases (gray-filled curve in the top), and three representative early warning methods (Naive, Single and Multiple Source) at county level.}
    \label{fig:14_state}
\end{figure} 

\begin{figure}
    \centering
    \includegraphics[width=0.9\textwidth,height=\textheight,keepaspectratio]{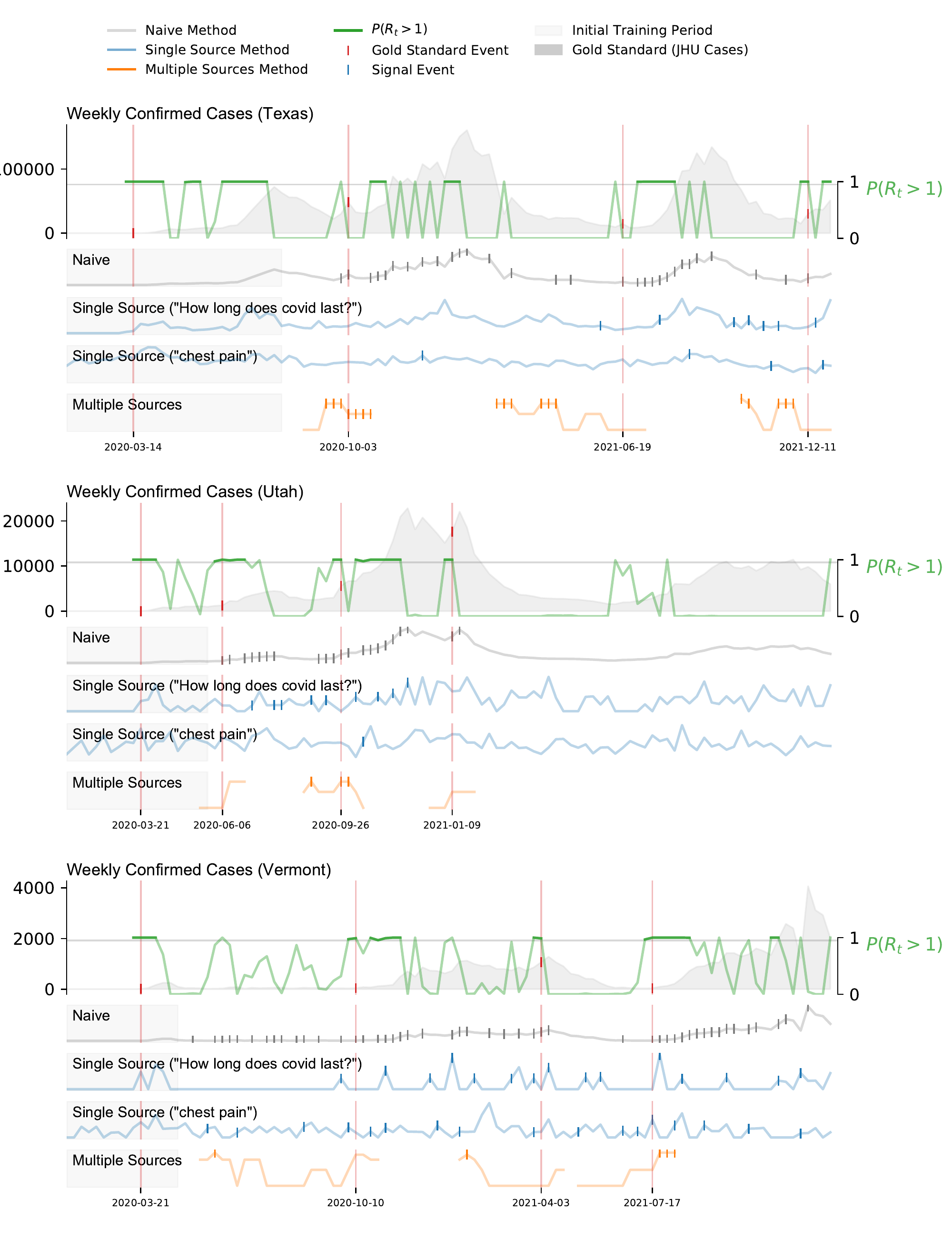}
    \caption{Graphical representation of the reproductive number, $R_t$, along with the weekly confirmed COVID-19 cases (gray-filled curve in the top), and three representative early warning methods (Naive, Single and Multiple Source) at county level.}
    \label{fig:15_state}
\end{figure} 

\begin{figure}
    \centering
    \includegraphics[width=0.9\textwidth,height=\textheight,keepaspectratio]{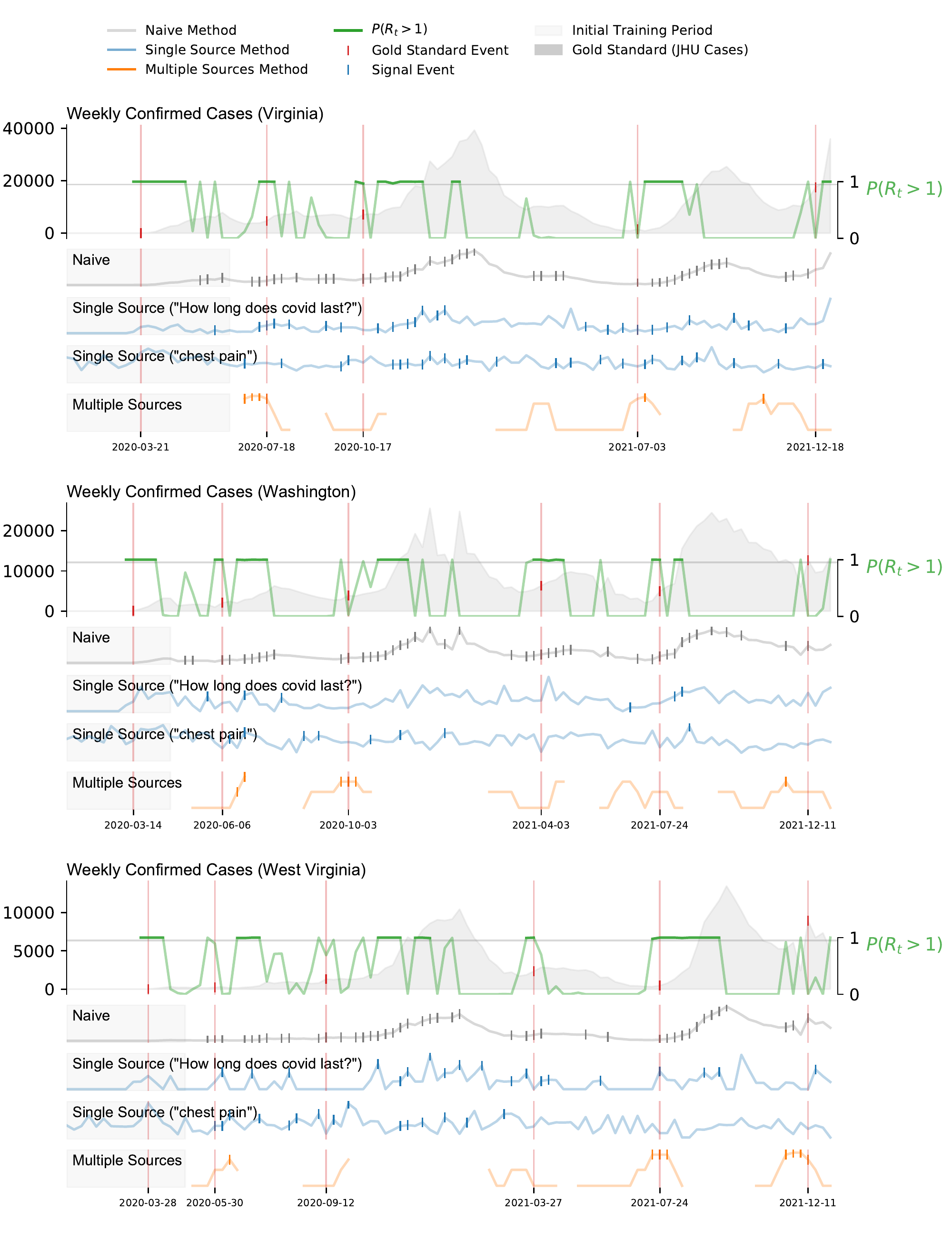}
    \caption{Graphical representation of the reproductive number, $R_t$, along with the weekly confirmed COVID-19 cases (gray-filled curve in the top), and three representative early warning methods (Naive, Single and Multiple Source) at county level.}
    \label{fig:16_state}
\end{figure}

\begin{landscape}
\begin{table}[]
\scriptsize
    \centering
   \begin{tabular}{lllllll}
\toprule
{} & Early Warning & Sync Warning & Late Warning & Missed Outbreaks & False Alarm, Increase  & False Alarm, No Increase \\
\midrule
(GT) Side effects of vaccine             &     176 (67\%) &       5 (2\%) &      12 (5\%) &         71 (27\%) &                                                 16 &         450 \\
(GT) Chest pain                          &     170 (64\%) &      20 (8\%) &      14 (5\%) &         60 (23\%) &                                                  8 &         273 \\
(GT) How long does covid last?           &     168 (64\%) &      14 (5\%) &      21 (8\%) &         61 (23\%) &                                                  6 &         365 \\
(GT) covid 19 who                        &     150 (57\%) &      12 (5\%) &     29 (11\%) &         73 (28\%) &                                                 11 &         353 \\
(GT), covid symptoms                     &     126 (48\%) &      11 (4\%) &     40 (15\%) &         87 (33\%) &                                                  3 &         249 \\
(GT) fever                               &     115 (44\%) &      14 (5\%) &     26 (10\%) &        109 (41\%) &                                                 10 &         265 \\
(GT) after covid vaccine                 &     112 (42\%) &      10 (4\%) &      20 (8\%) &        122 (46\%) &                                                 23 &         473 \\
COVID-19 Confirmed Deaths (county level) &      99 (38\%) &       8 (3\%) &      16 (6\%) &        141 (53\%) &                                                  3 &         222 \\
(GT) covid 19                            &      93 (35\%) &      10 (4\%) &     27 (10\%) &        134 (51\%) &                                                  9 &         258 \\
COVID-19 Confirmed Cases (state level)   &      91 (34\%) &      25 (9\%) &     49 (19\%) &         99 (38\%) &                                                  5 &         214 \\
COVID-19 Confirmed Cases (county level)  &      84 (32\%) &     27 (10\%) &     40 (15\%) &        113 (43\%) &                                                  5 &         200 \\
(GT) Effects of covid vaccine            &      78 (30\%) &       4 (2\%) &       7 (3\%) &        175 (66\%) &                                                 11 &         281 \\
(GT) UpToDate                            &      78 (30\%) &      16 (6\%) &      20 (8\%) &        150 (57\%) &                                                  0 &          99 \\
(GT) covid                               &      72 (27\%) &      16 (6\%) &     51 (19\%) &        125 (47\%) &                                                  9 &         254 \\
COVID-19 Confirmed Deaths (state level)  &      71 (27\%) &      10 (4\%) &      12 (5\%) &        171 (65\%) &                                                  4 &         158 \\
Twitter                                  &      56 (21\%) &      10 (4\%) &      17 (6\%) &        181 (69\%) &                                                  1 &          72 \\
\bottomrule
\end{tabular}
    \caption{Performance of selected terms for the Single Source method, at county level (N=263)}
    \label{tab:my_label}
\end{table}
\end{landscape}

\begin{table}[]
	\begin{adjustwidth}{-0.7in}{0in} 
	\vspace{-5.5em}
    \tiny
\begin{tabular}{lllllll}
\toprule
{} &                               &                                           &                                          &                               &                                   &                \\
\midrule
  &              Abdominal obesity &                             Abdominal pain &                                      Acne &              Actinic keratosis &                   Acute bronchitis &  Adrenal crisis \\
  &                        Ageusia &                                 Alcoholism &                   Allergic conjunctivitis &                        Allergy &                          Amblyopia &             \\
  &                     Amenorrhea &                                    Amnesia &                              Anal fissure &                    Anaphylaxis &                             Anemia &             \\
  &                Angina pectoris &                                 Angioedema &                         Angular cheilitis &                        Anosmia &                            Anxiety &             \\
  &                        Aphasia &                                    Aphonia &                                     Apnea &                     Arthralgia &                          Arthritis &             \\
  &                        Ascites &                          Asperger syndrome &                                  Asphyxia &                         Asthma &                        Astigmatism &             \\
  &                         Ataxia &                                   Atheroma &  Attention deficit hyperactivity disorder &         Auditory hallucination &                 Autoimmune disease &             \\
  &  Avoidant personality disorder &                                  Back pain &                       Bacterial vaginosis &               Balance disorder &                       Beau's lines &             \\
  &                   Bell's palsy &                              Biliary colic &                              Binge eating &                       Bleeding &                Bleeding on probing &             \\
  &                  Blepharospasm &                                   Bloating &                            Blood in stool &                 Blurred vision &                           Blushing &             \\
 &                           Boil &                              Bone fracture &                                Bone tumor &              Bowel obstruction &                        Bradycardia &             \\
 &     Braxton Hicks contractions &                      Breakthrough bleeding &                               Breast pain &                     Bronchitis &                             Bruise &             \\
 &                        Bruxism &                                     Bunion &                                      Burn &             Burning Chest Pain &             Burning mouth syndrome &             \\
 &                    Candidiasis &                                Canker sore &                            Cardiac arrest &         Carpal tunnel syndrome &                          Cataplexy &             \\
 &                       Cataract &                                    Chancre &                                 Cheilitis &                     Chest pain &                             Chills &             \\
 &                         Chorea &                               Chronic pain &                                 Cirrhosis &     Cleft lip and cleft palate &          Clouding of consciousness &             \\
 &               Cluster headache &                                    Colitis &                                      Coma &                    Common cold &                Compulsive behavior &             \\
 &            Compulsive hoarding &                                  Confusion &                   Congenital heart defect &                 Conjunctivitis &                       Constipation &             \\
 &                     Convulsion &                                      Cough &                                  Crackles &                          Cramp &                           Crepitus &             \\
 &                          Croup &                                   Cyanosis &                                  Dandruff &  Delayed onset muscle soreness &                           Dementia &             \\
 &        Dentin hypersensitivity &                          Depersonalization &                                Depression &                     Dermatitis &                       Desquamation &             \\
 &       Developmental disability &                                   Diabetes &                     Diabetic ketoacidosis &                       Diarrhea &                          Dizziness &             \\
 &               Dry eye syndrome &                               Dysautonomia &                                 Dysgeusia &                   Dysmenorrhea &                        Dyspareunia &             \\
 &                      Dysphagia &                                  Dysphoria &                                  Dystonia &                        Dysuria &                           Ear pain &             \\
 &                         Eczema &                                      Edema &                              Encephalitis &                 Encephalopathy &                    Epidermoid cyst &             \\
 &                       Epilepsy &                                   Epiphora &                      Erectile dysfunction &                       Erythema &         Erythema chronicum migrans &             \\
 &                    Esophagitis &               Excessive daytime sleepiness &                                  Eye pain &                     Eye strain &             Facial nerve paralysis &             \\
 &                Facial swelling &                              Fasciculation &                                   Fatigue &            Fatty liver disease &                 Fecal incontinence &             \\
 &                          Fever &                               Fibrillation &                Fibrocystic breast changes &                   Fibromyalgia &                         Flatulence &             \\
 &                        Floater &                              Focal seizure &                         Folate deficiency &                   Food craving &                   Food intolerance &             \\
 &             Frequent urination &            Gastroesophageal reflux disease &                             Gastroparesis &   Generalized anxiety disorder &   Generalized tonic–clonic seizure &             \\
 &                   Genital wart &                         Gingival recession &                                Gingivitis &               Globus pharyngis &                             Goitre &             \\
 &                           Gout &                                Grandiosity &                                 Granuloma &                          Guilt &                          Hair loss &             \\
 &                      Halitosis &                                  Hay fever &                                  Headache &               Heart arrhythmia &                       Heart murmur &             \\
 &                      Heartburn &                               Hematochezia &                                  Hematoma &                      Hematuria &                          Hemolysis &             \\
 &                     Hemoptysis &                                Hemorrhoids &                    Hepatic encephalopathy &                      Hepatitis &                     Hepatotoxicity &             \\
 &                         Hiccup &                                   Hip pain &                                     Hives &                      Hot flash &                      Hydrocephalus &             \\
 &                 Hypercalcaemia &                                Hypercapnia &                      Hypercholesterolemia &         Hyperemesis gravidarum &                      Hyperglycemia &             \\
 &                   Hyperkalemia &                             Hyperlipidemia &                             Hypermobility &              Hyperpigmentation &                        Hypersomnia &             \\
 &                   Hypertension &                               Hyperthermia &                           Hyperthyroidism &           Hypertriglyceridemia &                        Hypertrophy &             \\
 &               Hyperventilation &                              Hypocalcaemia &                           Hypochondriasis &                   Hypoglycemia &                       Hypogonadism &             \\
 &                    Hypokalemia &                                  Hypomania &                              Hyponatremia &                    Hypotension &                     Hypothyroidism &             \\
 &                      Hypoxemia &                                    Hypoxia &                                  Impetigo &          Implantation bleeding &                        Impulsivity &             \\
 &                    Indigestion &                                  Infection &                              Inflammation &     Inflammatory bowel disease &                       Ingrown hair &             \\
 &                       Insomnia &                         Insulin resistance &                   Intermenstrual bleeding &          Intracranial pressure &                    Iron deficiency &             \\
 &         Irregular menstruation &                                       Itch &                                  Jaundice &                 Kidney failure &                       Kidney stone &             \\
 &                      Knee Pain &                                   Kyphosis &                       Lactose intolerance &                     Laryngitis &                         Leg cramps &             \\
 &                         Lesion &                                 Leukorrhea &                           Lightheadedness &                  Low back pain &                    Low-grade fever &             \\
 &                     Lymphedema &                  Major depressive disorder &                             Malabsorption &               Male infertility &                     Manic Disorder &             \\
 &                        Melasma &                                     Melena &                                Meningitis &                    Menorrhagia &                   Middle back pain &             \\
 &                       Migraine &                                     Milium &                      Mitral insufficiency &                  Mood disorder &                         Mood swing &             \\
 &               Morning sickness &                            Motion sickness &                               Mouth ulcer &                 Muscle atrophy &                    Muscle weakness &             \\
 &                        Myalgia &                                  Mydriasis &                     Myocardial infarction &                      Myoclonus &                   Nasal congestion &             \\
 &                    Nasal polyp &                                     Nausea &                                 Neck mass &                      Neck pain &                  Neonatal jaundice &             \\
 &                   Nerve injury &                                  Neuralgia &                               Neutropenia &                   Night sweats &                       Night terror &             \\
 &             Nocturnal enuresis &                                     Nodule &                                 Nosebleed &                      Nystagmus &                            Obesity &             \\
 &                  Onychorrhexis &                           Oral candidiasis &                   Orthostatic hypotension &                     Osteopenia &                         Osteophyte &             \\
 &                   Osteoporosis &                                     Otitis &                            Otitis externa &                   Otitis media &                               Pain &             \\
 &                   Palpitations &                               Panic attack &                                    Papule &                       Paranoia &                        Paresthesia &             \\
 &    Pelvic inflammatory disease &                               Pericarditis &                       Periodontal disease &          Periorbital puffiness &              Peripheral neuropathy &             \\
 &                   Perspiration &                                   Petechia &                                    Phlegm &                Photodermatitis &                        Photophobia &             \\
 &                      Photopsia &                           Pleural effusion &                                  Pleurisy &                      Pneumonia &                           Podalgia &             \\
 &                   Polycythemia &                                 Polydipsia &                            Polyneuropathy &                       Polyuria &                       Poor posture &             \\
 &                Post-nasal drip &  Postural orthostatic tachycardia syndrome &                               Prediabetes &                    Proteinuria &                       Pruritus ani &             \\
 &                      Psychosis &                                     Ptosis &                           Pulmonary edema &         Pulmonary hypertension &                            Purpura &             \\
 &                            Pus &                             Pyelonephritis &                             Radiculopathy &                    Rectal pain &                    Rectal prolapse &             \\
 &                        Red eye &                                Renal colic &                    Restless legs syndrome &                          Rheum &                           Rhinitis &             \\
 &                     Rhinorrhea &                                    Rosacea &                       Round ligament pain &                     Rumination &                               Scar &             \\
 &                       Sciatica &                                  Scoliosis &                     Seborrheic dermatitis &                      Self-harm &               Sensitivity to sound &             \\
 &             Sexual dysfunction &                          Shallow breathing &                                Sharp pain &                      Shivering &                Shortness of breath &             \\
 &                        Shyness &                                  Sinusitis &                            Skin condition &                      Skin rash &                           Skin tag &             \\
 &                     Skin ulcer &                                Sleep apnea &                         Sleep deprivation &                 Sleep disorder &                            Snoring &             \\
 &                    Sore throat &                                 Spasticity &                              Splenomegaly &                         Sputum &                     Stomach rumble &             \\
 &                     Strabismus &                              Stretch marks &                                   Stridor &                         Stroke &                         Stuttering &             \\
 &              Subdural hematoma &                          Suicidal ideation &                                  Swelling &                   Swollen feet &                Swollen lymph nodes &             \\
 &                        Syncope &                                Tachycardia &                                 Tachypnea &                 Telangiectasia &                         Tenderness &             \\
 &                Testicular pain &                          Throat irritation &                          Thrombocytopenia &                 Thyroid nodule &                                Tic &             \\
 &                       Tinnitus &                                Tonsillitis &                                 Toothache &                         Tremor &                     Trichoptilosis &             \\
 &                          Tumor &                            Type 2 diabetes &                           Unconsciousness &                    Underweight &  Upper respiratory tract infection &             \\
 &                     Urethritis &                       Urinary incontinence &                   Urinary tract infection &                Urinary urgency &                Uterine contraction &             \\
 &               Vaginal bleeding &                          Vaginal discharge &                                 Vaginitis &                 Varicose veins &                         Vasculitis &             \\
 &       Ventricular fibrillation &                    Ventricular tachycardia &                                   Vertigo &                Viral pneumonia &                      Visual acuity &             \\
 &                       Vomiting &                                       Wart &                           Water retention &                       Weakness &                        Weight gain &             \\
 &                         Wheeze &                                  Xeroderma &                                Xerostomia &                           Yawn &                after covid vaccine &             \\
 &                        anosmia &                                 chest pain &                           chest tightness &                          covid &                          covid nhs &             \\
 &                 covid symptoms &                                   covid-19 &                              covid-19 who &       effects of covid vaccine &                  feeling exhausted &             \\
 &                  feeling tired &                                      fever &                  how long does covid last &                  hyperhidrosis &                      joints aching &             \\
 &                  loss of smell &                                 loss smell &                                loss taste &                     nose bleed &                       pancreatitis &             \\
\bottomrule
\end{tabular}
    \caption{Google Search term list used in the Multiple Source method}
    \label{tab:google_searches_list}
    	\end{adjustwidth}
\end{table}

\clearpage
\newpage
\bibliography{scibib}
\bibliographystyle{unsrt}

\end{document}